\documentclass[twocolumn,numberedappendix]{aastex62}

\usepackage{graphicx,ifthen,url,float,color,ulem,amsmath,ragged2e}
\newcommand {\kms}{km s$^{-1}$}

\newcommand{\scubaii}{\mbox{\sc Scuba-2}}

\def\ltsima{$\; \buildrel < \over \sim \;$}
\def\simlt{\lower.5ex\hbox{\ltsima}}
\def\gtsima{$\; \buildrel > \over \sim \;$}
\def\simgt{\lower.5ex\hbox{\gtsima}}

\newcommand {\um}{$\mu$m}

\def\um     {$\mu$m}
\def\ts     {\thinspace}
\def\kms    {\ifmmode{{\rm \ts km\ts s}^{-1}}\else{\ts km\ts s$^{-1}$}\fi}
\def\msol   {\ifmmode{{\rm M}_{\odot}}\else{M$_{\odot}$}\fi}
\def\lsol   {\ifmmode{{\rm L}_{\odot}}\else{L$_{\odot}$}\fi}
\def\zsol   {\ifmmode{{\rm Z}_{\odot}}\else{Z$_{\odot}$}\fi}
\def\etal   {{\rm et\ts al.}}
\def\ci     {\ifmmode{{\rm C}{\rm \small I}}\else{C\ts {\scriptsize I}}\fi}
\def\hi     {\ifmmode{{\rm H}{\rm \small I}}\else{H\ts {\scriptsize I}}\fi}
\def\hh     {\ifmmode{{\rm H}_2}\else{H$_2$}\fi}
\def\cone {\ifmmode{{\rm C}{\rm \small I}(^3\!P_1\!\to^3\!P_0)}
     \else{C\ts {\scriptsize I}{\small$(^3\!P_1\!\to\,^3\!P_0)$}}\fi}
\def\ctwo {\ifmmode{{\rm C}{\rm \small I}(^3\!P_2\!\to\,^3\!P_1)}
     \else{C\ts {\scriptsize I}{\small$(^3\!P_2\!\to\,^3\!P_1)$}}\fi}
\def\cij {\ifmmode{{\rm C}{\rm \small I}\,(^3P_i\to^3P_j)}\else{C\ts {\scriptsize I}\,{\small$(^3P_i\to^3P_j)$}}\fi}
\def\cii    {\ifmmode{{\rm C}{\rm \small II}}\else{C\ts {\scriptsize II}}\fi}
\def\tex {\ifmmode{{T}_{\rm ex}}\else{$T_{\rm ex}$}\fi}
\def\tmb {\ifmmode{{T}_{\rm mb}}\else{$T_{\rm mb}$}\fi}
\def\tkin {\ifmmode{{T}_{\rm kin}}\else{$T_{\rm kin}$}\fi}
\def\microns {\ifmmode{\mu{\rm m}}\else{$\mu$m}\fi}
\def\nhh   {\ifmmode{n({\rm H}_2)}\else{$n$(H$_2$)}\fi}

\newcommand{\msun}{{\rm\,M$_\odot$}}
\newcommand{\sfr}{{\rm\,M$_\odot$\,yr$^{-1}$}}
\newcommand{\lsun}{{\rm\,L$_\odot$}}
\newcommand{\lstar}{{\rm\,L$_\star$}}
\newcommand{\phistar}{{\rm\,$\Phi_\star$}}

\newcommand{\mum}{$\,\mu$m}

\newcommand{\lpeak}{$\lambda_{\rm peak}$}
\newcommand{\lir}{L$_{\rm IR}$}

\shorttitle{Dust Continuum at the EoR}
\shortauthors{C.~M. Casey et al.}
\begin{document}

\title{\sc The Brightest Galaxies in the Dark Ages: Galaxies' Dust Continuum Emission during the Reionization Era}
\correspondingauthor{Caitlin M. Casey}
\email{cmcasey@utexas.edu}

\author[0000-0002-0930-6466]{Caitlin M. Casey}
\affil{Department of Astronomy, The University of Texas at Austin, 2515 Speedway Blvd Stop C1400, Austin, TX 78712}

\author[0000-0002-7051-1100]{Jorge A. Zavala}
\affil{Department of Astronomy, The University of Texas at Austin, 2515 Speedway Blvd Stop C1400, Austin, TX 78712}

\author[0000-0003-3256-5615]{Justin Spilker}
\affil{Department of Astronomy, The University of Texas at Austin, 2515 Speedway Blvd Stop C1400, Austin, TX 78712}

\author[0000-0001-9759-4797]{Elisabete da Cunha}
\affil{Research School of Astronomy and Astrophysics, The Australian National University, Canberra ACT 2611, Australia}

\author{Jacqueline Hodge}
\affil{Leiden Observatory, Niels Bohrweg 2, 2333 CA Leiden, The Netherlands}

\author{Chao-Ling Hung}
\affil{Department of Physics, Manhattan College, 4513 Manhattan College Pkwy, Bronx, NY 10471}

\author[0000-0002-8437-0433]{Johannes Staguhn}
\affiliation{NASA Goddard Space Flight Center, Code 665, Greenbelt, MD 20771}
\affiliation{Bloomberg Center for Physics and Astronomy, Johns Hopkins University 3400 N. Charles Street, Baltimore, MD 21218}

\author[0000-0001-8519-1130]{Steven L. Finkelstein}
\affil{Department of Astronomy, The University of Texas at Austin, 2515 Speedway Blvd Stop C1400, Austin, TX 78712}

\author[0000-0003-3627-7485]{Patrick Drew}
\affil{Department of Astronomy, The University of Texas at Austin, 2515 Speedway Blvd Stop C1400, Austin, TX 78712}

\label{firstpage}

\begin{abstract}
Though half of cosmic starlight is absorbed by dust and reradiated at
long wavelengths (3\um--3\,mm), constraints on the infrared through
millimeter galaxy luminosity function (the `IRLF') are poor in
comparison to the rest-frame ultraviolet and optical galaxy luminosity
function, particularly at $z\simgt2.5$.  Here we present a backward
evolution model for interpreting number counts, redshift
distributions, and cross-band flux density correlations in the
infrared and submillimeter sky, from 70\um--2\,mm, using a model for
the IRLF out to the epoch of reionization.  Mock submillimeter maps
are generated by injecting sources according to the prescribed IRLF
and flux densities drawn from model spectral energy distributions that
mirror the distribution of SEDs observed in $0<z<5$ dusty star-forming
galaxies (DSFGs).  We explore two extreme hypothetical case-studies: a
dust-poor early Universe model, where DSFGs contribute negligibly
($<$10\%) to the integrated star-formation rate density at $z>4$, and
an alternate dust-rich early Universe model, where DSFGs dominate
$\sim$90\%\ of $z>4$ star-formation.  We find that current submm/mm
datasets do not clearly rule out either of 
 these extreme models. We suggest that
future surveys at 2\,mm will be crucial to measuring the IRLF beyond
$z\sim4$.  The model framework developed in this paper serves as a unique
tool for the interpretation of multiwavelength IR/submm extragalactic
datasets and will enable more refined constraints on the IRLF than can
be made from direct measurements of individual galaxies' integrated
dust emission.
\end{abstract}

\section{Introduction}

The census of cosmic star-formation out to the highest redshifts is a
central goal of galaxy evolution surveys and yet current measurements
are imbalanced, biased towards unobscured star-formation tracers
\citep{madau14a}.  Finding the most distant galaxies, formed less than
a billion years after the Big Bang, is of fundamental importance in
order to observationally test theories of galaxy assembly. This
includes constraining the Population~III stellar initial mass
function, the formation of early dust and metals, and the timescale of
dark matter halo collapse.  Significant effort and work has been
poured into taking census of galaxies detected via their rest-frame
ultraviolet emission
\citep[e.g.][]{schiminovich05a,dahlen07a,reddy09a,bouwens12a,schenker13a,ellis13a,coe13a,oesch13a,finkelstein13a,bouwens15a,finkelstein15a}.
The presence of a strong Lyman break has successfully been used for
redshift identification \citep{steidel96a} out to $z\sim11$
\citep{oesch16a}, revealing a peak in the cosmic star-formation rate
density from $z\sim2-4$ and values more consistent with the local
Universe at earlier times ($z\sim7-10$).  Debates as to the slope of
the cosmic star-formation rate density near the Epoch of Reionization
(EoR) are forming over ever increasing samples of early-Universe
Lyman-break galaxies \citep{oesch13a,oesch14a,mcleod15a}.

While this work in the rest-frame UV, redshifted into the near-IR at
$z>8$, has been pioneering, similar surveys of the early Universe at
long wavelengths have not kept pace.  And yet, this long
wavelength work is necessary in the census of cosmic star-formation,
not least because we know roughly half of the energy from the
extragalactic background radiation is output at long wavelengths.
This is because ultraviolet light from
young, massive stars is absorbed by dust and re-radiated.  And it is
clear that the conditions of the interstellar medium (ISM) and the
environments of star-formation have tremendous impact on whether or
not galaxies will appear largely unobscured or heavily obscured, thus
whether or not they are counted in existing surveys.  Due to their
very high star-formation rates and thus extreme levels of obscuration
\citep[e.g.][]{brinchmann04a,whitaker17a}, dusty star-forming galaxies
\citep*[DSFGs;][]{casey14a} are largely absent from the optical census
of cosmic star-formation. Though there are some DSFGs
  that may appear in optical surveys as LBGs, often their rest-frame
  UV colors imply very little dust, thus star-formation rates that are
  factors $\sim$100 times lower than implied by their long-wavelength
  emission \citep{casey14b}.  While locally the population of bright
DSFGs (SFRs$\simgt 100$\,\sfr) is negligible, at
$z\sim2-3$ the population is
over one-thousand-fold more common and becomes the dominant factories
of star-formation in the early Universe.  Therefore, taking census of
the Universe's star-formation history requires a bolometric approach,
analyzing galaxy populations detected via their direct starlight and
those via their dust emission.

Galaxy surveys at long wavelengths have naturally been more limited by
instrumentation and the additional hurdles involved in identifying
galaxies' redshifts -- a characteristic which, for the LBG samples, is
inferred directly from the observations used in their selection.  From
single-dish submillimeter and millimeter surveys, large beamsizes have
obfuscated the identification of precise multiwavelength counterparts
\citep{smail97a,hughes98a,barger98a,chapman03a}. Even when
multiwavelength counterparts are identified, redshift confirmation can
be extremely challenging with low yields
\citep{chapman05a,casey12b,casey12c,danielson17a,casey17a}.  Only
recent wide bandwidth receivers in the millimeter have made it
possible to spectroscopically confirm high-$z$ DSFGs without laborious
and observationally expensive multiwavelength campaigns
\citep{bradford09a,vieira13a}.  Spectroscopic follow-up in the mm of
the most luminous class of DSFGs has interestingly revealed a
prominent population of sources at $3<z<7$
\citep{weis13a,riechers13a,strandet17a,marrone17a}, but such surveys
have not yet become efficient for large samples of less luminous
(unlensed) DSFGs.

\begin{figure}
\includegraphics[width=0.99\columnwidth]{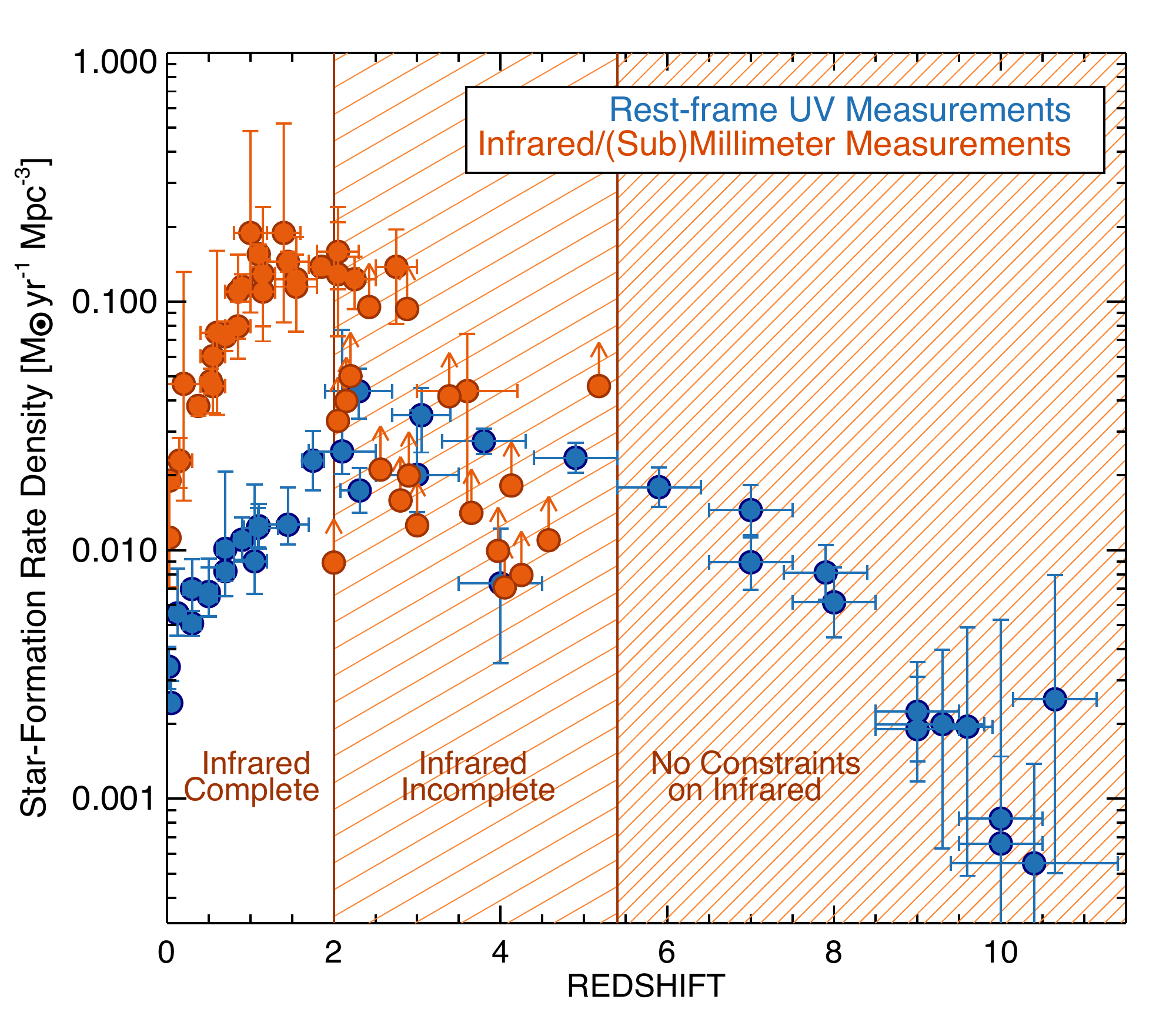}
\caption{The cosmic star-formation history of the Universe as measured
  at rest-frame UV wavelengths \citep[blue
    points;][]{schiminovich05a,dahlen07a,reddy09a,bouwens12a,schenker13a,ellis13a,coe13a,oesch13a,finkelstein13a,bouwens15a,finkelstein15a},
  and infrared through millimeter measurements \citep[orange
    points;][]{le-floch05a,magnelli11a,gruppioni13a,casey12a,casey12b,casey13a,barger12a,roseboom13a,chapman05a,wardlow11a}
  from facilities like the {\it Herschel Space Observatory}, SCUBA,
  and AzTEC.  While far-infrared/(sub)mm surveys (globally referred to
  as the IR) have mapped obscured star-formation with individual
  galaxy detections out to $z\sim7$ \citep{strandet17a,marrone17a},
  there are few constraints on their SFRD contribution at $z\simgt2.5$
  due to sample incompleteness.  Our current understanding of
  star-formation in the early Universe is severely limited by the lack
  of IR constraints, particularly beyond $z\sim4$.}
\label{fig:sfrd}
\end{figure}

Knowing the prevalence of dust-obscured star-formation is particularly
important at $z>4$, when cosmic time becomes a constraint on the
physical processes involved in producing dust, metals and stars seen
in galaxies.  For example, \citet{capak15a} show a marked difference
in the dust-to-gas ratio for a population of $z\sim5$ normal
star-forming galaxies; however, their sample was exclusively selected
via rest-frame UV and optical surveys, which are biased towards low
dust content.  Unfortunately, current deep-field {\it HST} surveys are
blind to $z>4$ sources with $\simgt$50\,\sfr\ due to pencil-beam sky
coverage limiting the dynamic range of observable galaxies. Existing
samples of DSFGs at these redshifts are extremely bright
($>$2000\,\sfr) and come from extremely shallow, biased
surveys. The South Pole Telescope sample of DSFGs
  \citep[e.g.][]{vieira13a}, though less biased with color or SFR, is
  dominated by gravitationally-lensed sources whose volume density is
  nearly impossible to measure. Therefore, there is almost no
constraint on the contribution of obscured star-formation to the
cosmic star-formation rate density at $z\simgt2$, and absolutely no
constraint beyond $z\sim5.5$ (see Figure~\ref{fig:sfrd}).

Identifying high-redshift obscured galaxies has proven to be
particularly challenging. IR color selection, like the technique used
to identify {\it Herschel} 500\um-risers
\citep{pearson13a,dowell14a,ivison16a}, seems to provide an effective
route to several exciting, high-$z$ discoveries
\citep[e.g.][]{oteo17a,zavala18a}, but the nature of source selection
and follow-up make it difficult to back out any information on
underlying population statistics.  On the other hand, deep blank-field
ALMA campaigns \citep{dunlop16a,walter16a}, which are
  not based on color
selection, have failed to yield a population of very high-redshift
sources.

In this paper, we describe a model for the far-infrared through
millimeter emission of galaxies from $z=0$ to $z\sim10$ to explain the
results of (sub)mm single-dish survey campaigns to-date.  An
accompanying paper presents results of analysis of the same models on
scales observable with sensitive interferometers like ALMA (Casey
\etal\ 2018, submitted).
This paper follows other literature works which present similar models
of the Universe's (sub)mm emission, including the Simulated Infrared
Dusty Extragalactic Sky (SIDES) and its predecessor models
\citep{bethermin12a,bethermin17a}, as well as the work of
\citet{zavala14a}, which explains the different redshift distributions
of (sub)mm-selected populations with a single underlying source
population.
We explore some of the strengths of each of these models by
analyzing differences in the {\it a priori} assumptions and
approaching from a different perspective focused on the total infrared
through millimeter galaxy luminosity function (henceforth referred to
as the IR luminosity function, or IRLF, in this paper).

We use existing measurements of submm number counts, redshift
distributions, and multi-band flux information collated from across
the literature to comment on the shape and behavior of the IR
luminosity function of galaxies (from dust continuum) out to the epoch
of reionization.  Two extreme case studies are used to frame this
discussion and outline goals of future work.  One case study assumes a
dust-poor early Universe, similar to existing models used by the
rest-frame UV community, while the other assumes a dust-rich early
Universe.  It is important to point out that here dust-rich does not
refer to the content of all galaxies uniformly, but rather, the
abundance of very dust-rich DSFGs relative to UV-bright galaxies.  The
construction of the model framework and its assumptions are described
in \S~\ref{sec:model},
and we compare our results
with literature datasets and other models in 
\S~\ref{sec:compare}.  The implications of our constraints are
discussed fully in \S~\ref{sec:discussion} and in
\S~\ref{sec:conclusion} we conclude.  We assume a {\it Planck}
cosmology throughout this paper, adopting
$H_{0}=67.7\,$\kms\,Mpc$^{-1}$ and $\Omega_{\lambda}=0.6911$
\citep{planck-collaboration16b}, and where SFRs are alluded to, we
assume a Chabrier IMF \citep{chabrier03a}.

\begin{figure}
\includegraphics[width=0.99\columnwidth]{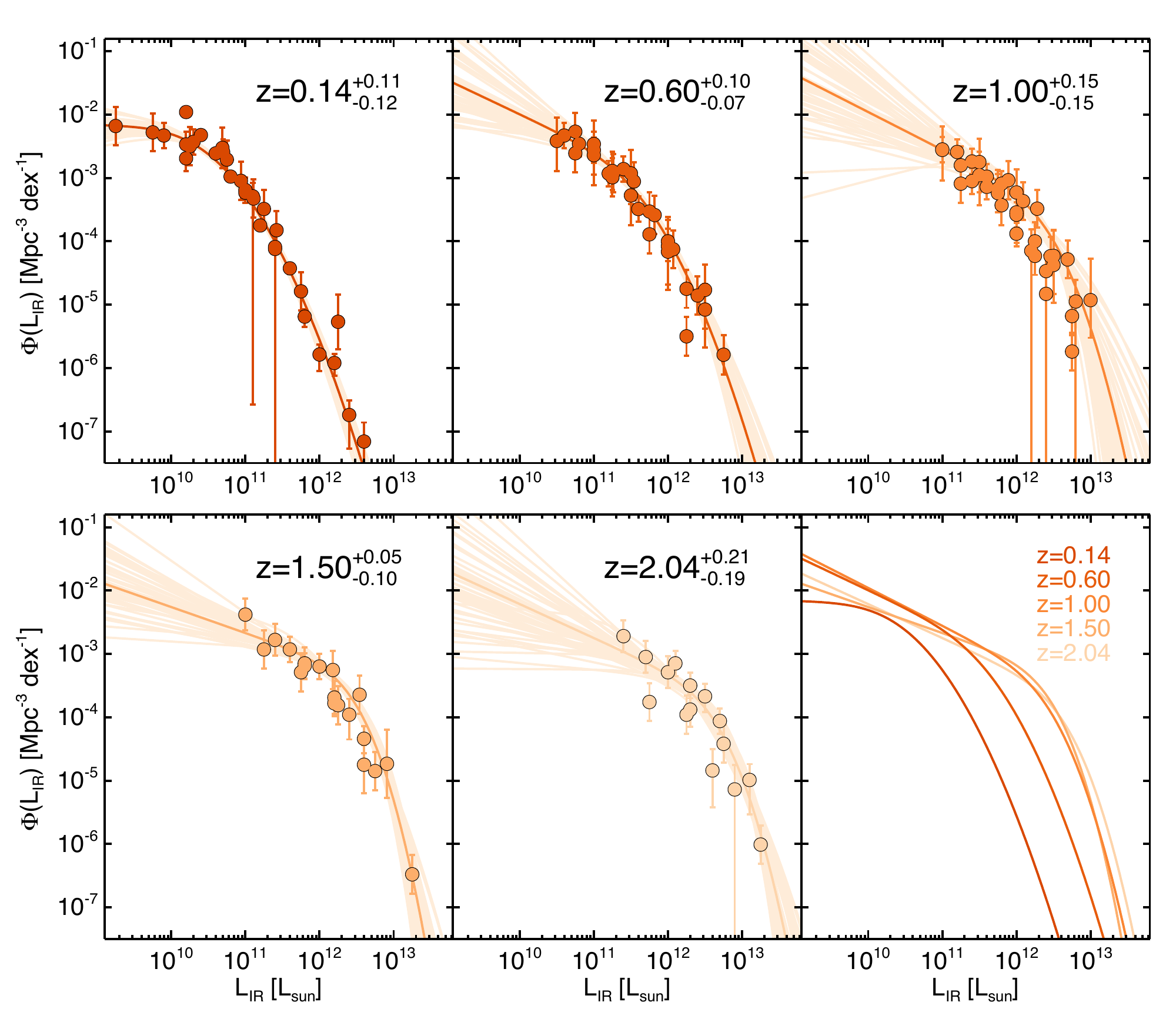}
\caption{ A summary of integrated L$_{\rm IR}$ luminosity functions in
  the literature, as collated in \citet*{casey14a}.  Original data is
  from:
  \citet{sanders03a,le-floch05a,casey12b,gruppioni13a,magnelli11a,magnelli13a}.
  Five redshift ranges are shown: $z=0.14^{+0.11}_{-0.12}$,
  $0.60^{+0.10}_{-0.07}$, $1.00\pm0.15$, $1.50^{+0.05}_{-0.10}$, and
  $2.04^{+0.21}_{-0.19}$, and a sixth panel shows the relative
  evolution between them.  Though these fits highlight a continuous double
  powerlaw form, the broken double powerlaw is statistically
  indistinguishable and we adopt it for its simplicity for the rest of
  this work.
\vspace{2mm}
 }
\label{fig:lf}
\end{figure}

\section{Model Construction}\label{sec:model}

Our backward-evolution model provides a prediction of far-infrared and
submillimeter flux number counts (from 70\um\ through 2\,mm), redshift
distributions and overlaps in populations, given a parameterized,
evolving galaxy luminosity function.  This model is empirically-driven
and motivated by existing measurements of the galaxy luminosity
function in the infrared and their measured SED characteristics.  
What follows here is a step-by-step detailed description of the model,
beginning with the luminosity function.
A table summarizing all of the model assumptions, including the
equations described below, is given in Table~\ref{tab:summary} at the
end of the description of each component.

\subsection{The IR Luminosity Function}\label{sec:lf}

Figure~\ref{fig:lf} shows a summary of measured galaxy luminosity
functions in the IR, as collated in \citet*{casey14a}.  There is
strong luminosity evolution evident in these data, with the
possibility of some minor evolution in galaxy number density.  The
shape of the IRLF is poorly constrained relative to
the rest-frame UV/optical luminosity function of galaxies.  We show a
continuous double powerlaw fit in Figure~\ref{fig:lf}, but emphasize
that there is no statistical difference between adopting a continuous
double powerlaw and a broken double powerlaw. 
A Schechter function is deemed inappropriate for these
IR-luminous galaxies because the bright-end falls off gradually and
not exponentially. Due to its simplicity and and intuitive nature, we
adopt a broken double powerlaw model for $\Phi$ which is a function of
both redshift $z$ and IR luminosity, which we will simply denote $L$:
\begin{equation}
\Phi(L,z) = \left\{
\begin{array}{lr}
\Phi_{\star} \left(\frac{L}{L_{\star}}\right)^{\alpha_{\rm LF}} & : L<L_{\star} \\
\Phi_{\star} \left(\frac{L}{L_{\star}}\right)^{\beta_{\rm LF}} & : L\ge L_{\star} \\
\end{array}
\right.
\label{eq:lf}
\end{equation}
It is clear that $L_\star$ evolves strongly with redshift (as shown in
Figure~\ref{fig:lf}), and it is possible that \phistar, $\alpha_{\rm LF}$ and
$\beta_{\rm LF}$ also have some redshift dependence, although there is little
data to constrain this currently (though our accompanying ALMA-focused
paper addresses possible evolution in $\alpha_{\rm LF}$ with redshift and
implications on detections in small ALMA deep fields).  The units of
$\Phi$ are Mpc$^{-3}$ dex$^{-1}$ and $L$ and $L_\star$ are in \lsun.
We discuss the values and nature of the redshift evolution of these
parameters in \S~\ref{sec:zevolution}.

\begin{figure*}
\includegraphics[width=1.06\columnwidth]{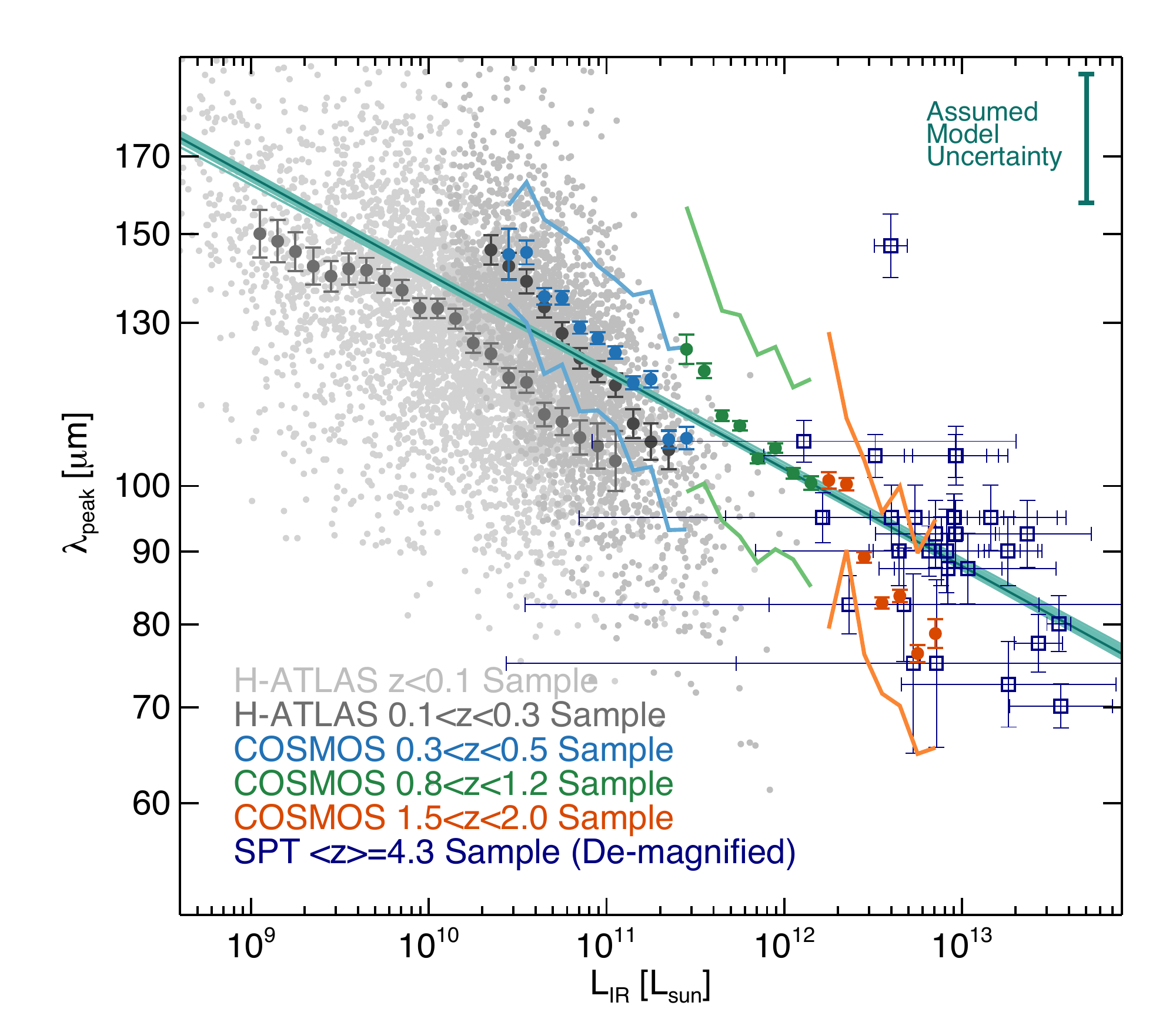}
\includegraphics[width=0.93\columnwidth]{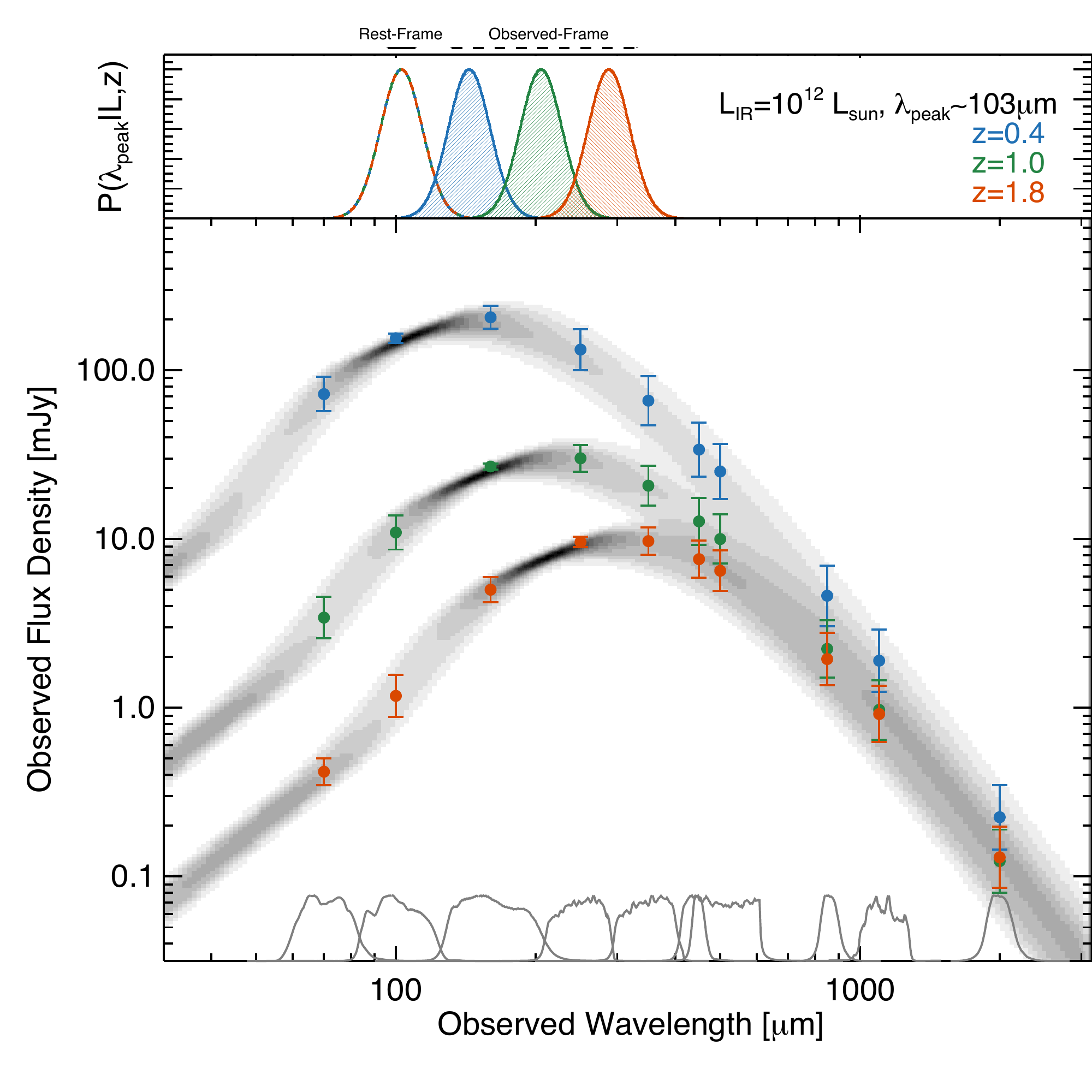}
\caption{{\sc Left:} The relationship between luminosity and dust
  temperature, shown here in observable quantities: \lir\ and
  rest-frame peak wavelength \lpeak\ of S$_\nu$.  The local sample
  \citep{valiante16a} is shown as gray points in two redshift bins
  $z<0.1$ (lightest gray) and $0.1<z<0.3$ (light gray).  Darker gray
  points are the median values of \lpeak\ at a given \lir\ for each
  sample.  Higher redshift galaxies \citep{lee13a} sit on the
  extension of this relationship toward higher luminosities; 1$\sigma$
  scatter is shown as light blue, green and orange lines. Overplotted
  are the sample of South Pole Telescope DSFGs with well-measured SEDs
  and constrained magnification factors \citep{strandet16a} with a
  median redshift of $\langle z\rangle=4.3$.  The adopted model is
  overplotted as a teal line with associated fit uncertainty, and the
  scatter about that model used to generate a diversity of SEDs at all
  redshifts and shown in the upper right.
{\sc Right:} Given a redshift and \lir, here we show an example of how
we estimate far-IR through millimeter flux densities to inject into
our model maps.  The example sources have a fixed \lir=10$^{12}$\lsun,
but sit at different redshifts, $z=0.4$, $1.0$ and $z=1.8$.  The range
of rest-frame peak wavelengths is represented by the open Gaussian
distribution at top and the observed-frame \lpeak\ by the filled
distributions at each redshift.  Below, we generate 1000 SEDs for each
probability distribution in \lpeak\ given \lir, and overplot
the median predicted flux densities across the far-IR/mm bands.  The
filter profiles of the bands we use for this simulation are inset and
described in the text.}
\label{fig:tdust}
\end{figure*}

\subsection{Galaxies' IR Spectral Energy Distributions}\label{sec:sed}

Modeling the multi-wavelength (sub)millimeter emission of galaxies
requires a keen understanding of their spectral energy distributions
(SEDs) in addition to the underlying galaxy luminosity function.  Dust
radiative transfer models \citep{silva98a,dopita05a,siebenmorgen07a}
and observations of local IR-luminous galaxies \citep{u12a} show that
the far-infrared/submillimeter SEDs of galaxies are well-represented by
a single modified blackbody, with additional emission in the
mid-infrared representing the emission of less massive and more
concentrated pockets of warm to hot dust in the galaxy's ISM.
Emission from polycyclic aromatic hydrocarbons (PAHs) can also
dominate this mid-infrared regime, contributing as much as 10\%\ to
the total integrated IR luminosity of galaxies.  While many works in
the literature use detailed empirically-driven templates
\citep{chary01a,dale01a,dale02a,draine07a,rieke09a} or energy-balance
techniques \citep{burgarella05a,da-cunha08a,da-cunha13a,noll09a} to
model the emission of high-$z$ submm-detected galaxies, the detail of
these models goes beyond the constraints of existing data for large
DSFG samples.  

For the purposes of this paper, we adopt a very simple four parameter
mid-infrared powerlaw + modified blackbody \citep{blain03a} fit as
described in \citet{casey12a}.  The free parameters of the model are
the luminosity $L$ (the integral under the curve, roughly scaling to
its normalization), the dust temperature $T_{\rm dust}$ (related to
the wavelength where the SED peaks, $\lambda_{\rm peak}$), the
mid-infrared powerlaw slope $\alpha_{\rm MIR}$, and the emissivity
spectral index $\beta_{E}$ (we give it the subscript to distinguish
with $\beta_{\rm LF}$, the bright-end slope of the IRLF).  For the
purposes of our model we fix the latter two parameters to $\alpha_{\rm
  MIR}=2.0$ and $\beta_{\rm E}=1.8$ in line with the average
constraints from well-characterized galaxies both in the nearby and
distant Universe \citep[e.g.][]{paradis10a}.  The adoption of
$\alpha_{\rm MIR}=2.0$ measured as the median mid-IR slope for GOALS
galaxies \citep{u12a} accounts for both hot dust emission and PAH
emission via its integral, though does not directly spectrally model
the PAH features because we determine this to only have a significant
effect on galaxy observability at rest-frame wavelengths $<$10\,\um,
which makes up a negligible fraction of the total power in the bands
analyzed in this paper.

Note that the dust temperature, $T_{\rm dust}$, represents the
temperature of the cold dust in the ISM, and the temperature input for
the dominating cold-dust modified blackbody component of the SED.  Its
relationship to the peak wavelength of the SED, \lpeak, depends on the
adopted dust opacity model; in this work we assume that $\tau=1$ at
$\lambda_{\rm rest}=100$\um\ \citep{conley11a}, whereby the blackbody
is optically thick at shorter wavelengths and optically thin at longer
wavelengths.  This is consistent with observations of DSFGs in the
local Universe; there is little evidence to suggest that this would
not also hold for DSFGs in the early Universe.  Though a different
assumption of opacity will have a dramatic impact on the relationship
between $T_{\rm dust}$ and \lpeak\ \citep*[see Figure 20
  of][]{casey14a}, we choose to parameterize our model using
\lpeak\ instead of $T_{\rm dust}$. This choice makes our SEDs
insensitive to different adopted opacity models.  Hence the rest of
this paper discusses the idea of temperature only through the
measurable quantity \lpeak.

Figure~\ref{fig:tdust} shows the existing empirical constraints on the
peak SED wavelength of IR-luminous galaxies as a function of $L$ and
$z$.  It has been known for some time that there is a direct
correlation between galaxies' intrinsic IR luminosity (or total
star-formation rate) and their observed peak of the IR SED, or dust
temperature \citep{sanders03a,chapman03c}.  Galaxies with higher IR
luminosities have intrinsically hotter luminosity-weighted dust
temperatures, or lower values of \lpeak.  We show the extensive data
from the H-ATLAS survey \citep{valiante16a} largely encompassing
galaxies from $0<z<0.5$ and with {\it Herschel} SPIRE-detected galaxies in
the COSMOS field extending out to $z\sim2$ \citep{lee13a}.  Though
samples at higher redshifts are sparse, we draw on the
\citet{strandet16a} compilation of statistics on the South Pole
Telescope (SPT)-detected DSFGs that have well-constrained SEDs and a
median redshift of $\langle z\rangle=4.3$; these high-redshift DSFGs
seem to follow the same broad trend, where higher luminosity galaxies
have hotter SEDs.

All SEDs of galaxies from the literature are refit using the
\citet{casey12a} SED fitting method to provide uniform analysis of
their characteristics. We test for biases introduced by limited band
coverage, only including galaxies with sufficiently robust photometric
measurements above detection limits where selection biases are
negligible (more details are provided in \S~\ref{sec:appendsed}).
The \lir-\lpeak\ relationship can be modeled by a powerlaw
such that:
\begin{equation}
\langle\lambda_{\rm peak}(L)\rangle=\lambda_{\rm 0}\Big(\frac{L}{L_{t}}\Big)^{\eta} 
\label{eq:lirlpeak}
\end{equation}
We measure $\lambda_{\rm 0}=102.8\pm0.4$\um\ at
$L_{\rm t}\equiv10^{12}$\,\lsun\ and $\eta=-0.068\pm0.001$ from the
aggregate samples plotted on Figure~\ref{fig:tdust}.  We do note that
the local galaxy sample, particularly at $z<0.1$ appears offset from
the best-fit \lir-\lpeak\ relationship toward slightly warmer
temperatures. We discuss this deviation further in
Appendix~\ref{sec:appendsed} but argue here that the shift seen in
low-redshift galaxies does not impact either our number counts or our
inferred redshift distributions for sources found on 1\,deg$^2$
scales.  This is primarily because of the relative rarity of
$z\simlt0.3$ dust-obscured galaxies relative to $z>1$ sources.

Physically, this \lir-\lpeak\ relationship can be thought of as a
galaxy-scale Stefan-Boltzmann law for the cold ISM.  While a direct
translation of Stefan-Boltzmann to \lir-\lpeak\ space would imply a
value of $\eta=-0.25$, variations in galaxy shape (which are certainly
not spherical or emitting isotropically), as well as a correlation of
galaxies' sizes (or effective surface areas) with luminosity, and the
fact that galaxies' dust emission is not a perfect blackbody, lead to
the shallower value of $\eta=-0.068\pm0.001$.
While this provides some context for shallower $\eta$ slopes, we
emphasize that it is an empirically measured quantity as taken from
the samples in Figure~\ref{fig:tdust}.  Folding in the
  assumed dust emissivity index, $\beta_{E}=1.8$, $\eta$ would change
  to $-0.17$, and radiative transfer modeling suggest similarly
  intermediate slopes of $\eta$, for example $\eta\approx-0.16$ found
  in \citet{siebenmorgen07a}, who use a spherically symmetric model
  accounting for different star-formation rates, sizes and dust
  masses.  A more in depth analysis of the \lir-\lpeak\ relationship
is needed to understand the physical drivers behind the observed
trends and possible evolution, but that is beyond the scope of this
work. The exact value of $\eta$ becomes less important
  when considering the observed scatter of real galaxies about the
  relation.

The average model uncertainty adopted, $\sigma_{\langle\log\lambda_{\rm
    peak}\rangle}=0.045$ (corresponding to $\Delta{\rm \lambda_{\rm
    peak}}/\lambda_{\rm peak}\approx10$\%), is shown in the upper left
of Figure~\ref{fig:tdust} and is derived from the average deviation of
individual galaxies about the median \lir-\lpeak\ relationship.  Each
galaxy in the model is assigned a dust temperature, or \lpeak\ value,
according to the probability density function in $\log\lambda_{\rm
  peak}$ for that galaxy's redshift and $L$.

\subsection{Relating luminosity to the SED}

The next step in building our model is to generate fake maps of the
sky at a variety of far-IR/submm wavelengths, and to do that we must
generate a list of input sources drawn from our luminosity functions
and use our data constraints to assign a best-guess far-infrared SED.
For example, if we are to inject one source into the map with
\lir=10$^{12}$\,\lsun, we can predict that its rest-frame peak
wavelength is close to $\langle$\lpeak(L$|$z)$\rangle=$103\,\um.
Thus, each injected source is `assigned' a peak wavelength (and thus
far-IR SED) after drawing from a normal probability distribution in
$\log\lambda_{\rm peak}(L)$ with width $\sigma$.  Such a distribution
in rest-frame \lpeak\ is shown in the top right panel of
Figure~\ref{fig:tdust} (unfilled).  The implied distributions in
observed-frame peak wavelength are shown as filled histograms at each
redshift.  With a peak wavelength drawn from the probability
distribution in $\log\lambda_{\rm peak}$, an SED is constructed using
the \citet{casey12a} analytic approach and flux densities are measured
across the far-infrared through millimeter bands with their filter
bandpasses.  Our initial phase of modeling includes the following
filters: {\it Herschel} PACS at 70\um, 100\um, and
160\um\ \citep{poglitsch10a}, {\it Herschel} SPIRE at 250\um, 350\um,
and 500\um\ \citep{griffin10a}, SCUBA-2 at 450\um\ and
850\um\ \citep{holland13a}, AzTEC at 1.1\,mm \citep{wilson08a}, a
hypothetical TolTEC filter at 1.4\,mm, and GISMO at 2\,mm
\citep{staguhn14a}.  We also note that ongoing 2\,mm surveys at the
IRAM\,30\,m telescope have begun with the NIKA-2 instrument
\citep{catalano16a}; the 2\,mm beamsize at IRAM is 16.5$''$.  This is
not sufficiently different than the beamsize with the LMT as to cause
differences in the measured 2\,mm number counts, but of course does
matter in the identification of multiwavelength counterparts.  An
accompanying paper models emission in ALMA bands 3, 4, 6, and 7.
Given the lack of instrumentation available at 3\,mm on single-dish
facilities to map large areas, plus the need to push deeper at 3\,mm
to detect galaxies of matched luminosity as those found in 1\,mm or
2\,mm, we do not model 3\,mm single-dish continuum number counts in
this paper.  However, the accompanying paper analyzing modeled ALMA
datasets does find that the 3\,mm channel can be quite useful in
constraining source densities at high-redshift.


\subsection{Impact of the CMB at Long Wavelengths}\label{sec:cmb}

With the goal of estimating the dust continuum emission of galaxies
near the epoch of reionization, it is important to consider the effect
of heating from the cosmic microwave background (CMB).
\citet{da-cunha13a} explore the impact of the CMB on dust continuum
and CO observations in detail and we refer the reader to their paper
for more contextual background.  Towards higher redshifts, the
temperature of the CMB itself was sufficiently large to heat the
internal ISM of galaxies forming during that epoch (where $T_{\rm
  CMB}\sim T_{\rm dust}$).  This causes a boost in the submillimeter/millimeter
output of the ISM, however, observationally this results in a net loss
in flux density (compared to the absence of the CMB) because galaxies
must always be detected in contrast to the CMB thermal background.  To
summarize the discussion in \citet{da-cunha13a}, we must alter the
fitted dust temperature for sources according to this effect, first by
adjusting their internal dust-temperature:
\begin{equation}
T_{\rm dust}(z) = \big( (T_{\rm dust}^{z=0})\!^{4+\beta_{E}}
 + T_{\rm
  CMB}^{z=0}\!^{4+\beta_{E}}[(1+z)^{4+\beta_{E}}-1]\big)^{\frac{1}{4+\beta_{E}}}
\label{eq:cmb1}
\end{equation}
Here $T_{\rm dust}^{z=0}$ represents the dust temperature the galaxy
would have at $z=0$ or in the absence of the CMB, the emissivity
spectral index is taken to be $\beta_{E}=1.8$, and $T_{\rm
  CMB}^{z=0}=2.73$\,K.  Then the fraction of the flux density that is
observable against the CMB background is:
\begin{equation}
f(z,T_{\rm dust}) = 1 - \frac{B_{\nu}[T_{\rm CMB}(z)]}{B_{\nu}[T_{\rm dust}(z)]}
\label{eq:cmb2}
\end{equation}
In other words, this is the ratio of the galaxy's observed flux
density against the CMB versus what the galaxy's flux density would be
in the absence of the CMB.  Procedurally, we do this by first
computing a galaxy's SED as it would be in the absence of the CMB, and
then we fold in this effect by multiplying that flux density by the
factor in Eq.~\ref{eq:cmb2} that effectively shifts the peak towards
warmer temperatures and lower flux densities.  Note that this
adjustment {\it does} depend on the input dust temperature of our
model, $T_{\rm dust}$, and not the observable peak wavelength,
$\lambda_{\rm peak}$; this means that our assumptions about opacity --
that SEDs are optically thick to rest-frame $\sim$100\um-- impact the
perceived impact of the CMB at high-$z$.  For a galaxy that peaks at
rest-frame 100\um, the difference between an optically thick blackbody
and optically thin blackbody is $\sim$10\,K, which translates to about
a 10\%\ difference in anticipated impact of the CMB on that galaxy's
SED.

The impact of the CMB is most prominent at $z>5$ and $\lambda_{\rm
  obs}>1$\,mm.  The effect is not uniform for all galaxies at this
epoch, however, as some will have intrinsically warmer temperatures
than others.  If the \lir-\lpeak\ trend seen in Figure~\ref{fig:tdust}
holds (in some form) at high-redshift, then this will result in the
lowest luminosity galaxies falling below the detection limit out of
our survey, leaving only the brighter galaxies with intrinsically
warmer temperatures to be detected.  This has some important
implications on the search for dust continuum emitters towards the
EoR, which is discussed more in the context of our results in
\S~\ref{sec:discussion}.

Because this paper primarily focuses on galaxies above \lstar,
detectable with single-dish submillimeter facilities on deg$^2$
scales, we favor the  opacity model that
  includes self-absorption on the Wien tail as indicated
 in Table~\ref{tab:summary}. We note that an optically thin
  assumption would only alter the resulting flux densities (after
  correction for the CMB) a small amount for these characteristically
  luminous sources, since they are likely to be significantly hotter
  than the CMB at most redshifts.   For example, the CMB will
    result in a $\approx$30\%\ flux density reduction for
    10$^{12}$\,\lsun\ sources at $z\sim6$, and the difference in the
    deficit between the optically thin model and general opacity model
    is of order 10\%.

\subsection{Impact of AGN Dust Heating and Synchrotron Emission}

One real effect that is not explicitly baked into the model is the
impact of AGN.  The shortest-wavelength bands, 70--160\,\um, are
significantly dominated by DSFGs containing AGN at $z\approx1-2$
\citep{kartaltepe12a,kocevski12a,brown18a}.  Additional dust-heating
by AGN in the vicinity of the central dust torus to temperatures of a
few $\sim$100\,K typically flatten out the mid-infrared spectrum, to
$\alpha_{\rm MIR}\approx1-1.5$.  This additional emission is not added
into our model directly, but needs to be accounted for after the fact.
To do this, we use measurements of $0<z<2$ AGN luminosity functions
\citep{lacy15a} as measured in the mid-infrared, and randomly draw
sources at the same redshifts and source densities and reassign their
flux densities to account for shallower mid-infrared slopes (which we
assign to be $\alpha_{\rm MIR}=1.5$ at $L_{\rm IR}=10^{11}$\,\lsun, up
to $\alpha_{\rm MIR}=1$ at $L_{\rm IR}=10^{13}$\,\lsun).  This
effectively provides a boost of order 1.1--2.0$\times$ to the flux
densities in the {\it Herschel} {\sc Pacs} bands, and does not impact
any of the longer-wavelength bands.

While AGN might also be thought to possibly contribute to sources
detected at long-wavelengths ($>$1\,mm) through radio-loud synchrotron
emission, the number counts generated from such sources should be
quite low in surveys $\sim$1\,deg$^2$ of the depths we explore
\citep{de-zotti05a,tucci11a}.  Such radio-loud quasars would become
much more dominant at higher flux densities covering much larger
areas, like those explored by the South Pole Telescope.  We exclude
such sources from our model since we are primarily focused on
exploring the prevalence of DSFGs in the $\sim$mJy flux density
regime. 

\subsection{Redshift Evolution of the model}\label{sec:zevolution}

We build the majority of uncertainty of our model into the galaxy
luminosity function, such that its evolution with redshift is
unconstrained beyond $z\simgt2.5$, but it must be modeled in order to
reproduce millimeter deep field number counts, redshift distributions,
and correlation of flux densities for sources between different
selection wavelengths.  

\begin{figure}
\includegraphics[width=0.99\columnwidth]{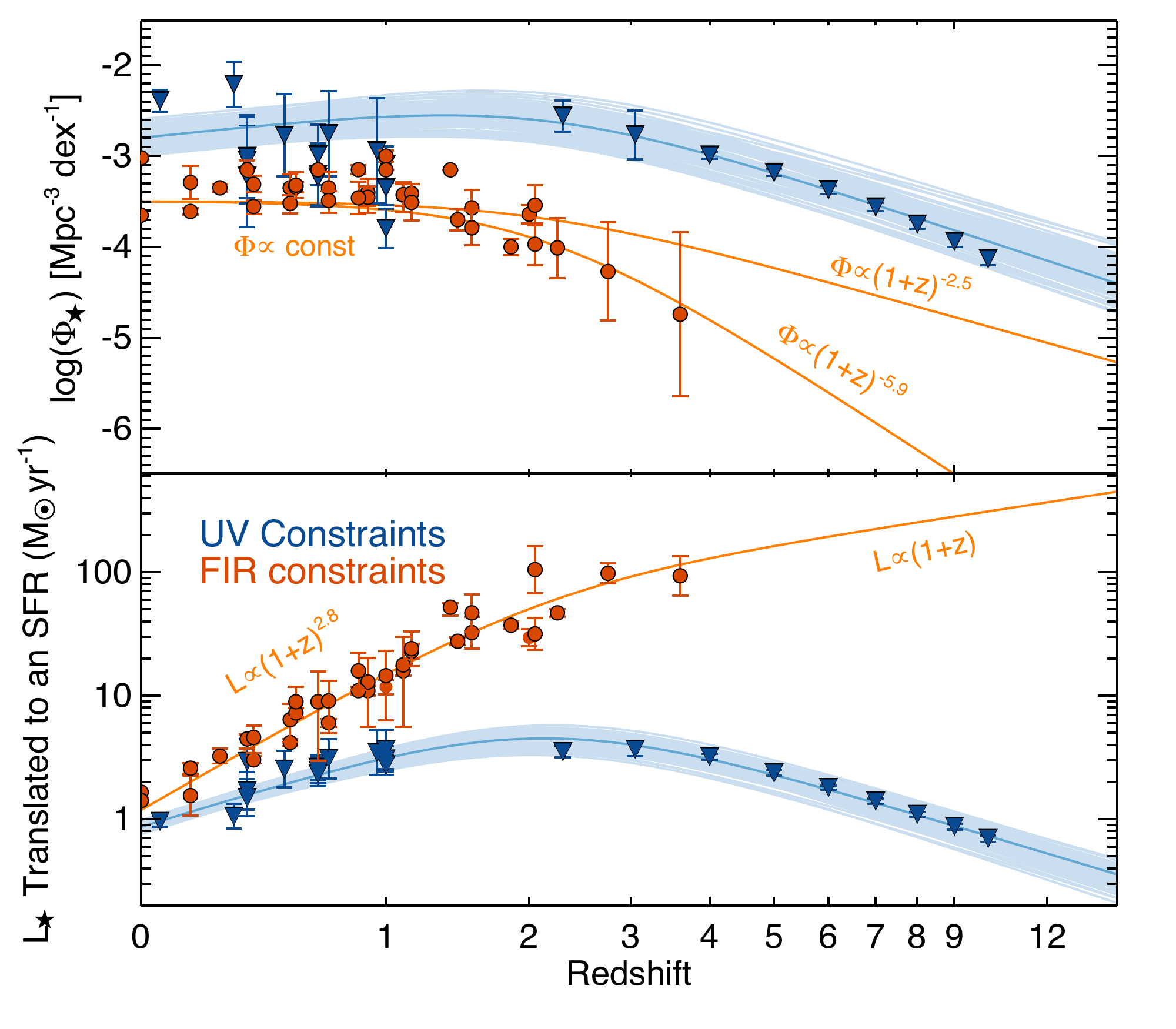}
\caption{A comparison of the luminosity function parameters
  \lstar\ and \phistar\ from the rest-frame UV community, and from the
  FIR/submm community.  Data relevant to the rest-frame UV luminosity
  function is gathered from \citet{arnouts05a}, \citet{reddy09a}, and
  \citet{finkelstein16a}, where M$_\star$ has been translated from a
  magnitude to a SFR for direct comparison to the IR data.  Data from
  the FIR/submm community comes from \citet{le-floch05a},
  \citet{caputi07a}, \citet{goto10a}, \citet{magnelli11a},
  \citet{magnelli13a}, and \citet{gruppioni13a}. Values for IR data
  have been renormalized to match at $z\sim0$.  The shaded blue region
  shows a range of plausible models for the UV LF of the form shown in
  Eq's~\ref{eq:lstarevol} \&~\ref{eq:phistarevol} using $z_{\rm
    turn}=3.5$.  The adopted parameterizations in this paper are shown
  in light orange.  The primary difference in proposed outcomes is the
  high-redshift evolution of \phistar, either evolving steeply
  $\propto(1+z)^{-5.9}$ (Model A) or more gradually
  $\propto(1+z)^{-2.5}$ (Model B).}
\label{fig:lfevol}
\end{figure}

Our model posits that the evolution of \lstar\ follows:
\begin{equation}
L_{\star}(z) \propto \left\{
\begin{array}{lr}
(1+z)^{\gamma_{1}} & : z \ll z_{\rm turn} \\
(1+z)^{\gamma_{2}} & : z \gg z_{\rm turn} \\
\end{array}
\right.
\label{eq:lstarevolform}
\end{equation}
And similarly, that the evolution of \phistar follows:
\begin{equation}
\Phi_{\star}(z) \propto \left\{
\begin{array}{lr}
(1+z)^{\psi_{1}} & : z \ll z_{\rm turn} \\
(1+z)^{\psi_{2}} & : z \gg z_{\rm turn} \\
\end{array}
\right.
\label{eq:phistarevolform}
\end{equation}
To achieve these conditions with a smooth transition at a `turnover'
redshift, $z_{\rm turn}$, we gradually transition from one redshift
dependence to the other over a redshift interval that has thickness
$z_{\rm w}$. For example, \lstar\ might evolve like
$(1+z)^{\gamma_{1}}$ up to $z\sim1.5$, and then transition to
$(1+z)^{\gamma_{2}}$ gradually by a redshift of $z\sim3.5$ (in this
example, $z_{\rm turn}=2.1$ and $z_{\rm w}=2.0$ are adopted as
appropriate ballpark estimates for one of our two models).  We
parameterize this in terms of $x$ such that:
\begin{equation}
\begin{array}{l}
x\equiv\log_{10}(1+z)\\
x_{t}\equiv\log_{10}(1+z_{\rm turn})\\
x_{w}\equiv \frac{z_{w}}{\ln(10)(1+z_{\rm turn})}\\
\end{array}
\label{eq:def}
\end{equation}
and then \lstar\ evolves with $x$ like:
\begin{equation}
\begin{split}
\log_{10}L_\star(x) = & \frac{(\gamma_2-\gamma_1)x_{w}}{2\pi}  \Big[\ln\big(\cosh(\pi\frac{x-x_{t}}{x_{w}})\big)\\
 &  - \ln\big(\cosh(-\pi\frac{x_{t}}{x_{w}})\big)\Big]\\
& + \frac{(\gamma_2-\gamma_1)}{2}x + \log_{10}(L_{0})\\
\end{split}
\label{eq:lstarevol}
\end{equation}
Similarly, 
\begin{equation}
\begin{split}
\log_{10}\Phi_\star(x) = & \frac{(\psi_2-\psi_1)x_{w}}{2\pi}  \Big[\ln\big(\cosh(\pi\frac{x-x_{t}}{x_{w}})\big)\\
 &  - \ln\big(\cosh(-\pi\frac{x_{t}}{x_{w}})\big)\Big]\\
& + \frac{(\psi_2-\psi_1)}{2}x + \log_{10}(\Phi_{0})\\
\end{split}
\label{eq:phistarevol}
\end{equation}
This functional form follows the same structure as is often adopted by
the rest-frame UV community in analyzing the luminosity function for
Lyman-break galaxies (LBGs).  Figure~\ref{fig:lfevol} shows measured
constraints for both \lstar\ and \phistar\ in the rest-frame UV with
best-fit values of L$_{0}$, $\Phi_{0}$, $\gamma_2$, $\gamma_1$,
$\psi_2$, $\psi_1$, $z_{\rm turn}$, and $z_{\rm width}$ for those UV
measurements as parameterized above.  In contrast, we show various
measurements of the IR values of \lstar\ and \phistar\ from the
literature, which show a very different (and less well-constrained)
evolutionary path.  We overplot the adopted evolutionary curves for
the models in this paper in orange.  

\begin{table}
\caption{Characteristics of Observational Setup}
\centering
\begin{tabular}{c@{ }c@{ }cc}
\hline\hline
{\sc Passband} & {\sc Instrument}/ & {\sc Beamsize} & {\sc RMS} \\
               & {\sc Telescope}   &  FWHM [$''$]   & [mJy]     \\
\hline
70\um\ & {\sc pacs} ({\it Herschel}) & 5 & 0.4 \\
100\um\ & {\sc pacs} ({\it Herschel}) & 7 & 0.4 \\
160\um\ & {\sc pacs} ({\it Herschel}) & 12 & 0.9 \\
250\um\ & {\sc spire} ({\it Herschel}) & 18 & 5.8 \\
350\um\ & {\sc spire} ({\it Herschel}) & 25 & 6.3 \\
500\um\ & {\sc spire} ({\it Herschel}) & 36 & 6.8 \\
450\um\ & {\sc \scubaii} (JCMT) & 7 & 1.0 \\
850\um\ & {\sc \scubaii} (JCMT) & 15 & 0.8 \\
1100\um\ & {\sc AzTEC} (32\,m LMT) & 8.5 & 0.3 \\
1400\um\ & {\sc TolTEC}(50\,m LMT) & 6.9 & 0.3 \\
2000\um\ & {\sc GISMO}/{\sc TolTEC} (50\,m LMT) & 9.9 & 0.1 \\
\hline\hline
\end{tabular}
\label{tab:obs}

\justify {\small {\bf Notes.} This table summarizes the different
  observational setups we test for our 1\,deg$^2$ simulations from
  70\um--2\,mm using various past/existing instruments.  The
  simulations at 1.4\,mm do not explicitly simulate observations from
  any existing instrument, though will be analogous to future surveys
  from the TolTEC instrument at the LMT; in this paper, they serve as
  a good analogue to unlensed South Pole Telescope-detected
  1.4\,mm-selected sources, which are magnified by factors of
  $\mu=$\,5--20.}
\vspace{2mm}
\end{table}

The redshift dependence of SED characteristics, or the
\lir-\lpeak\ relationship shown in Figure~\ref{fig:tdust}, is a bit
more difficult to constrain given the lack of complete samples in the
early Universe, and the introduction of potential dust-temperature
biases.  Given the consistency of SEDs across $0.3<z<5$ we proceed
with a non-evolving \lir-\lpeak\ relationship though we discuss
possible caveats of this assumption in \S~\ref{sec:appendsed}.  This
is, by design, open to revision if it is later determined that
high-$z$ DSFG SEDs do evolve with redshift or exhibit some other bulk
characteristics or trends with higher quality data.

\begin{table*}
\caption{Summary of Model Assumptions}
\centering
\begin{tabular}{lcl}
\hline\hline
Name & Equation & Description \\
\hline
\begin{minipage}{0.6in}
(Eq~\ref{eq:lf})\\ Galaxy Luminosity Function
\end{minipage}
 & \begin{math}
\Phi(L,z) = \left\{
\begin{array}{lr}
\Phi_{\star}(z) \left(\frac{L}{L_{\star}(z)}\right)^{\alpha_{\rm LF}}(z) & : L<L_{\star}(z) \\
\Phi_{\star}(z) \left(\frac{L}{L_{\star}(z)}\right)^{\beta_{\rm LF}}(z) & : L\ge L_{\star}(z) \\
\end{array}
\right.
\end{math}
& 
\begin{minipage}{3.5in}
{\footnotesize
\vspace{2mm} We adopt a luminosity function model that is a broken
powerlaw with a faint-end slope, $\alpha_{\rm LF}$, a bright-end
slope, $\beta_{\rm LF}$, the characteristic luminosity at the knee of
the luminosity function, $L_\star$ (given in \lsun) and characteristic
number density \phistar\ (given in Mpc$^{-3}$\,dex$^{-1}$).  In
principle, all four parameters of the luminosity function
($\alpha_{\rm LF}$, $\beta_{\rm LF}$, $L_\star$, and \phistar) can
be redshift dependent.}
\vspace{2mm}
\end{minipage} \\

\begin{minipage}{0.6in}
\begin{flushleft}
(Casey\,2012 Eq~3) Form of Dust SED
\end{flushleft}
\end{minipage}
&
\begin{math}
\begin{aligned}
S_{\nu}(T_{\rm dust})=C_{1}\frac{(1-e^{-\tau(\nu)})\nu^3}{e^{h\nu/kT_{\rm dust}}-1}+ \\C_{2}v^{-\alpha_{\rm MIR}}e^{-(\nu_c/\nu)^2}\\
\end{aligned}
\end{math}
&
\begin{minipage}{3.5in}{\footnotesize
\vspace{2mm} Analytic approximation for a sources' flux density (in
mJy) as a function of dust temperature ($T_{\rm dust}$) and frequency ($\nu$), in
the form of a modified blackbody added to a mid-infrared powerlaw,
following the methodology given in \citet{casey12a}.  Here,
$\tau(\nu)=(\nu/\nu_{0})^{\beta_{E}}$, where $\beta_{E}$ is the spectral
emissivity index, and $\nu_{0}\approx$ 3\,THz.  The slope of the
mid-infrared powerlaw is $\alpha_{\rm MIR}$, and the coefficients
$C_1$ and $C_2$ are fixed with respect to one another, and set so that
the integral under this curve between 8-1000\um\ is $L$ in \lsun.
$\nu_c$ is the frequency at which the powerlaw and modified blackbody
contribute equally, and is a fixed function of $T_{\rm dust}$, $\nu_{0}$, and
$\beta_{E}$.}
\vspace{2mm}
\end{minipage} \\

\begin{minipage}{0.6in}
\begin{flushleft}
(Eq~\ref{eq:lirlpeak}) $\lambda_{\rm peak}$ of SED
\end{flushleft}
\end{minipage} & 
\begin{math}
\langle\lambda_{\rm peak}(L)\rangle=\lambda_{\rm 0}\Big(\frac{\,L}{L_{t}}\Big)^{\eta}
\end{math} &
\begin{minipage}{3.5in}{\footnotesize
\vspace{2mm} The {\it median} rest-frame wavelength at which a dust
SED will peak given its luminosity, $L$.  We measure no significant redshift evolution in this relation beyond $z\sim0.3$, and this
is based on the observed empirical relationship shown in
Figure~\ref{fig:tdust}.  Here, $\lambda_{\rm peak}$ relates to the
model's input dust temperature, $T$, via $\lambda_{\rm peak}\approx
b/T^{0.9}$, where $b=2.898\times10^{3}$\,\um\,K, Wien's Displacement
Constant.  Note this is an approximation and not exact (and not
=$b/T$) because the opacity of the model shifts the peak of the SED
towards longer wavelengths than the peak of a perfect blackbody
\citep[see][ Figure 20]{casey14a}.  
The $\lambda_{\rm peak}$ for any one galaxy is assigned assuming a
Gaussian probability distribution in $\log_{10}(\langle\lambda_{\rm
  peak}\rangle)$ with width $\sigma$.  $L_{t}$ is fixed to
10$^{12}$\,\lsun\ and holds no physical meaning.
\vspace{2mm}}
\end{minipage} \\
\begin{minipage}{0.6in}
\begin{flushleft}
(Eq~\ref{eq:cmb1}) $T$ change due to CMB
\end{flushleft}
\end{minipage} & 
\begin{math}
\begin{aligned}
T_{\rm dust}^\prime(z) = & \big( (T)\!^{4+\beta_{E}} \\
& + T_{\rm
  CMB}\!^{4+\beta_{E}}[(1+z)^{4+\beta_{E}}-1]\big)^{\frac{1}{4+\beta_{E}}}\\
\end{aligned}
\end{math}
&
\begin{minipage}{3.5in}{\footnotesize
\vspace{2mm} Here $T_{\rm dust}$ is the intrinsic dust temperature of the galaxy
as it would be at $z=0$, i.e. the same as $T_{\rm dust}$ from Casey\,2012 Eq~3
above.  $T_{\rm CMB}^{z=0}=2.725$\,K, $\beta_{E}$ is the emissivity
spectral index, and $T_{\rm dust}(z)$ is the adjusted temperature of
the galaxy taking into consideration heating from the CMB.  This
temperature is then used to infer the fraction of flux at any
frequency $\nu$ that would be observable at the given redshift.
\vspace{2mm}}
\end{minipage}
\\
\begin{minipage}{0.6in}
\begin{flushleft}
(Eq~\ref{eq:cmb2}) $S_{\nu}(T)$ change due to CMB
\end{flushleft}
\end{minipage} & 
\begin{math}
f(z,T_{\rm dust}^\prime) = 1 - \frac{B_{\nu}[T_{\rm CMB}(z)]}{B_{\nu}[T_{\rm dust}^\prime(z)]}
\end{math}
&
\begin{minipage}{3.5in}{\footnotesize
\vspace{2mm} The fraction of flux density $S$ of $S_{\nu}(T_{\rm dust})$ as given
in Casey\,2012 Eq~3, i.e. $S_{\rm obs}=f(z,T_{\rm dust}^\prime)S_{\nu}(T_{\rm dust})$, that would be
detectable by an observer at frequency $\nu$, redshift $z$, and
adjusted dust temperature $T_{\rm dust}$ from Eq~\ref{eq:cmb1}.  $B_{\nu}$ is
the Planck Function dependent on temperature.  Here, the CMB
temperature follows the redshift dependence $T_{\rm CMB}(z)=T_{\rm
  CMB}^{z=0}(1+z)$, where $T_{\rm CMB}^{z=0}=2.725$\,K.
\vspace{2mm}}
\end{minipage}
\\
\begin{minipage}{0.6in}
(Eq~\ref{eq:def})\\Clarifying\\ Definitions
\end{minipage} &
\begin{math}
\begin{array}{l}
x\equiv\log_{10}(1+z)\\
x_{t}\equiv\log_{10}(1+z_{\rm turn})\\
x_{w}\equiv \frac{z_{w}}{\ln(10)(1+z_{\rm turn})}\\
\end{array}
\end{math} & 
\begin{minipage}{3.5in}{\footnotesize
\vspace{2mm} Simple definitions to clarify the evolution of
$\log_{10}$(\lstar) and $\log_{10}$(\phistar) in Eq~\ref{eq:lstarevol}
and Eq~\ref{eq:phistarevol}.  $z_{\rm turn}$ is the adopted turnover
redshift while $z_{\rm w}$ is the width in redshift over which the
transition from one state to the other happens.  $x_{\rm t}$ is a
direct mapping of $z_{\rm turn}$, while $x_{\rm w}$ is a mapping of
$z_{\rm w}$.
\vspace{2mm}}
\end{minipage}\\
\begin{minipage}{0.6in}
(Eq~\ref{eq:lstarevol}) $L_\star$\\ evolution
\end{minipage} &
\begin{math}
\begin{array}{ll}
\log_{10}L_\star(x) = & \frac{(\gamma_2-\gamma_1)x_{w}}{2\pi}  \Big[\ln\big(\cosh(\pi\frac{x-x_{t}}{x_{w}})\big)\\
 &  - \ln\big(\cosh(-\pi\frac{x_{t}}{x_{w}})\big)\Big]\\
& + \frac{(\gamma_2-\gamma_1)}{2}x + \log_{10}(L_{0})\\
\end{array}
\end{math} & 
\begin{minipage}{3.5in}{\footnotesize
\vspace{2mm} The evolution of the knee of the luminosity function
$L_\star$ with redshift is assumed to evolve as $(1+z)^{\gamma_{1}}$,
with a possible redshift turnover or `reversal' happening at a
redshift of $z_{\rm turn}$ such that at higher redshifts the relation
evolves with a different slope, $\gamma_{2}$.
\vspace{2mm}}
\end{minipage}\\
\begin{minipage}{0.6in}
(Eq~\ref{eq:phistarevol}) \phistar\\ evolution
\end{minipage} &
\begin{math}
\begin{array}{ll}
\log_{10}\Phi_\star(x) = & \frac{(\psi_2-\psi_1)x_{w}}{2\pi}  \Big[\ln\big(\cosh(\pi\frac{x-x_{t}}{x_{w}})\big)\\
 &  - \ln\big(\cosh(-\pi\frac{x_{t}}{x_{w}})\big)\Big]\\
& + \frac{(\psi_2-\psi_1)}{2}x + \log_{10}(\Phi_{0})\\
\end{array}
\end{math} & 
\begin{minipage}{3.5in}
\vspace{2mm}{\footnotesize
The evolution of the characteristic number density of the luminosity function
\phistar\ with redshift is assumed to evolve as $(1+z)^\psi_{1}$,
with a possible redshift turnover or `reversal' happening at a
redshift of $z_{\rm turn}$ such that at higher redshifts the relation
evolves with a different slope, $\psi_{2}$.
\vspace{2mm}}
\end{minipage}\\
\hline\hline
\end{tabular}
\label{tab:summary}
\end{table*}

\begin{table*}
\caption{Parameter Definitions and Adopted Values}
\centering
\begin{tabular}{llccc}
\hline\hline
{\sc Name} & {\sc Description} & {\sc Model A} & {\sc Model B} & {\sc Quality of} \\
 & & & & {\sc Constraints}$^{a}$ \\
\hline
\multicolumn{4}{c}{\underline{\it $-$ Luminosity Function Parameters $-$}}\\
$L_{0}$ & {\footnotesize Knee of the IR luminosity function at $z=0$, in \lsun.} & 1.3$\times$10$^{11}$ & 1.3$\times$10$^{11}$ & {\sc Secure} \\
$\Phi_{0}$ & {\footnotesize Characteristic Number Density of the IR luminosity function at $z=0$, } & 3.2$\times10^{-4}$ & 3.2$\times10^{-4}$ & {\sc Secure} \\
 & {\footnotesize in Mpc$^{-3}$\,dex$^{-1}$.} & & \\
$\alpha_{\rm LF}$ & {\footnotesize Best-fit faint-end slope of the IR luminosity function from $z=0$ to $z=2.5$.} & --0.6 & --0.6 & {\sc Minor Impact} \\
$\beta_{\rm LF}$ & {\footnotesize Best-fit bright-end slope of the IR luminosity function from $z=0$ to $z=2.5$.} & --3.0 & --3.0 & {\sc Secure} \\
\multicolumn{4}{c}{\underline{\it $-$ Rest-Frame SED Parameters $-$}}\\
$\alpha_{\rm MIR}$ & {\footnotesize Mid-Infrared Powerlaw Slope.} & 2.0 & 2.0 & {\sc Secure} \\
$\beta_{E}$ & {\footnotesize Emissivity Spectral Index.}     & 1.8 & 1.8 & {\sc Secure} \\
\multicolumn{4}{c}{\underline{\it $-$ Peak of SED Parameters $-$}}\\
$\log\lambda_{\rm 0}$ & {\footnotesize $\lambda_{0}$ is the average rest-frame wavelength of $S_{\nu}$ at $L_{t}=10^{12}$\,\lsun.} & 2.012 & 2.012 & {\sc Secure} \\
$\eta$ & {\footnotesize The slope of the \lir-\lpeak\ relation, as shown in Figure~\ref{fig:tdust}.} & --0.068 & --0.068 & {\sc Minor Impact} \\
$\sigma$ & {\footnotesize Standard deviation of $\log(\lambda_{\rm peak})$ at any given luminosity $L$.} & 0.045 & 0.045 & {\sc Minor Impact} \\
\multicolumn{4}{c}{\underline{\it $-$ Parameters describing Redshift Evolution $-$}}\\
$\gamma_{1}$ & {\footnotesize At $z\ll z_{\rm turn}$, $\gamma_1$ describes the redshift evolution of $L_\star$, such that $L_\star\propto(1+z)^{\gamma_{1}}$} & 2.8 & 2.8 & {\sc Secure}\\
$\gamma_{2}$ & {\footnotesize At $z\gg z_{\rm turn}$, $\gamma_2$ describes the redshift evolution of $L_\star$, such that $L_\star\propto(1+z)^{\gamma_{2}}$} & 1.0 & 1.0 & {\sc Unknown}\\
$\psi_{1}$ & {\footnotesize At $z\ll z_{\rm turn}$, $\psi_1$ describes the redshift evolution of \phistar, such that \phistar$\propto(1+z)^{\psi_{1}}$      } & 0.0 & 0.0 & {\sc Secure}\\
$\psi_{2}$ & {\footnotesize At $z\gg z_{\rm turn}$, $\psi_2$ describes the redshift evolution of \phistar, such that \phistar$\propto(1+z)^{\psi_{2}}$      } & --5.9 & --2.5 & {\sc Unknown} \\
$z_{\rm turn}$ & {\footnotesize The `turning point' redshift at which $L_\star$ and \phistar\ are transitioning in their evolution.                      } & 2.1 & 1.8 & {\sc Secure}\\
$z_{\rm w}$ & {\footnotesize The redshift interval over which the evolution shifts exponents (e.g. $\gamma_{1}$ to $\gamma_{2}$).                        } & 2.0 & 2.0 & {\sc Secure}\\
\hline\hline
\end{tabular}
\label{tab:params}

\justify {\small {\bf Notes.}  $^{a}$ We classify the level at which a
  parameter is already constrained by data in three classes: secure,
  minor impact or unknown.  Secure means that the parameter is
  directly measurable with existing data.  Minor impact means that the
  parameter is perhaps not very well known, but that changes to this
  variable (within reason) would not dramatically impact our measured
  results in this paper.  Variables that are unknown are those which
  have no constraints.  Appendix~\ref{sec:appendA} expands on how well
  each of these parameters is known and how changes to their values
  impact the results.\\
Parameters which are considered fixed as part of the SED ($\nu_{c},
C_1$ and $C_2$) are fixed functions of $\alpha_{\rm MIR}$, $T_{\rm
  dust}$, and $L$ and therefore not given in this table.  See
\citet{casey12a} for details.  }
\end{table*}

\subsection{Generating Source Maps}

\begin{figure*}
\centering
\includegraphics[width=2.05\columnwidth]{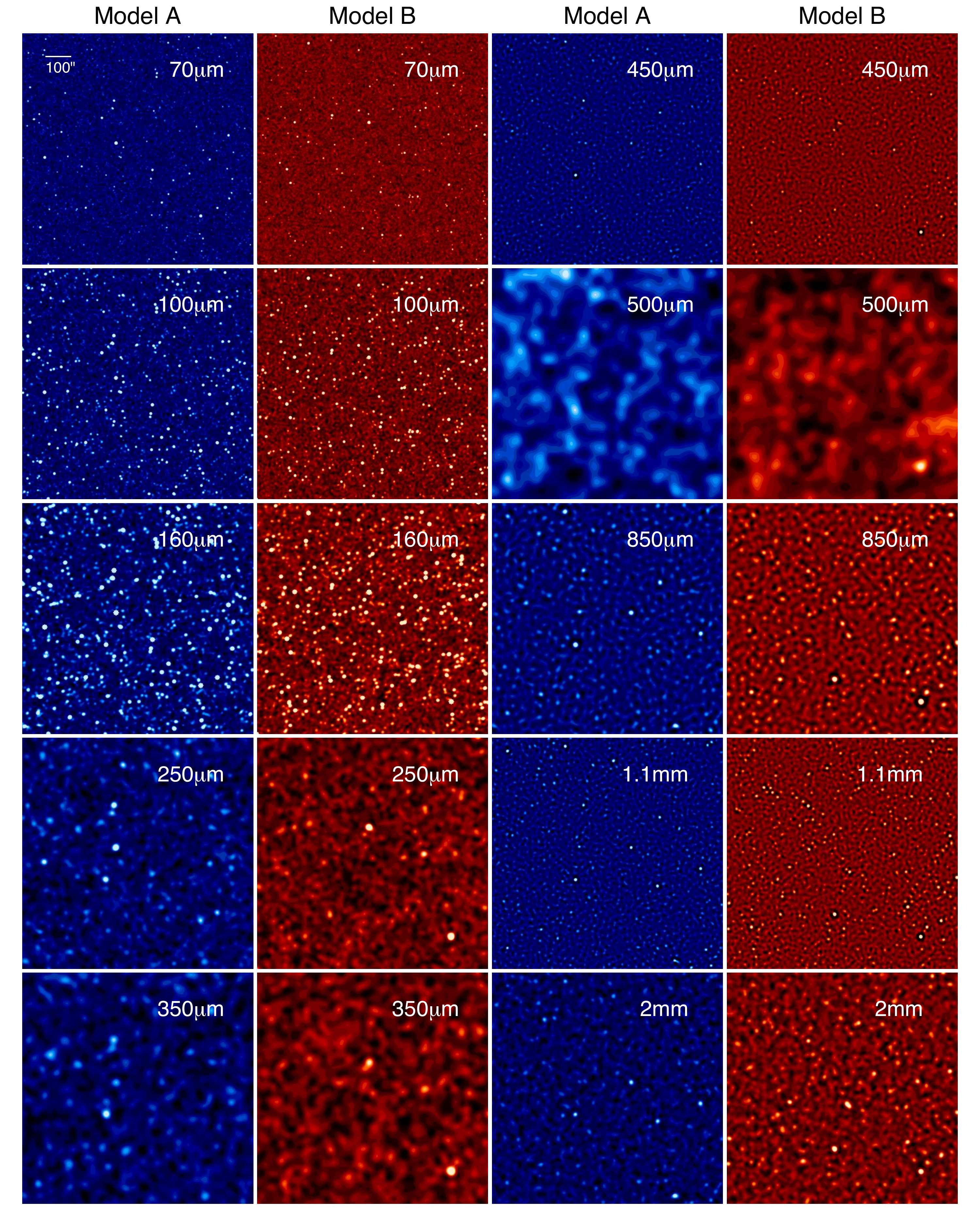}
\caption{ Simulated 900$\arcsec\times$900$\arcsec$ signal-to-noise map
  cutouts of mock 1\,deg$^{2}$ simulations, following the luminosity
  prescriptions described for Model A (the dust-poor early Universe)
  and Model B (the dust-rich early Universe). Our cutouts include mock
  {\it Herschel} {\sc Pacs} 70--160\,\um, {\sc SPIRE} 250--500\mum,
  {\sc Scuba-2} 450\um\ and 850\um, AzTEC 1.1\,mm (with a 32\,m
  diameter LMT) and GISMO 2.0\,mm (with a 50\,m diameter LMT). The
  hypothetical 1.4\,mm TolTEC maps are not shown but are similar to
  the 1.1\,mm and 2.0\,mm maps.}
\label{fig:cutouts}
\end{figure*}

Sources are injected into a series of maps of fixed solid angle; for
this paper we generate 1\,deg$^2$ maps with a 0.5$''$ pixel scale but this
is easily adjusted to test observational setups different from those
described herein.  Sources are injected with uniformly random
positions and with a surface density determined by the projection of
the galaxy luminosity function and flux densities from inferred SEDs.
The effect of the CMB heating of high-$z$ galaxies' ISM is taken into
account, impacting the injected sources' final observed flux
densities.  Each filter has its own map, and though the positions are
conserved from wavelength to wavelength, no clustering is taken into
account; we compare our model predictions to the SIDES project, which
does incorporate clustering from semi-analytic models in
\S~\ref{sec:compare}. Input flux densities, positions and redshifts
are recorded for later use.  After sources have been injected, the
maps are convolved with the filter beam. The beam is taken from real
data maps by stacking hundreds of significant detections at each
wavelength observed with each facility \citep[note that the beams are
  not well represented by a 2D Gaussian, as described further
  in][]{coppin15a}. We then generate a noise map by convolving the
beam with a standard normal distribution of pixel values, and
rescaling the resulting noise map to the appropriate RMS.  This noise
map is then added to the beam-convolved map with source injections.
Maps are then renormalized so the mode is equal to zero; this
adjustment is only significant for the mock {\it Herschel} {\sc SPIRE}
maps but is in line with the instrument's flux calibration procedure
\citep{griffin10a}.  The details of the noise and beam
characterization are given in Table~\ref{tab:obs} which summarizes
observationally-driven model inputs. Example cutouts from the fake
maps are shown in Figure~\ref{fig:cutouts} (the differences between
the two models highlighted in the figure are described in the next
section).

Sources are identified in the mock maps by first constructing
signal-to-noise maps (`SNR' map), by dividing the simulated map by the
instrumental noise as quoted in Table~\ref{tab:obs}.  All significant
peaks in the SNR map are then identified with a ``region grow''
algorithm in IDL.  Sources' positions and flux densities are then
reported as corresponding to the point of their peak signal-to-noise.
The threshold for detection, or the lower limit of SNR is initially
set to 3.5, although we conservatively limit our analysis to
$>$5$\sigma$ sources when discussing source redshift distributions.
No adjustments for confusion or Eddington boosting are made, as all
comparisons with the literature are made against raw quantities.

\subsection{Two Case Studies: Impact of Parameters}\label{sec:models}

Table~\ref{tab:params} provides a list of all of the tunable
parameters of the model; the reader should consult this table in
conjunction with Table~\ref{tab:summary} for a complete understanding
of the model construction and parameter space.  Fifteen different
parameters are listed, and though all could theoretically be left
open, most are already constrained well by existing datasets, while
others are relatively unconstrained and are the focus of our study.
Those that are well constrained are so noted in the table; the
justification of their choice values and the impact of changing their
values is discussed further in Appendix~\ref{sec:appendA}.

\begin{figure*}
\centering
\includegraphics[width=0.99\columnwidth]{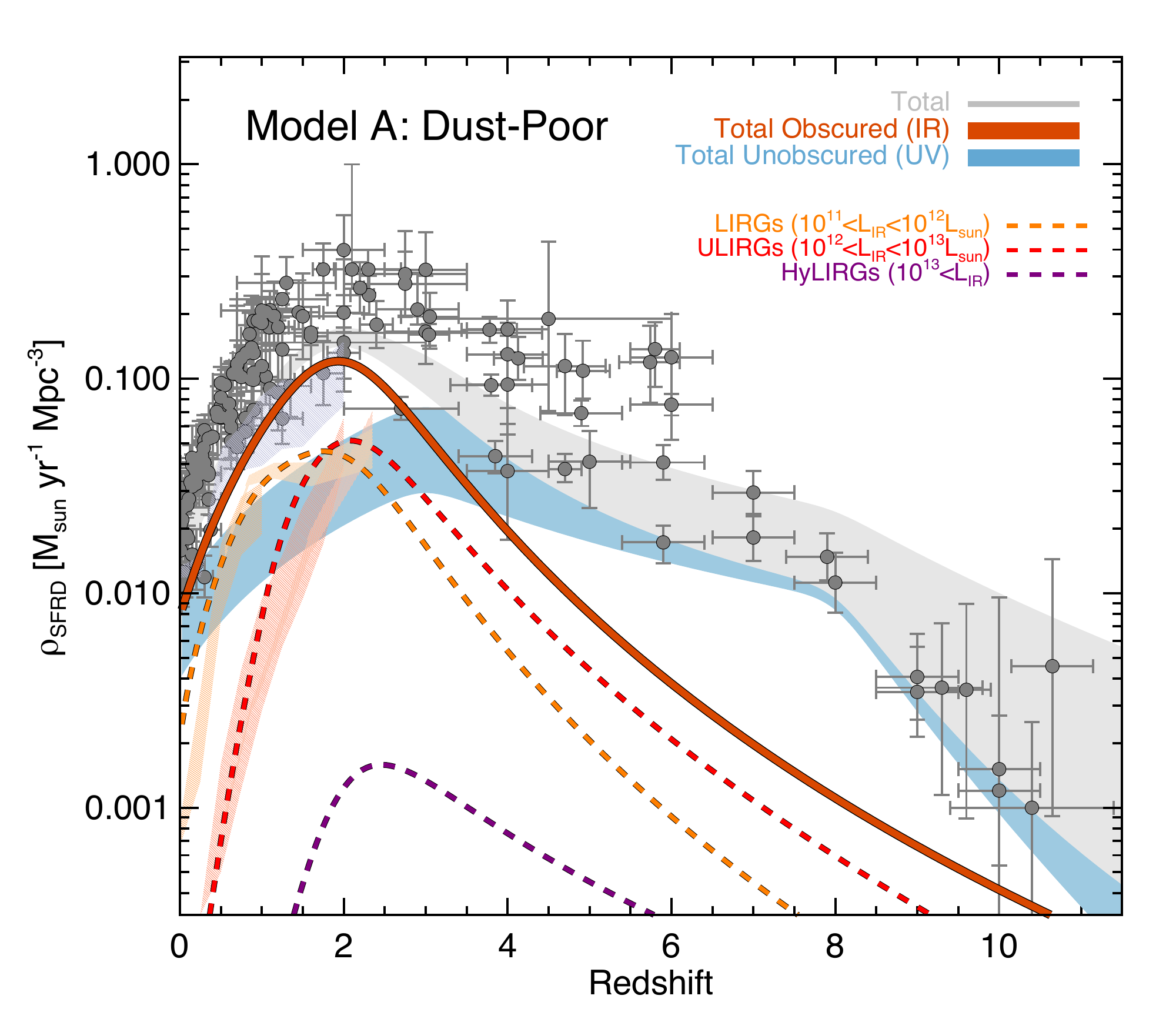}\includegraphics[width=0.99\columnwidth]{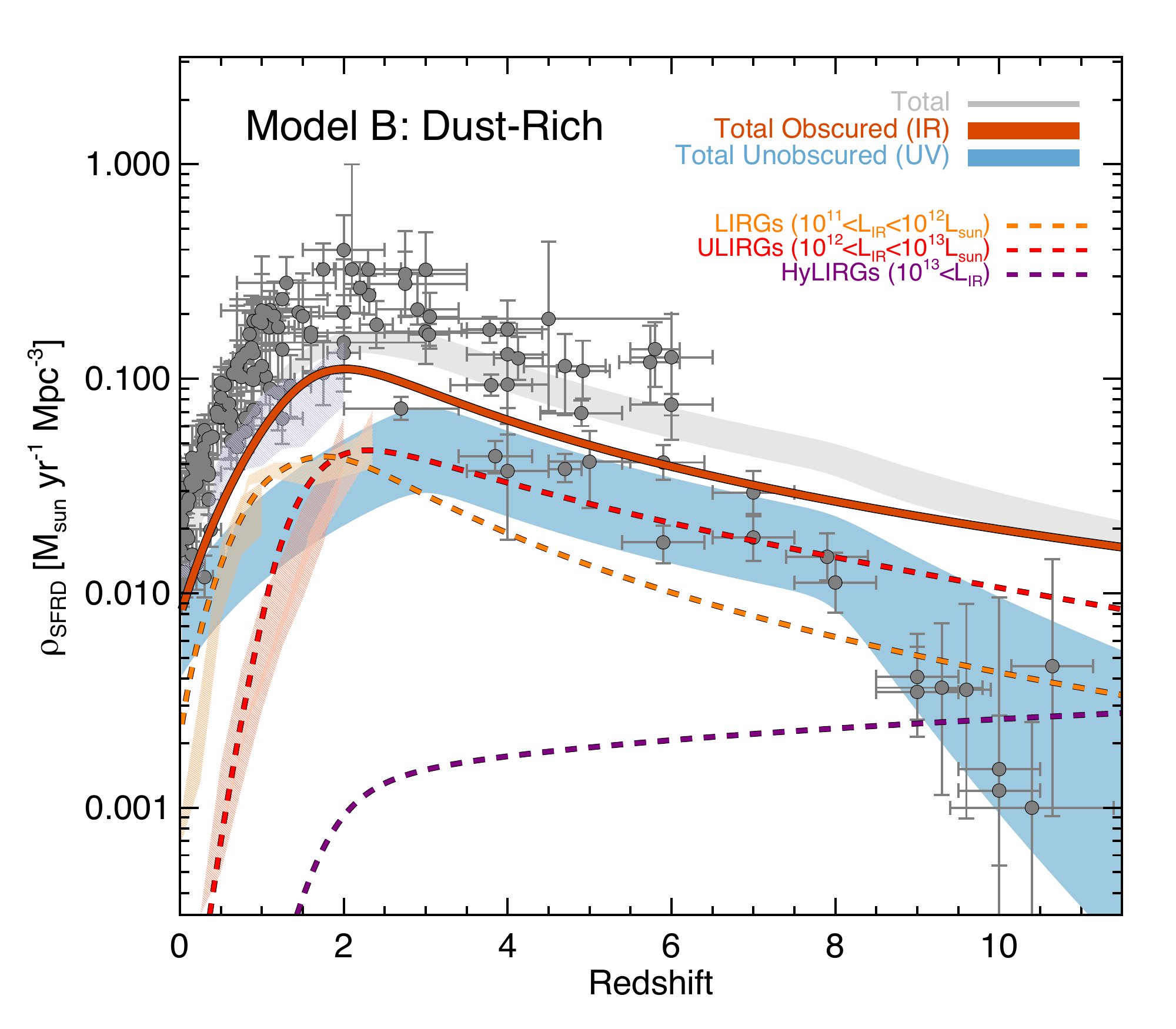}
\caption{The implied star-formation rate densities for our two model
  universes.  The thick lines denote the adopted models while the
  points and shaded regions represent existing measurements.  Model A,
  the ``dust-poor'' model (described in \S~\ref{sec:A}), asserts that
  the contribution of heavily obscured galaxies to cosmic
  star-formation is negligible in comparison to the contribution of
  UV-bright, unobscured galaxies at $z\simgt4$.  Model B, the
  ``dust-rich'' model (described in \S~\ref{sec:B}), posits a very
  different history of cosmic star-formation, whereby dust-obscured
  galaxies dominate cosmic star-formation at $z\simgt4$ over UV-bright
  galaxies by over a factor of ten.  Measurements from the literature
  are shown as gray points \citep{madau14a}, though only UV-based
  measurements have been made beyond $z\approx3.5$.  The total
  contribution from unobscured sources is noted as a blue shaded
  region, the model IR contributions are shown as a solid orange line,
  and the total of these two components shown as a gray shaded region.
  The breakdown in contribution from LIRGs ($10^{11}<L_{\rm
    IR}<10^{12}$\,\lsun), ULIRGs ($10^{11}<L_{\rm
    IR}<10^{12}$\,\lsun), and HyLIRGs ($L_{\rm IR}>10^{13}$\,\lsun)
  are shown as dashed gold, red, and purple lines, respectively.  The
  light orange, red and purple shaded regions represent some of the
  best to-date measurements of the LIRG, ULIRG and total IR
  contributions to the SFRD
  \citep{murphy11a,magnelli11a,magnelli13a}.}
\label{fig:sfrdmodels}
\end{figure*}

Some of the most impactful parameters that are fixed for this model
are $\gamma_{1}=2.8$, which traces the evolution of \lstar\ towards
much higher luminosities from $z=0$ to $z\sim2$ as shown in
Figure~\ref{fig:lfevol}.  Similarly, the corresponding number density
\phistar\ does not evolve over the same interval, so we fix
$\psi_{1}=0$.  A simple set of tests -- sampling different potential
values of $z_{\rm turn}$ and $z_{\rm w}$ -- reveal that the turnover
redshift must be close to $z=2$, otherwise the measured number counts
comparison will be off substantially, underestimated if $z_{\rm
  turn}<<2$ and overestimated if $z_{\rm turn}>>2$.  We explicitly
choose the values of $z_{\rm turn}$ (either =1.8 or =2.1, depending on
high-$z$ evolutionary parameters), and $z_{\rm w}=2.0$ so that the
measured number density evolution of LIRGs, ULIRGs, and the total IR
contribution to the star-formation rate density is well matched to
data (see Figure~\ref{fig:sfrdmodels}).

The most highly uncertain quantities ($\gamma_{2}$ and $\psi_{2}$)
describe the evolution of \lstar\ and \phistar\ beyond $z\simgt 2$,
where measurements are sparse.  In this paper we present two case
studies, adopting dramatically different values for $\psi_{2}$,
signifying either a dust-poor early Universe, or extremely dust-rich
early Universe.  Both of these models adopt $\gamma_{2}=1$, asserting
that \lstar\ continues to evolve upwards toward higher redshifts.  A
positive value of $\gamma_{2}$ is chosen for three reasons: adopting
$\gamma_{2}\le0$ underpredicts IR number counts above $\sim$1\,mJy
regardless of adopted evolution of source number density ($\psi_{2}$),
a reversal might also imply evolution back towards warmer dust
temperatures at high-redshift which is not seen for SPT-detected
galaxies \citep[contradicting the claims of][]{faisst17a}, and
adopting $\gamma_{2}=1$ neatly results in \lstar\ consistent with
\lstar\ of the quasar luminosity function at $z>4$ \citep{hopkins07a}.
This positive value for $\gamma_{2}$ is also consistent with reports
in the literature of a dramatically-bright \lstar\ value towards the
epoch of reionization \citep{cowie17a}.
The latter result is in line with what might be expected from
hierarchical formation and cosmic downsizing, suggesting a close
relationship between the most massive starbursting galaxies, the
supermassive black holes that grow at their centers, and the assertion
that both live in some of the most overdense regions of the early
Universe.  A convergence of \lstar\ values between IR luminosity
functions and quasar luminosity functions, but a lack of agreement
between number densities at those luminosities, hints at the possible
factor of $\sim$10 difference in the quasar and DSFG lifetimes.
Below we discuss the physical context of our adopted values of
$\psi_{2}$ for our case studies: models A and B, and we illustrate the
differences in implied star-formation rate density between them in
Figure~\ref{fig:sfrdmodels}.  A more thorough discussion of alternate
values of $\gamma_{2}$ are given in \S~\ref{sec:appendLF}.  It should
be noted that both models are in agreement with the measured total
energy output of the cosmic infrared background
\citep[CIB;][]{puget96a,fixsen98a,dwek98a}.  The CIB is dominated by
sources at $z<2$ by nature of the dominant source of emission at
$\lambda\simlt500$\,\um\ \citep{viero15a}, where the integrated
background is unconstrained at longer wavelengths due to the
brightness of the CMB.

\subsubsection{Model A: A Dust-Poor Early Universe}\label{sec:A}

Model A adopts $\psi_{2}=-5.9$, suggesting a steep number density
evolution for IR-luminous galaxies from the epoch of reionization to
$z\sim2$.  In other words, this model suggests DSFGs are extremely
rare in the early Universe ($z>4$) such that their contribution to
cosmic star-formation is negligible compared to much more numerous
Lyman-break galaxies at the same epoch.  Model A suggests DSFGs only
come to dominate cosmic star-formation globally for a billion years or
so near $z\sim2$ and are very rare in both the early Universe and the
local Universe.  The adopted value of $\psi_{2}=-5.9$ originates from
the measured number density evolution of bright $M_{\rm UV}\approx -
21$ galaxies from $4<z<8$, which follows $\Phi\propto(1+z)^{-5.9}$
\citep{finkelstein15a}.  Adopting the same type of evolution for
UV-luminous galaxies and IR-luminous galaxies might be quite
appropriate if they occupy dark matter halos of similar masses and
grow on similar timescales.  Note that model A, or slight variants
thereof, represent the currently accepted paradigm where the
dust-formation timescale  (primarily via AGB stars but
  also via supernovae) is longer than the formation timescale of the
first UV-bright galaxies; it is often the 
  adopted evolutionary scenario in the rest-frame UV
literature
\citep{bouwens09a,bouwens15a,bouwens16a,finkelstein15a,oesch13a}.

\vfill
\subsubsection{Model B: A Dust-Rich Early Universe}\label{sec:B}
Model B adopts $\psi_{2}=-2.5$, asserting a gradual evolution in the
number density of DSFGs from very high-redshifts up to $z\sim2$.  The
implications of this value are that DSFGs would play an increasingly
important role in cosmic star-formation towards higher redshifts, and
this star-formation would be distributed far less `evenly' than is
suggested by rest-frame UV surveys.  In other words, this model
suggests that most of cosmic star-formation at early times was
isolated to rare starbursts with very high star-formation rates,
rather than more homogeneously distributed in lower-luminosity UV-bright
galaxies.  As shown in Figure~\ref{fig:sfrdmodels}, this model suggests that
DSFGs would dominate cosmic star-formation (at roughly $\simgt$1/2 of
total) from $1.5<z<6.5$.  Furthermore, at $z>6.5$, DSFGs might
dominate all star-formation by factors $>$10 higher than UV-luminous
galaxies.

Note that this model is designed to be somewhat extreme and thus
provocative, because there has been no evidence to-date that DSFGs
dominate cosmic star-formation at these very high redshifts.  In fact,
fewer than a dozen DSFGs have yet been found at these epochs ($z>5$)!
Despite this, we adopt this extreme case to illustrate a few points.
The first is that such an extreme history has not yet been ruled out
by existing surveys.  The lack of DSFG identifications at high-$z$ is,
in large part, an observational limitation \citep*[as discussed
  extensively in Chapters 4 and 6 of][also see
  e.g. \citealt{marrone17a}]{casey14a}.  The second point worth
illustrating -- as the next section will detail -- is that such
dramatically different number density evolution at high-$z$ (as
constructed by Models A and B) result in largely the same observable
characteristics in IR datasets from 70\um--1\,mm, with only subtle
differences.  As we will show, the way of distinguishing between
high-redshift DSFG models relies almost exclusively on the 1.4\,mm and
2\,mm bands, for which there is only sparse data existing to-date.

One obvious caveat to Model B is the lack of clarity on when the first
DSFGs might turn on. This is not a problem in Model A because that
model asserts that DSFGs are extremely rare at early times. In our
implementation of model B we have not explicitly set a cutoff redshift
above which DSFGs do not exist, though it is likely the case that such
a limit exists in reality.  We find that this has a relatively minor
impact on the conclusions we reach for 1\,deg$^2$ fields as the number
of very high-$z$ sources is not a dominant source of emission at any
wavelength.  However, it could impact results for $>$1\,mm surveys on
larger scales ($\gg$1\,deg$^2$), and a more elaborate model could be
implemented to account for this.

\section{Comparison with Data \&\ Models}\label{sec:compare}

We quantify the appropriateness of models A and/or B using all
available data constraints, including number counts, redshift
distributions, and relative detection rates and measured flux
densities of sources across the many IR/submm bands.  We also describe
alternate models that simulate the submm sky to provide some context
in interpreting the dominant sources of emission and relative
importance and constraints (or lack thereof) of galaxies in the early
Universe.

\subsection{Comparison with Alternate Models}

The SIDES model \citep[or the Simulated Infrared Dusty Extragalactic
  Sky;][]{bethermin17a} is an update of the 2SFM \citep[two
  star-formation modes][]{bethermin12a,sargent12a} galaxy evolution
model to analyze the impact of clustering on IR map analysis.  Our
model differs from the 2SFM \citep{bethermin12b} model at a
fundamental level in its treatment of the underlying galaxy
population: 2SFM builds galaxies' SEDs from their position on the
``main sequence'' relation \citep[the relationship between galaxies'
  stellar mass and star-formation rate][]{noeske07a} and our model
builds them up from their IR luminosity (or SFR) only.  This
difference would not necessarily result in a discrepancy, but the
former assumes galaxies SEDs are linked more fundamentally to their
specific SFR in two modes, either on the main sequence, or in a
starbursting phase;  the SIDES model has slightly
  warmer dust in $z>2$ galaxies than the 2SFM model.  Galaxies with
higher specific SFR are asserted to have much warmer dust SEDs which
is traced to harder radiation fields ($\langle U\rangle$).  In this
paper, we argue against a bi-modal population, and that luminosity, or
SFR alone, is more fundamentally linked to a galaxy's SED (and thus
the nature of their IR flux densities) than their specific SFR.  When
separated by stellar mass, for example, there is little evidence that
galaxies sitting on the high-mass end of the main sequence have colder
temperatures than those at similar SFRs that are elevated above the
main sequence at lower masses \citep{u12a}, and instead, their dust
temperatures track very tightly with IR luminosity.

Beyond the scope of 2SFM, SIDES \citep{bethermin17a} incorporates
clustering and its effects in source multiplicity by using abundance
matching to populate dark matter halos in a
 dark-matter only large volume simulation
using stellar mass abundance matching.  Then, those galaxies are represented by IR
SEDs according to the 2SFM model in order to explain discrepancies
between high- and low-resolution IR number counts.  This is not within
the scope of this paper, though we do incorporate the effects of
angular resolution.  Our model does not account for source clustering.
This makes it difficult to characterize the entire nature of DSFG
multiples \citep[sources that break into several components with
  higher resolution, and whether or not those multiples are physically
  associated, e.g.][]{hodge13a,hayward13a,hill17a}; though it would,
in principle be possible to estimate the multiple fraction of DSFGs
caused by chance projections in this model, it is beyond the scope of
this paper.  Of course the issue of multiplicity is important for our
understanding of the physical drivers of DSFGs and thus provides
important insight which the SIDES model addresses.

At a more fundamental level, the luminosity function in SIDES/2SFM is
fixed at high-redshift according to the evolution of \phistar\ modeled
in \citet{sargent12a}, which leans heavily on galaxy luminosity
functions measured at wavelengths shortward of 24\um.  Specifically,
at high-redshifts, the \citet{bethermin12a} model assumes the IR
contribution to the SFRD follows the shape of the measured rest-frame
UV evolution to $z\sim6$, that the main sequence (or galaxies'
specific star-formation rates) is fixed at $z>2.5$, and that the
decreasing contribution of IR luminous galaxies at very early times is
due entirely to increased rarity (i.e. a drop in \phistar).
 The updated SIDES model has a similar drop at the
  highest redshift despite continued luminosity function at $z>2.5$
  \citep[following][]{schreiber15a}, which is a direct consequence of
  the evolving stellar mass function in the model
  \citep{davidzon17a}. In this manner, the SIDES model follows the
high-redshift evolution of our Model A, the dust-poor early Universe.
Because the goals of our work are to place constraints on the
evolution of the IR luminosity function itself, our model is, and
needs to be, constructed in a very different way to provide insight
for quantities that are otherwise fixed in SIDES.

\begin{figure*}
\includegraphics[width=0.66\columnwidth]{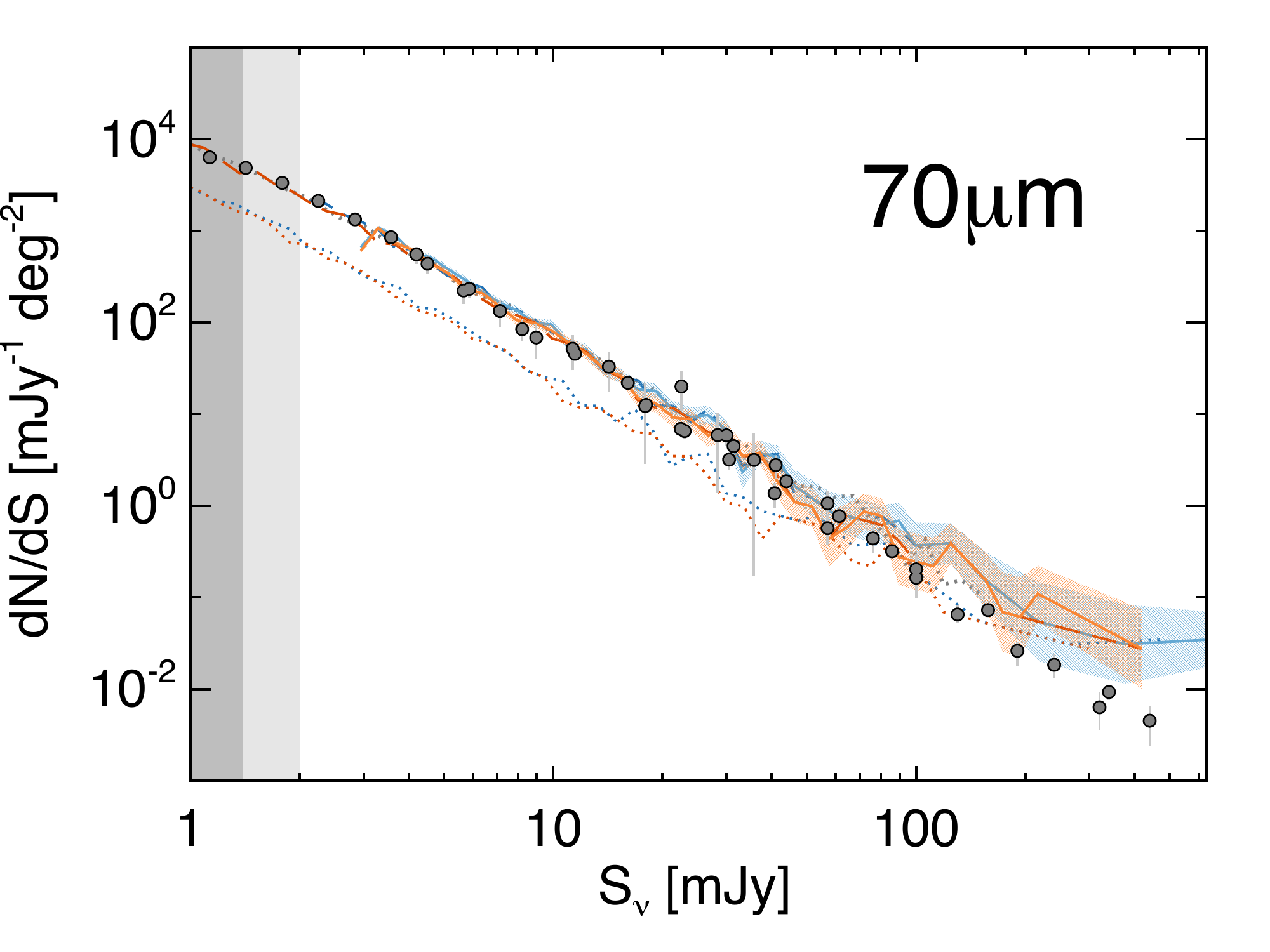}
\includegraphics[width=0.66\columnwidth]{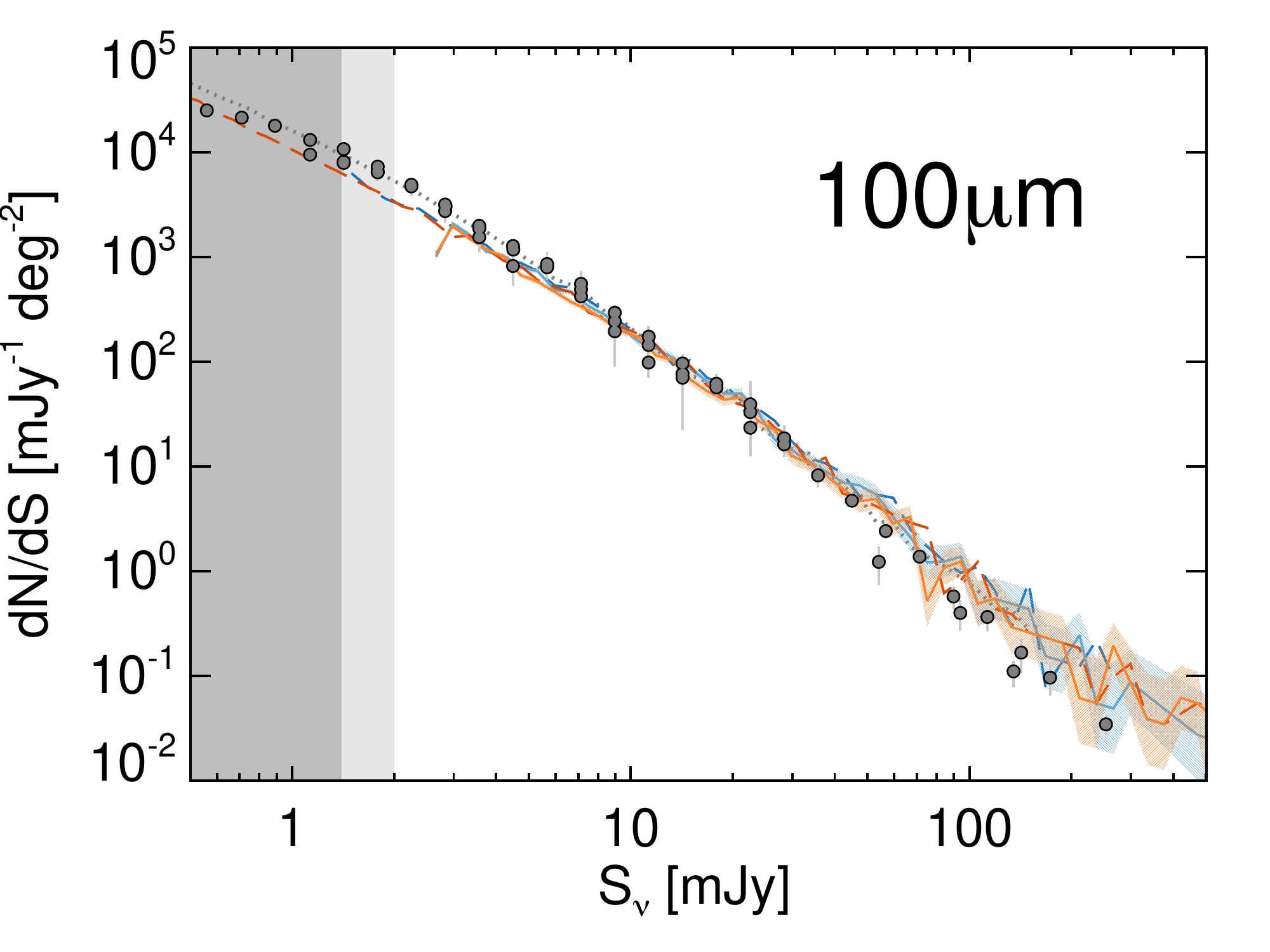}
\includegraphics[width=0.66\columnwidth]{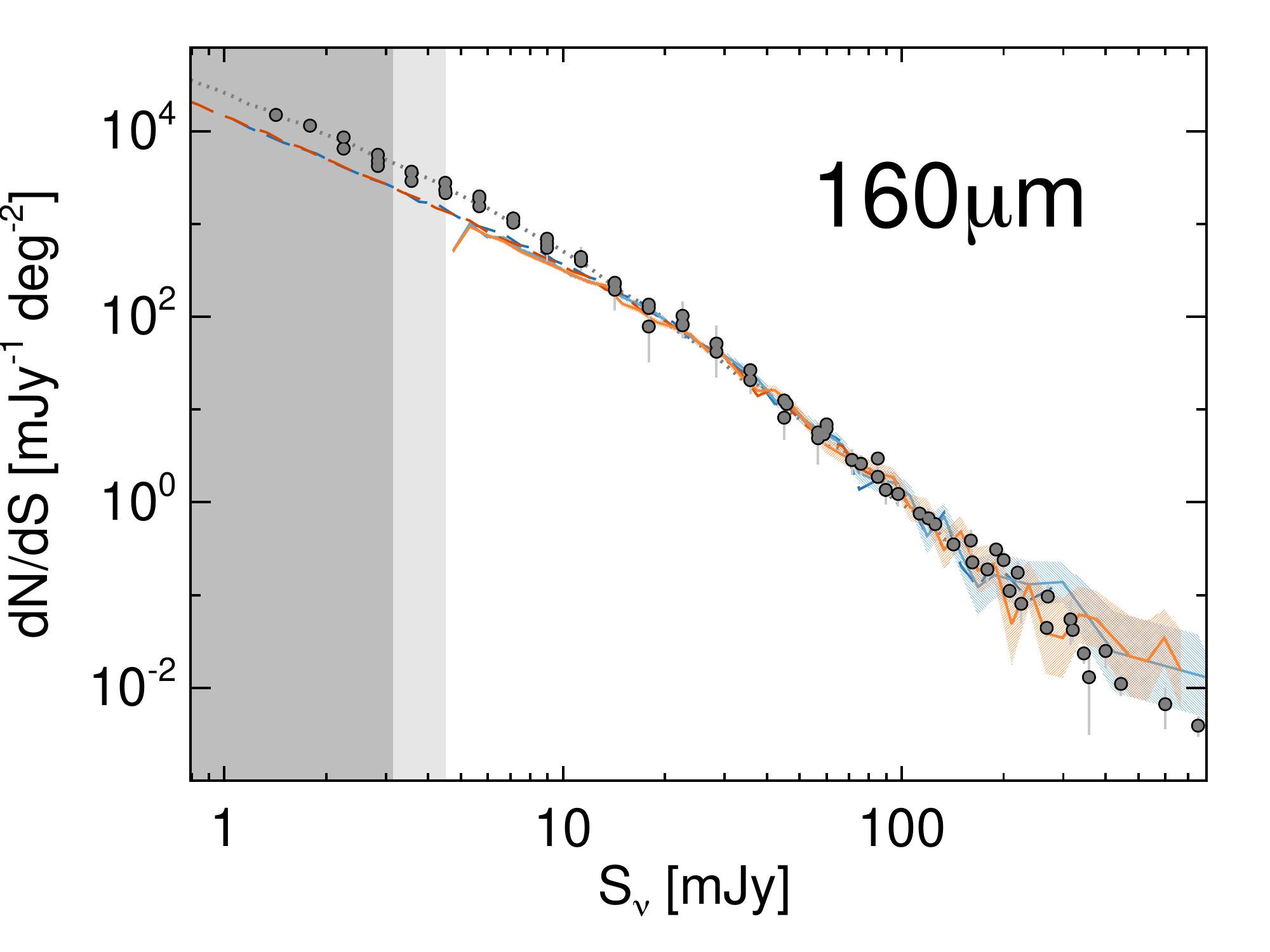}\\
\includegraphics[width=0.66\columnwidth]{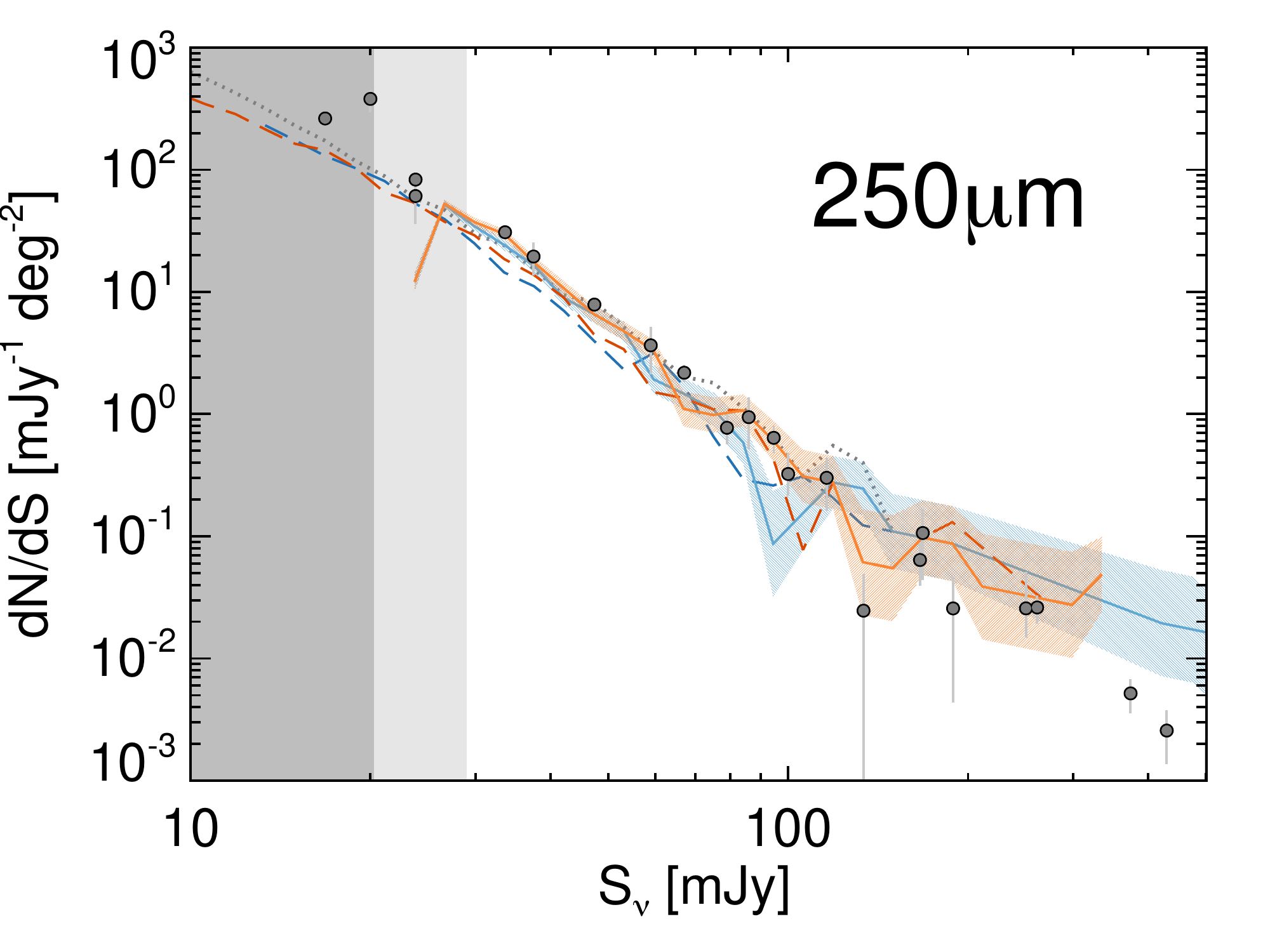}
\includegraphics[width=0.66\columnwidth]{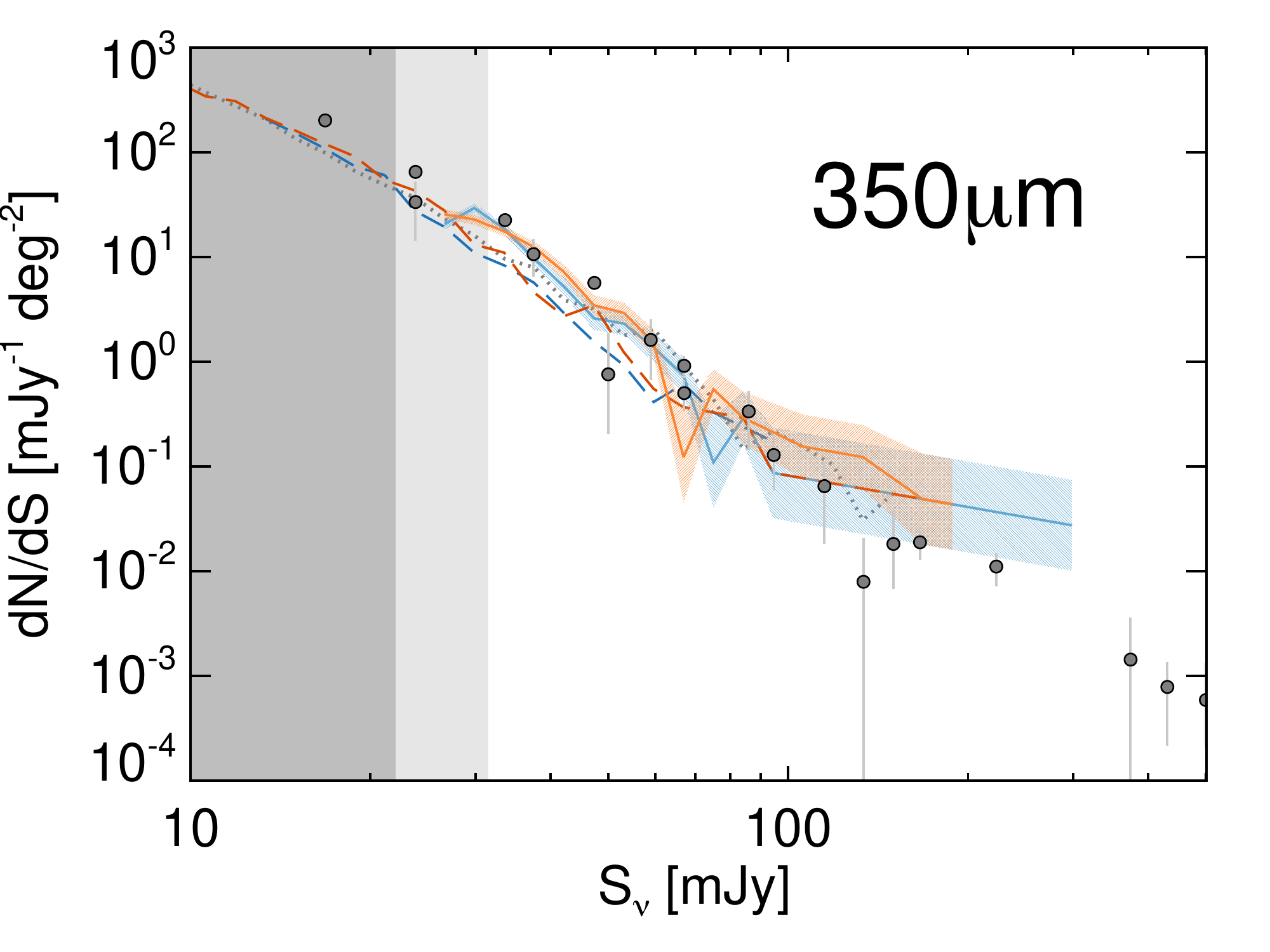}
\includegraphics[width=0.66\columnwidth]{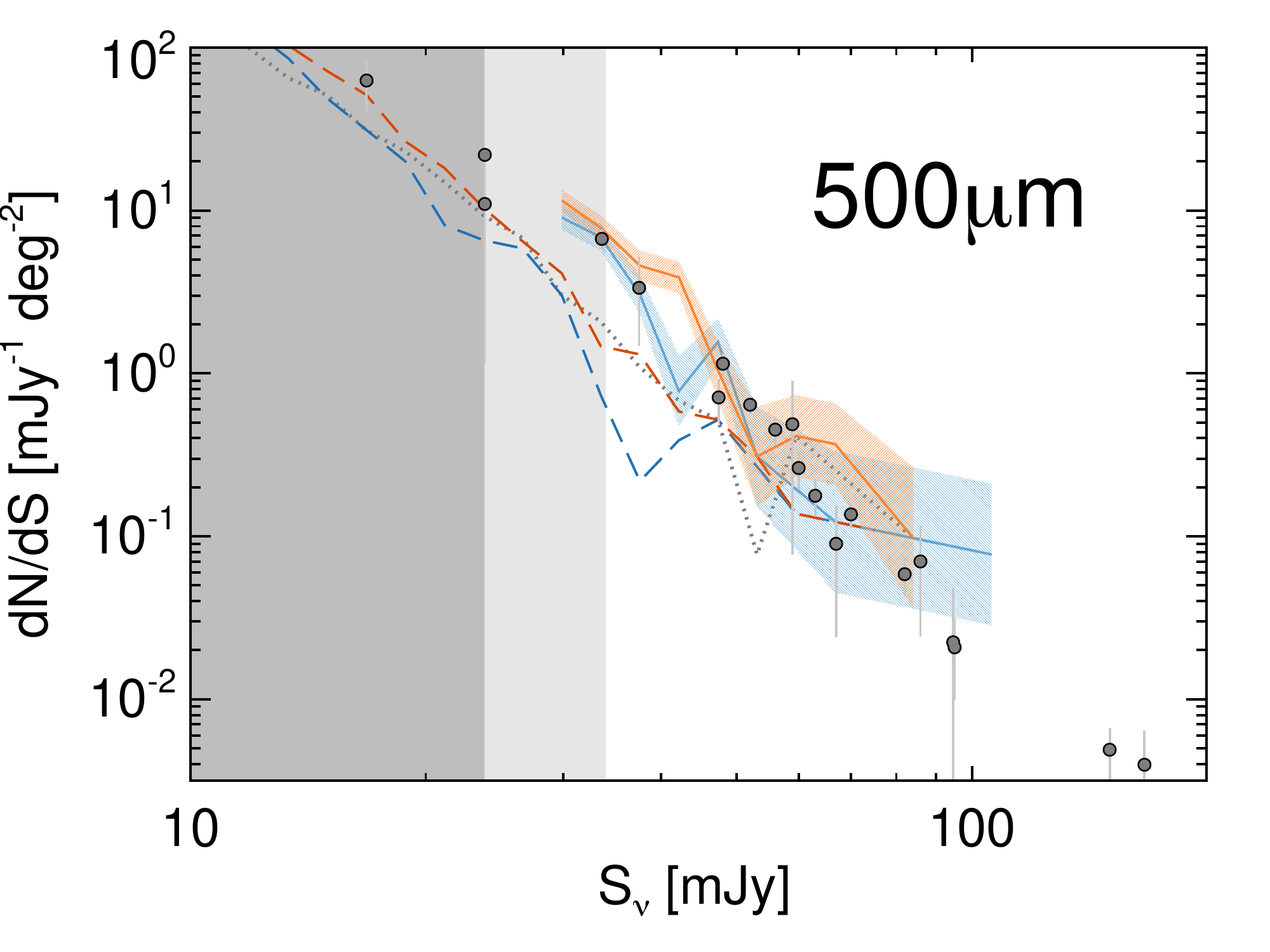}\\
\includegraphics[width=0.99\columnwidth]{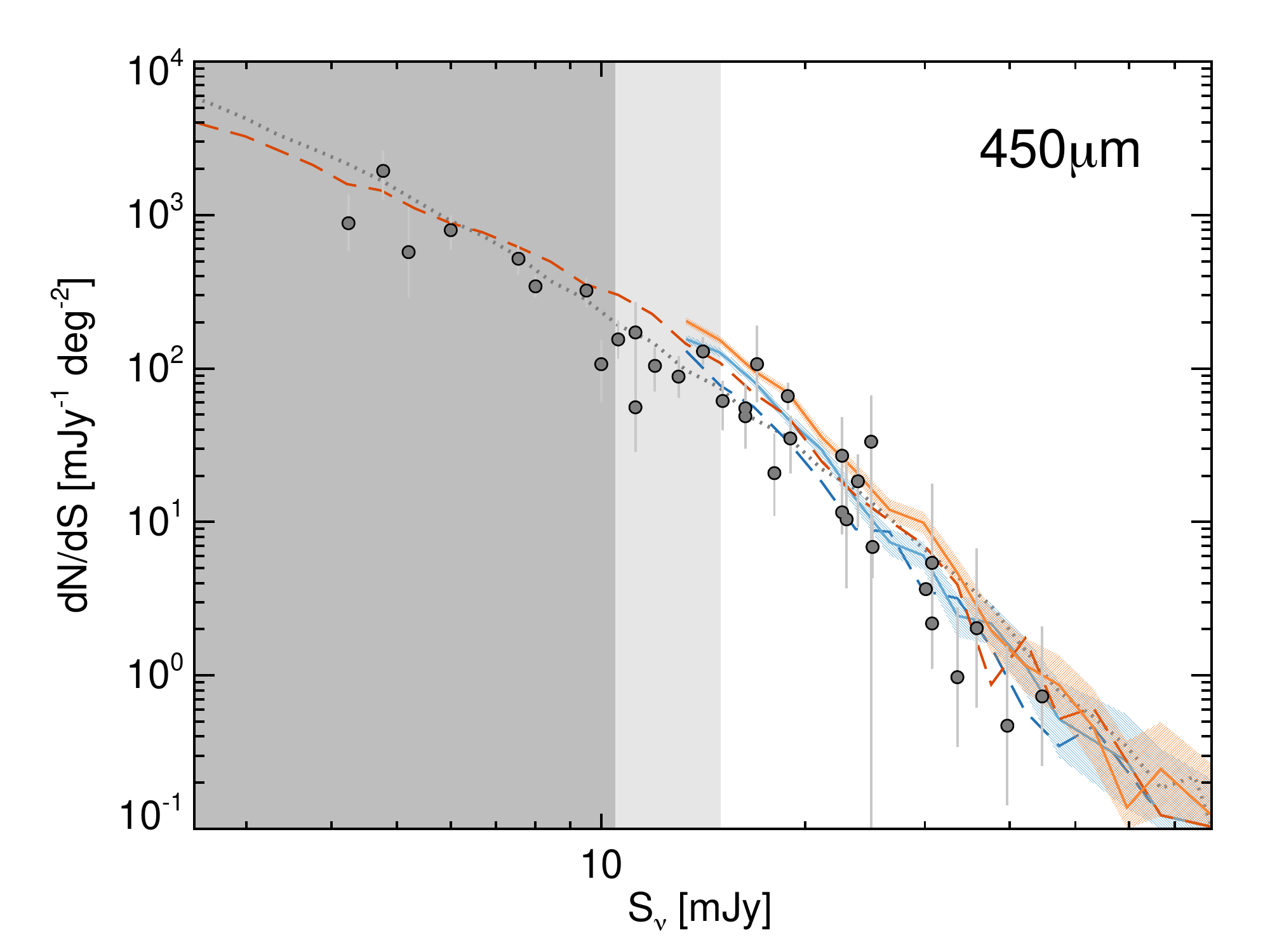}
\includegraphics[width=0.99\columnwidth]{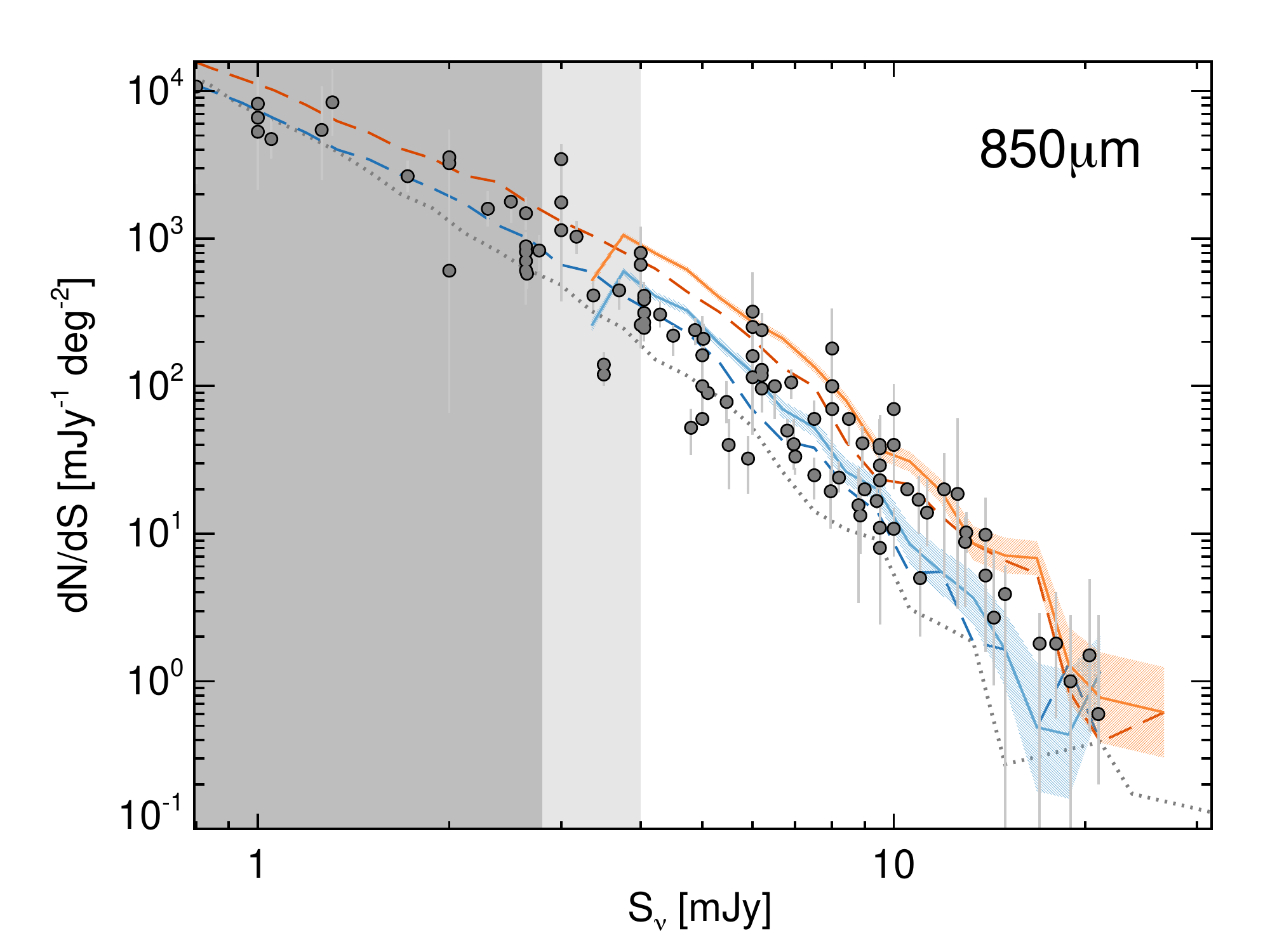}\\
\includegraphics[width=0.99\columnwidth]{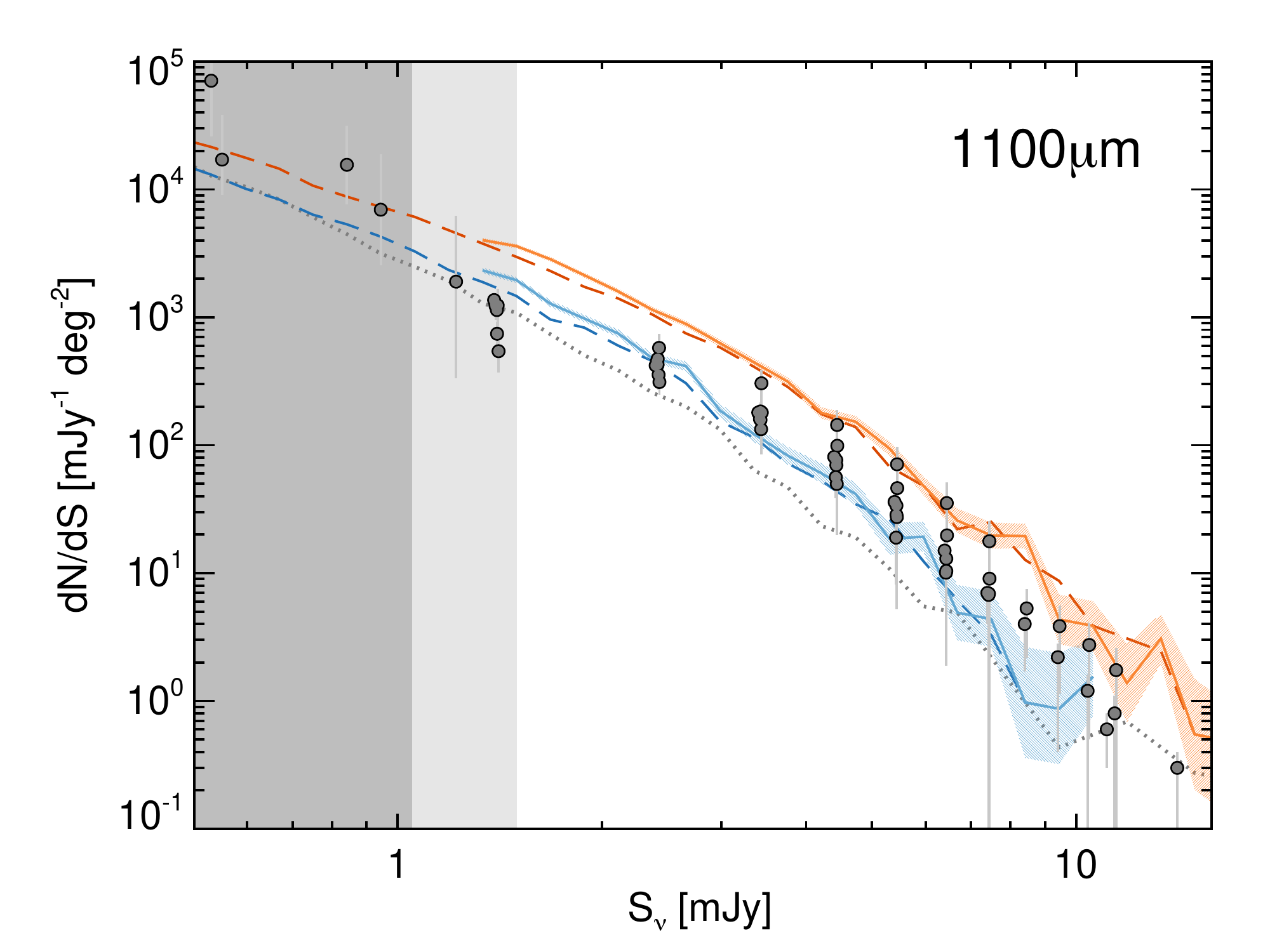}
\includegraphics[width=0.99\columnwidth]{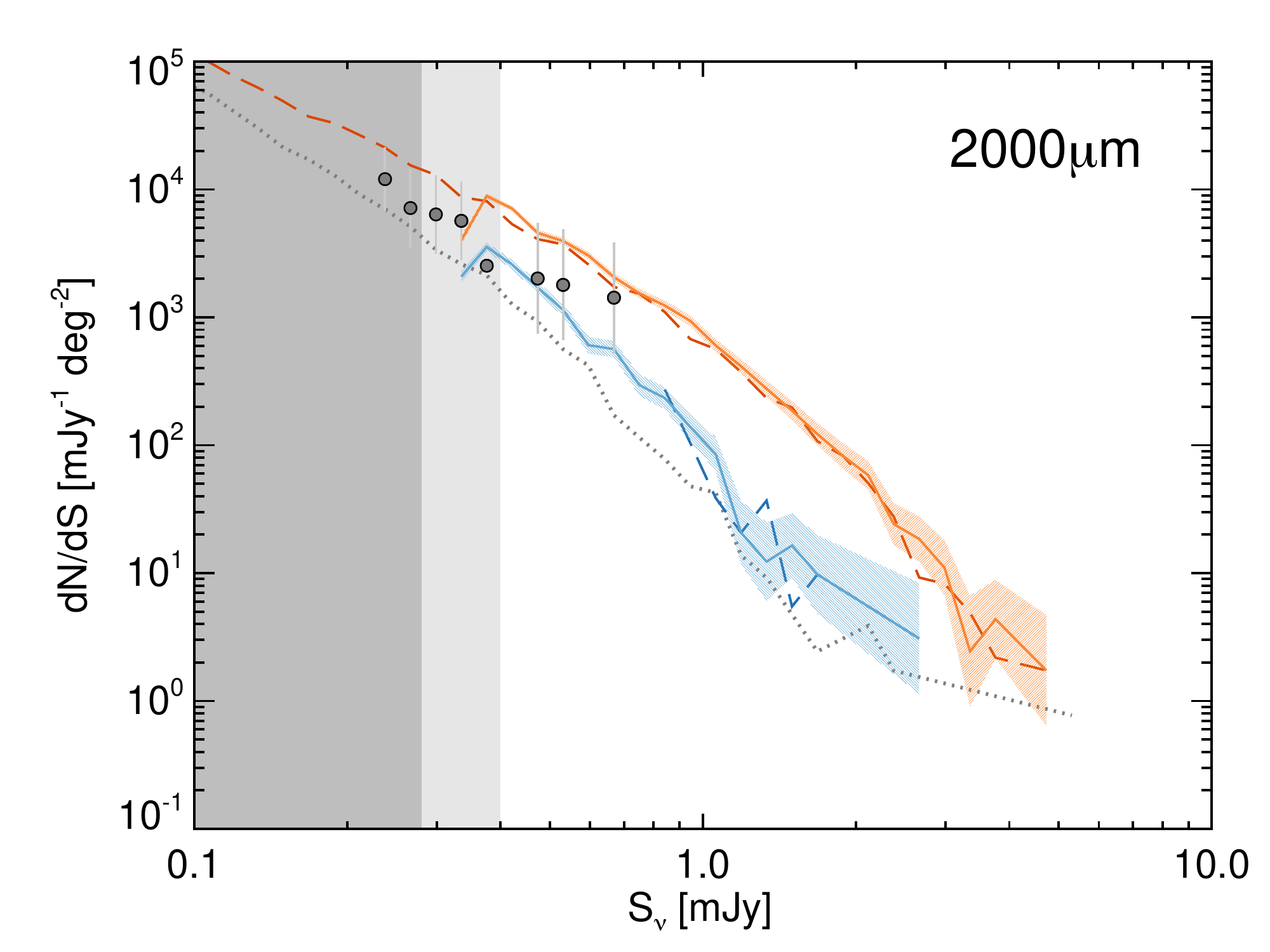}
\caption{The resulting differential number counts of our two
  case-study models from 70\um--2\,mm.  Data points (gray) from the
  literature are pooled from many sources referenced in the text and
  summarized in \citet{casey14a}.  The results of the SIDES model
  \citep{bethermin17a} are shown as a dotted gray line while the
  results of the dust-poor model (Model A) are shown in blue, and the
  dust-rich model (Model B) are shown in orange.  The injected source
  counts are the darker, long-dashed curves without uncertainty.  The
  extracted source counts, after degrading to the beamsize and RMS of
  typical observations, are shown as the lighter blue and orange, with
  shaded uncertainty as measured from a 1\,deg$^2$ simulated map.
  Dark gray background denotes flux densities at $<$3.5$\sigma$ for
  our simulation while light gray background denotes flux densities at
  3.5$<\sigma<$5.0.  The 70\um\ panel also shows the
    model predictions without the included AGN component (dotted blue
    and orange lines).} 
\label{fig:nc}
\end{figure*}

We also compare redshift distributions with the \citet{zavala14a}
model, which focuses exclusively on reproducing (sub)mm galaxy
redshift distributions from a single underlying high-$z$ population
that is detected at 1.1\,mm, but does not go so far as to model the
underlying IR luminosity function.

\vfill\eject

\subsection{Number Counts Comparison}

The data used as a baseline for comparison come from the existing rich
literature of FIR/submm number counts.  Specifically, we draw on
datasets at
70\um\ \citep{dole04a,bethermin10a,berta11a},
100\um\ \citep{heraudeau04a,rodighiero04a,kawara04a,berta11a,magnelli13a},
160\um\ \citep{dole04a,kawara04a,bethermin10a,berta11a,magnelli13a},
250\um, 350\um\ and 500\um\ \citep{patanchon09a,oliver10a,clements10a,bethermin10a,bethermin12a},
450\um\ \citep{smail02a,geach13a,casey13a,chen13a,wang17a,zavala17a},
850\um\ \citep{blain99a,scott02a,chapman02a,smail02a,cowie02a,borys03a,webb03a,barnard04a,coppin06a,scott06a,knudsen08a,beelen08a,weis09a,karim13a,casey13a,chen13a,simpson15a,geach16a}, and
1.1\,mm
\citep{perera08a,austermann10a,scott10a,hatsukade11a,aretxaga11a,scott12a,oteo16a,aravena16a,fujimoto16a}.
Though there have been some initial estimates of number counts at
2\,mm \citep[e.g.][]{staguhn14a}, they are not yet robust enough to
place useful constraints on our model.  Nevertheless, we include them
for a qualitative comparison.

Figure~\ref{fig:nc} shows differential number counts from 70\um--2\,mm
taken from the above literature measurements against both of our
models' number counts output, as well as the output from the SIDES
model.  There is generally good order-of-magnitude agreement of all
models in all bands, despite the very different model assumptions on
which each is built.

At shorter wavelengths, $\lambda_{\rm obs}<500$\,\um, the output of
our two case studies (blue as model A and orange as model B) are
indistinguishable from one-another.  This is largely due to the fact
that all emission at these bands comes from galaxies at $z<2$ where
our models are identical.  In both cases, the {\it Herschel} {\sc
  Pacs} wavelengths, 70\,\um--160\,\um, have required the adjustment
for AGN contribution to the mid-infrared powerlaw, as discussed in
\S~\ref{sec:models}.  No such adjustment is needed for the mock {\it
  Herschel} {\sc Spire} bands or ground-based submm wavelengths where
AGN are not a dominant emission mechanism.  

At wavelengths beyond $\lambda>200$\um, both of our models match the
galaxy number counts well, though as selection wavelength increases,
the separation between Model A and Model B become more distinct, with
Model A systematically providing lower number count predictions than
Model B.  Both 850\um\ and 1.1\,mm provide the most discriminating
power between the models while still having sufficient data available
to constrain number counts.  However, the spread on data measurements
is somewhat extraordinary, varying by up to an order of magnitude for
$\sim$mJy sources.  Both Model A and B predict 850\um\ and 1.1\,mm
number counts well within the constraints of existing data (though
perhaps model A is slightly favored at 1.1\,mm if you consider the
extension towards the faint end, sub-mJy sources).  The measurements
of 2\,mm number counts to-date are too few to meaningfully comment on
which model is favored; the GISMO Deep Field \citep{staguhn14a} only
contains seven sources due to its small size. However, it is clear
that of all the wavelength regimes, 2\,mm has the most discriminating
power between hypothetical high-$z$ models and should be a priority
for future observational efforts to constrain the high-$z$ IR
luminosity function.

We note that at these long wavelengths, the SIDES simulation (dotted
line) underpredicts the number density of $\sim$1\,mJy 850\um--1.1\,mm
sources in comparison to both our Models A and B and literature
measurements.  We attribute this to their adoption of a UV-like
\lstar\ and \phistar\ evolution towards the highest-redshift epochs.

\begin{figure}
\centering
\includegraphics[width=0.99\columnwidth]{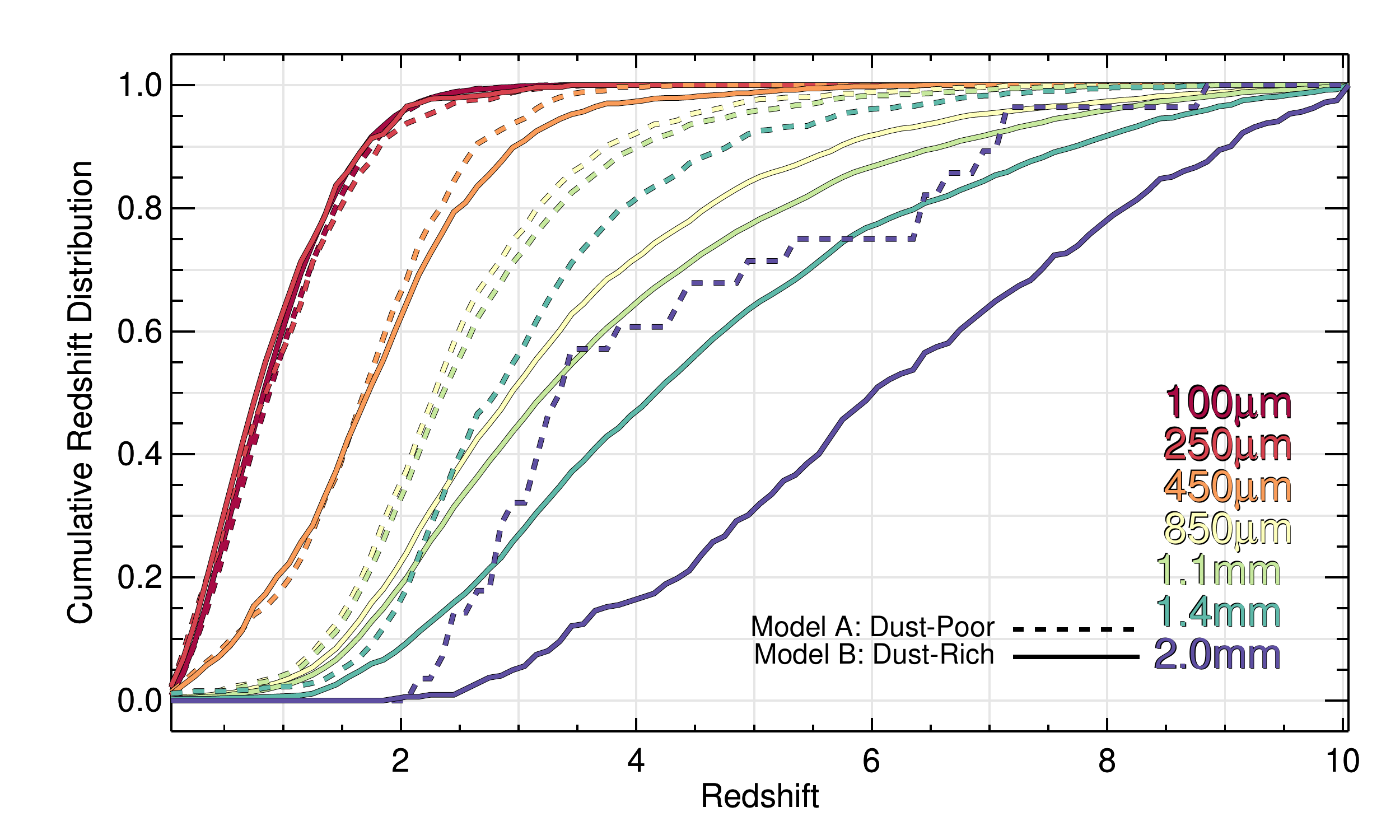}
\caption{ The predictive cumulative redshift distribution of modeled
  selection wavelengths for our two case studies. Predicted
  distributions for a dust-poor Universe (Model A) are shown as dashed
  lines and a dust-rich Universe (Model B) are shown as solid lines.
  The sources included on this plot are only those detected above a
  5$\sigma$ detection threshold in each of our 1\,deg$^2$ simulated
  maps, with noise and beamsize characteristics given in
  Table~\ref{tab:obs}.  Median redshift for the samples track with
  selection wavelength, and the two models become sufficiently
  distinct at selection wavelengths long-ward of $\sim$1\,mm.}
\label{fig:nz}
\end{figure}

\begin{figure*}
\centering
\includegraphics[width=1.1\columnwidth]{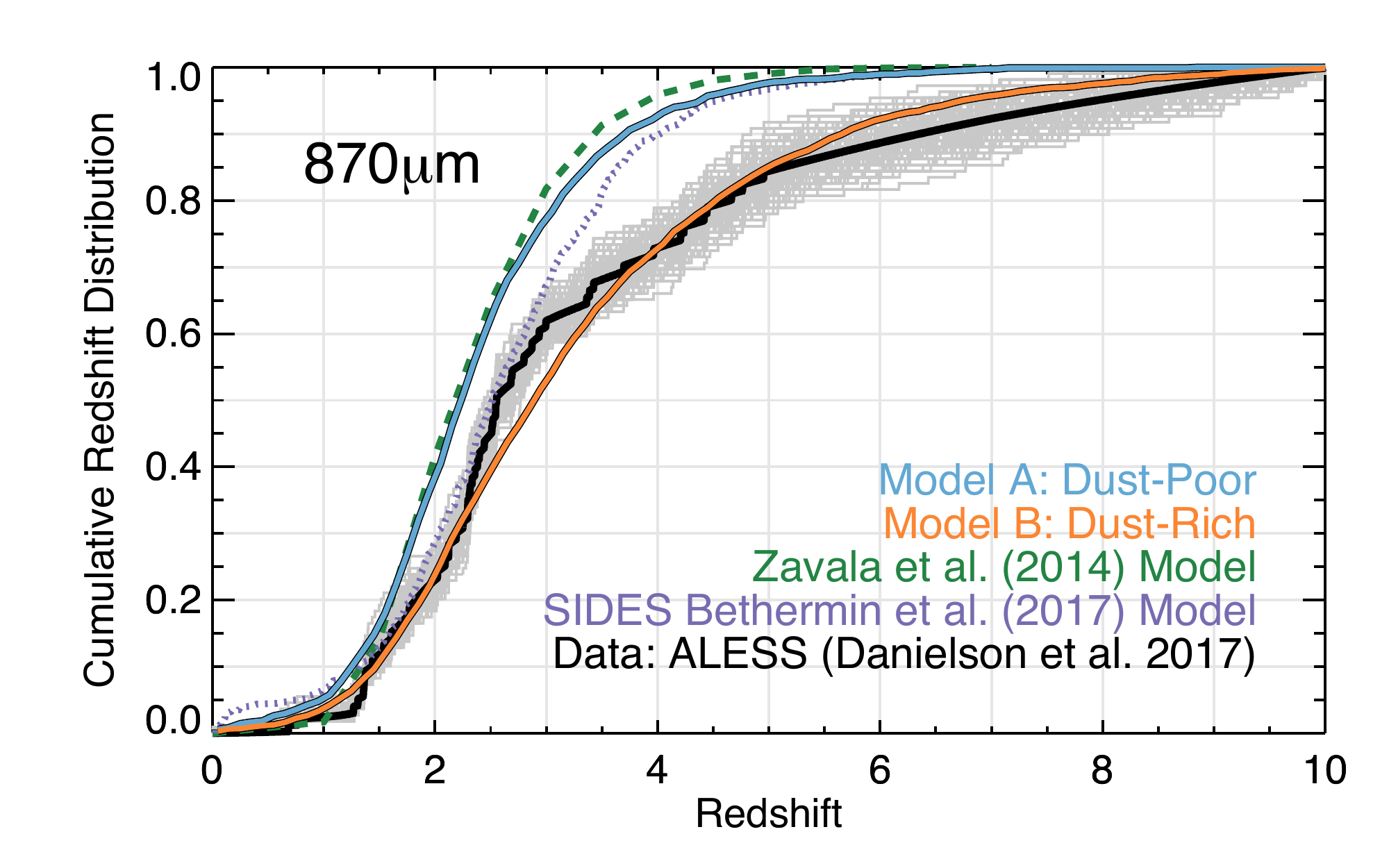}\\
\includegraphics[width=0.9\columnwidth]{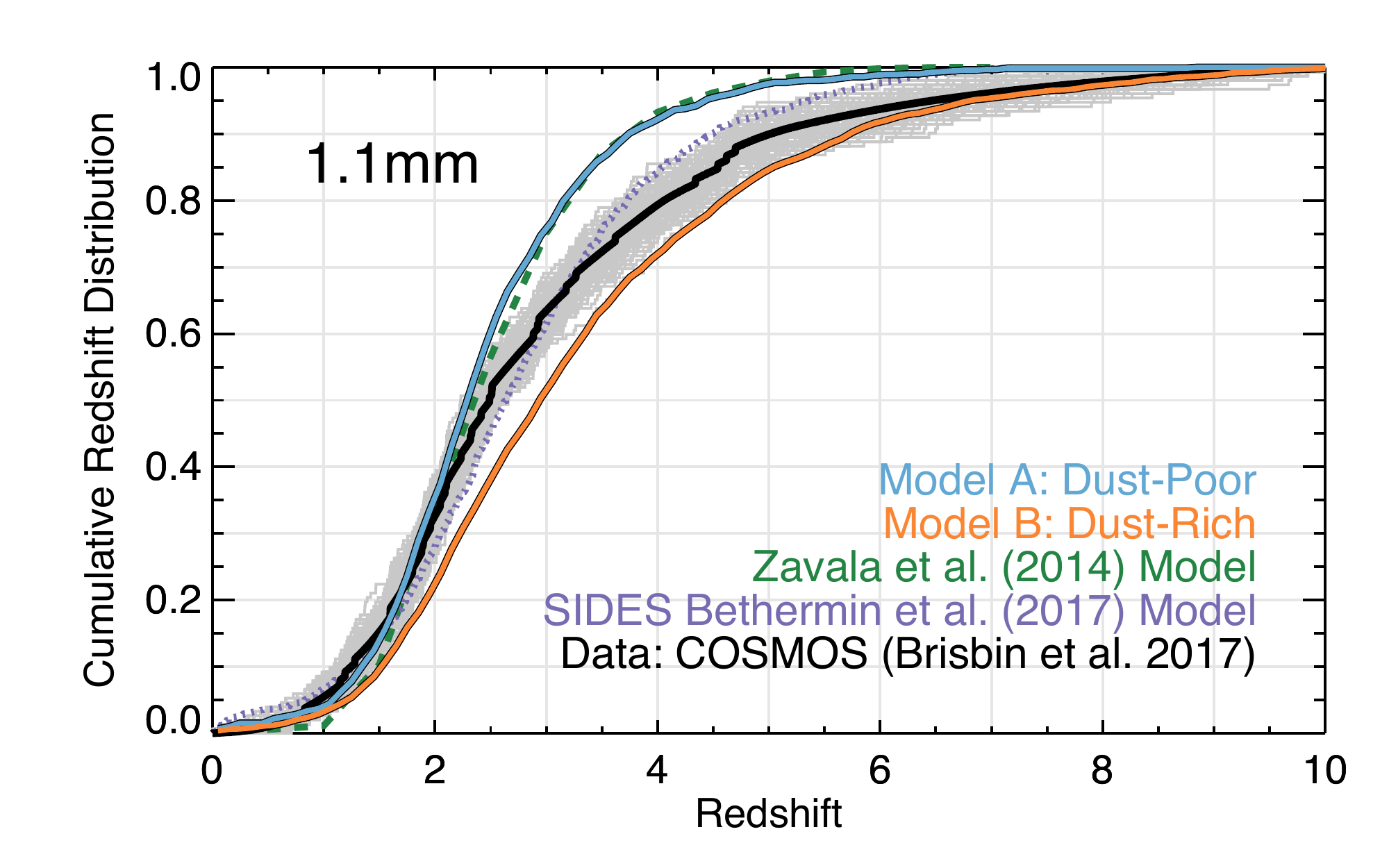}\includegraphics[width=0.9\columnwidth]{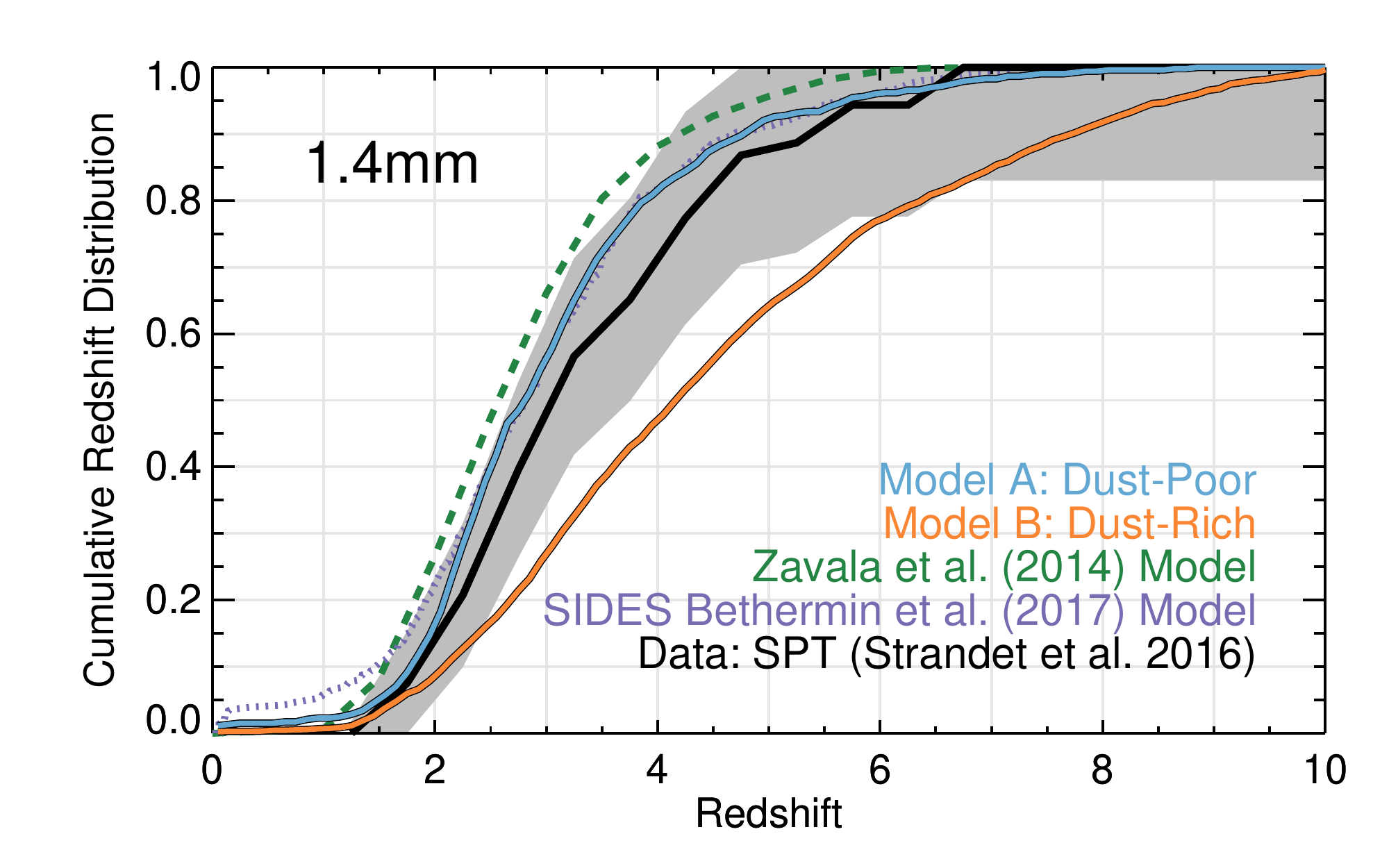}
\caption{Direct comparisons of our model-predicted redshift
  distributions and data.  Here we compare against three independent
  and robust datasets on redshifts of (sub)millimeter galaxies that
  are the most complete to-date: the ALESS spectroscopic survey of
  LABOCA-selected 870\um\ sources \citet{danielson17a}, the combined
  spectroscopic and photometric survey of AzTEC/ALMA-followed up
  sources in COSMOS at 1.1\,mm from \citet{brisbin17a}, and the South
  Pole Telescope 1.4\,mm-selected SMGs described in
  \citet{strandet16a}.  Gray regions denote uncertainties in the
  measured distributions.  Model A is denoted by the blue line and
  Model B is the orange line (note that these curves are not identical
  to those in Figure~\ref{fig:nz} due to different flux density
  limits).  The dashed green line shows a comparison to the
  \citet{zavala14a} model redshift distribution, and the predictions
  of SIDES \citep{bethermin17a} are shown as a dotted purple line.}
\label{fig:nzdata}
\end{figure*}

\subsection{Redshift Distributions}\label{sec:nz}

Due to the very negative K-correction in the submillimeter
\citep*{casey14a}, sources' flux densities are largely
independent of redshift, and so redshift distributions break down
another important dimension of our mock data, giving an independent
measurement of the quality of our models that is distinct from the
comparison of number counts.  Redshifts for extracted sources are
determined by first identifying all possible sources within a beamsize
(as stated in Table~\ref{tab:obs}) that could contribute flux to a
given identified source in the output map.  The injected source that
contributed the most input flux density at the selection wavelength
within the beamsize is then marked as the primary source.  This method
is imperfect, as it assumes that the redshift of the brightest source
within a beam provides the best representation of a given detected
source, and it neglects the impact of source multiplicity caused by
line-of-sight projections \citep{hayward13a}.  However, we find that
this method provides a fair representation of the statistical
populations of galaxies detected at each of the selection wavelengths
(in particular the observational setups with higher angular
resolution).  This method is also most analogous to the observational
methods used in the majority of redshift survey work carried out
to-date on single-dish submm galaxy populations
\citep{chapman03a,chapman05a,swinbank04a,wardlow11a,casey12b,casey12c,casey17a,danielson17a}.

Figure~\ref{fig:nz} illustrates the predicted cumulative redshift
distributions from 70\um--2\,mm for models A (dashed lines) and B
(solid lines). The importance of the long-wavelength regime
($\lambda_{\rm obs}\simgt1$\,mm) for picking out high-$z$ DSFGs
becomes quite clear on this plot.  As might be self-evident, a
dust-poor early Universe should have far fewer high-redshift
detectable sources than a dust-rich early Universe, and so Model A's
results are more skewed toward lower-redshift distributions.  For
example, at 870\um\ \citep[the selection wavelength of the ALESS
  sample, one of the best-studied, uniformly-selected samples DSFGs in
  the literature;][]{hodge13a,danielson17a}, the predicted median
redshift for the brightest, most robust subset ($S_{\rm
  870}>3.5$\,mJy) is $\langle z_{870}\rangle=2.4$ for Model A vs
$\langle z_{870}\rangle=3.1$ for Model B.  The predicted median
redshift for a $S_{\rm 1.1\,mm}>1.5$\,mJy sample is $\langle
z_{1.1}\rangle=2.4$ for Model A and $\langle z_{1.1}\rangle=3.4$ for
Model B.

Intriguingly, measurements of redshift distributions from the
literature do not completely favor model A and rule out model B.  The
direct comparisons of redshift distributions for measured samples (at
matching flux density cuts) is shown in Figure~\ref{fig:nzdata}.  The
three data samples given are the most robust, most complete
sub-samples of spectroscopically-confirmed DSFGs in the literature.
At 870\um, we compare against the LABOCA-selected \citep{weis09a},
ALMA followed-up \citep{hodge13a} sample whose redshift survey is
described in \citet{danielson17a}.  At 1.1\,mm, we compare against
AzTEC-selected \citep{aretxaga11a} sources in the COSMOS field that
have ALMA or other interferometric followed-up
\citep{smolcic12a,brisbin17a}.  Both the 870\um\ and 1.1\,mm samples
are hybrid combinations of spectroscopic and photometric redshifts,
with a few sources lacking redshifts completely.  To accurately model
the uncertainty on the aggregate redshift distribution for the whole
sample, we assign each source a probability density function in
redshift according to measurement uncertainties (either spectroscopic
or photometric); sources without constraints are assumed to sit at
$z>1$ with probability density function mirroring a step-function. We
then generate many realizations of the measured cumulative redshift
distribution by drawing from each source's individual probability
density function.

The last data sample we use for comparison is the SPT,
1.4\,mm-selected sample of lensed SMGs discussed in
\citet{strandet16a}.  While this sample is the most spectroscopically
complete of any DSFG sample, it has the added complication in that
relatively small and it is almost entirely comprised of
strongly-lensed systems, which is a natural consequence of its high
flux density cut, $S_{\rm 1.4}>20$\,mJy \citep{weis13a}.  The measured
median magnification factor for the SPT sample is $\langle
\mu\rangle\approx 6$ \citep{hezaveh13a,spilker16a}, and so the
equivalent unlensed flux density limit would be around
$S_{1.4}\simgt3.3$\,mJy.  To mimic this selection in our models, we
select 1.4\,mm-detected sources assuming a 6.9$''$ beamsize FWHM above
a flux density of 2\,mJy (which accounts for additional uncertainty in
the magnification factor).  The SPT sample also has a redshift bias
that excludes sources at $z\ll 1.5$ \citep{hezaveh11a} due to the low
probability of galaxy-galaxy lenses.  While this second effect could
be substantial, the lack of $z<1.5$ sources in our 1.4\,mm-selected
sample to begin with implies this effect is negligible.

\begin{figure*}
\centering
\includegraphics[width=0.99\columnwidth]{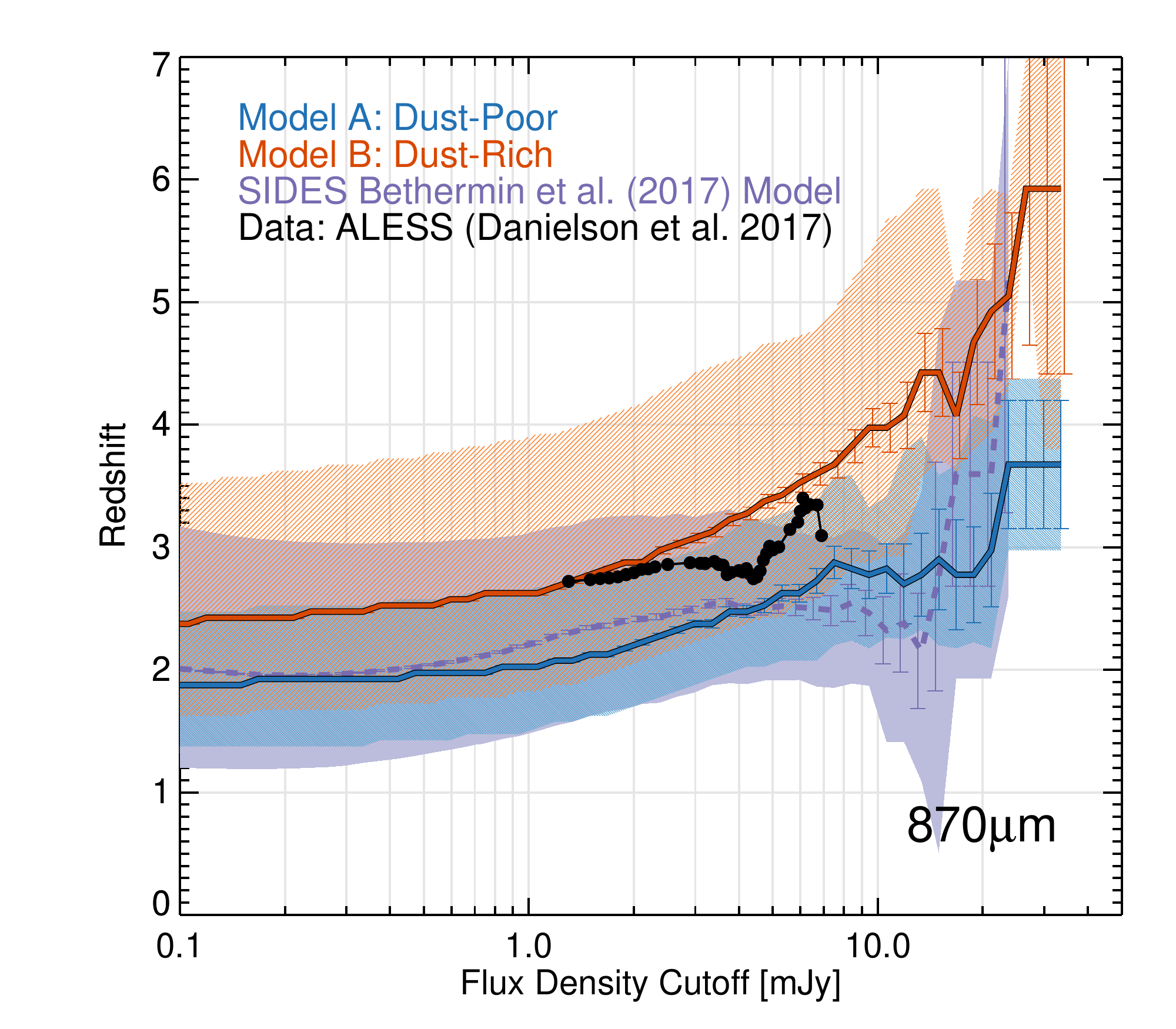}\includegraphics[width=0.99\columnwidth]{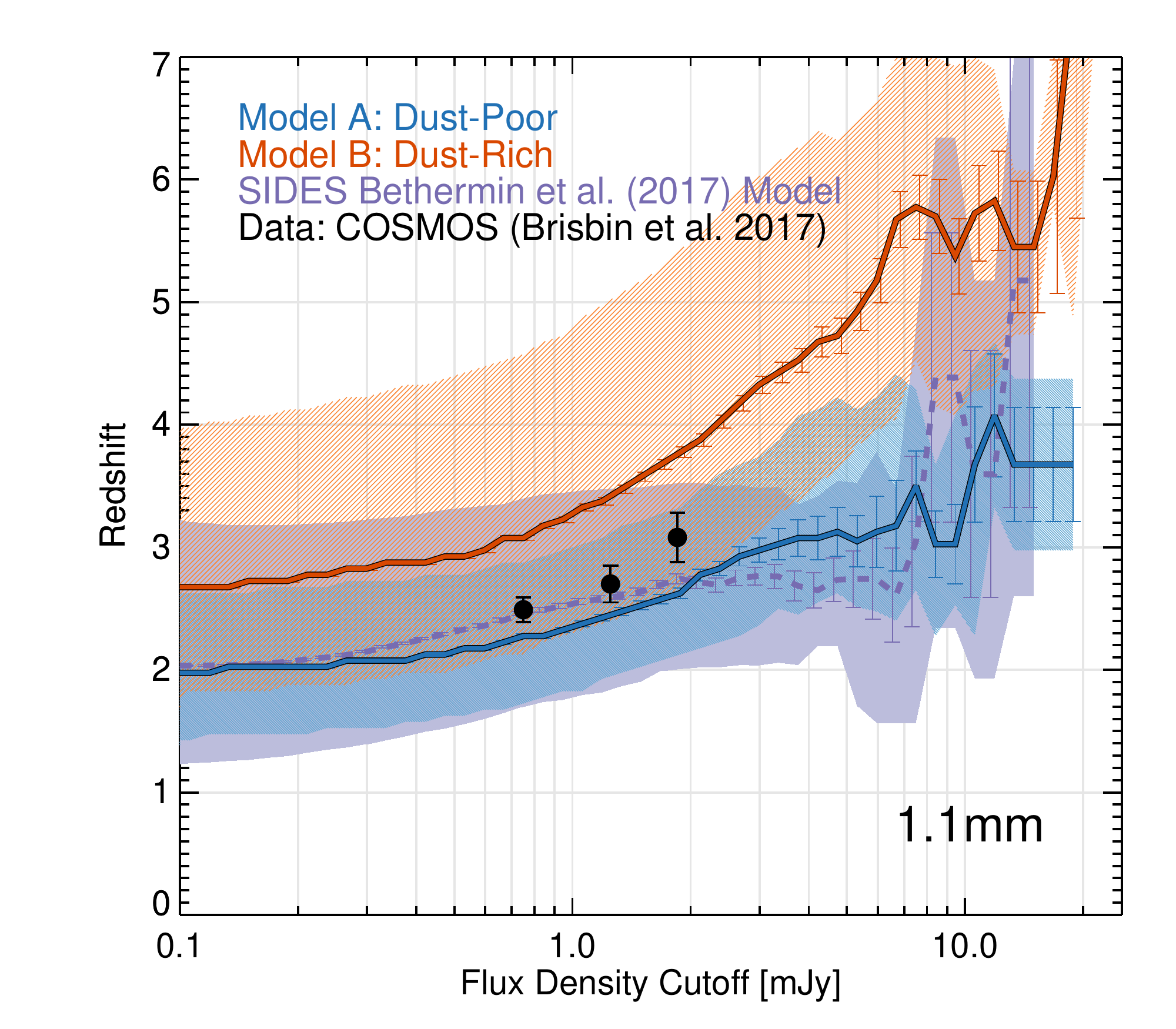}
\caption{The dependence on flux density cutoff and that galaxy
  sample's median redshift for our Model A (blue), Model B (orange),
  and the SIDES model (purple).  The shaded regions enclose the inner
  50\%\ of sources from each model, and the bootstrap-measured
  uncertainty on the median is shown via the colored error bars; these
  become dominant in the flux density regime where there are very few
  sources per square degree.  We compare to data from both ALESS
  \citep{danielson17a} and COSMOS \citep{brisbin17a} at 870\um\ and
  1.1\,mm, respectively.}
\label{fig:bet}
\end{figure*}

At all three wavelengths, we also include comparisons to the SIDES
model output and the model redshift distributions of
\citet{zavala14a}.  While none of the models -- including those from
this paper -- provide perfect matches to the redshift distributions
found by the data, the comparison is illuminating as to the nature of
high-$z$ DSFG prevalence.  For example, the measured median redshift
for ALESS 870\um-selected galaxies is $\langle
z_{870}\rangle=2.4\pm0.1$, or 2.6$^{+0.2}_{-0.1}$ after accounting for
unconfirmed sources and likely high-$z$ solutions for those sources.
Both the \citeauthor{zavala14a} and our dust-poor Model A have median
redshifts of 2.2 falling short of the median measured redshift.  While
the median of SIDES output is the closest to data, the SIDES model
fails to catch the measured high-redshift tail of the population.
Model B, while overestimating the median of the full sample,
accurately catches this high-redshift tail of the population.
A similar phenomenon is seen in the comparisons of data and models at
1.1\,mm, though more subtle.  As pointed out in \citet{brisbin17a},
the measured redshift distribution at 1.1\,mm is very dependent on the
exact flux cutoff (they measure $\langle z\rangle=2.18\pm0.09$ for
sources with $S_{1.2}<1.25$\,mJy and $\langle z\rangle=3.08\pm0.17$
for sources with $S_{1.2}>1.8$\,mJy).  Both the ALESS and COSMOS
samples exhibit significant high-$z$ tails, with 23\%\ of
870\um\ sources above $z>3$ and 27\%\ of 1.1\,mm sources above $z>3$.
At 1.4\,mm, the small number statistics on the measured redshift
distribution implies more uncertainty and difficulty in distinguishing
between favored models, though model A is slightly more favored (with
model B in less than 1$\sigma$ tension with data).

Figure~\ref{fig:bet} shows a more detailed breakdown of the
flux-density dependence on samples' median redshifts.  As a function
of the lower-limit flux cutoff of a survey, we compute the median
redshift for all galaxies in our model detected above that threshold,
and also compare against the SIDES model input.  The
\citeauthor{danielson17a} and \citeauthor{brisbin17a} samples are used
for comparison.  The interpretation of this analysis is not
straightforward: the data are in slight tension with the SIDES output
and fall between our two extreme model case studies, likely pointing
to a true value of $-5.9<\psi_{2}<-2.5$.  We wish to highlight that
these results, instead of adequately distinguishing between
conflicting models, fail to rule out even extreme models for the
prevalence of DSFGs in the early Universe.  We discuss a possible path
forward with 2\,mm observations in \S~\ref{sec:future}.

\subsection{Correlation of Flux Densities across Bands}\label{sec:corr}

Beyond number counts and redshift distributions, we can also compare
flux densities across (sub)mm bands by measuring their relative
correlation.  For example, are sources' detected at 850\um\ also
detected at 450\um, and are their flux densities proportional to one
another?  To measure such a correlation for any two bands, we
construct lists of corresponding flux densities from the union of
detected sources in either band.  For sources detected in both bands,
the extracted peak flux densities in each band is recorded with a
positional match accuracy less than the smallest beamsize; for
single-band detections, we measure the flux density in the other band
at the position of the extracted source.  From this sample, we measure
 the Pearson's correlation coefficient
$\rho(\lambda_{1},\lambda_{2})$ such that:
\begin{equation}
\rho(\lambda_{1},\lambda_{2}) = \frac{cov(\lambda_{1},\lambda_{2})}{\sigma_{\lambda_{1}}\sigma_{\lambda_{2}}}
\end{equation}
where $cov(\lambda_{1},\lambda_{2})$ is the covariance of measured
flux densities at observed $\lambda_{1}$ and $\lambda_{2}$, and
$\sigma_{\lambda}$ is the standard deviation of flux densities for
either sample.  Values near one represent perfect correlation,
zero represent no correlation, and negative one represents perfect
anti-correlation.  One could imagine that a universe with very diverse
dust SEDs could result in lower cross-band correlation than a universe
where galaxies' dust SEDs are remarkably similar.  In this sense,
measuring the correlation across bands in real data samples and in
models gives a measure of how homogeneous or diverse galaxies' dust
SEDs are.  For the models, this makes use of extracted sources
only, after maps have been degraded to the spatial resolution and
instrumental noise of real observations.

Extracting such correlations from real data is somewhat challenging
for large samples, since very few regions of the sky have been mapped
deeply in many bands across the far-infrared/submm.  For its superb
multiwavelength coverage, we draw on maps already in hand in the
COSMOS field, from {\it Herschel} {\sc PACS}, {\sc SPIRE}, {\sc
  SCUBA-2}, and {\sc AzTEC} data.  We use the compilation of data from
\citet{lee13a}, \citet{casey13a} and \citet{aretxaga11a}, in addition,
drawing on the {\sc SCUBA-2} data for the EGS described in
\citet{zavala17a}.  

\begin{figure}
\centering
\includegraphics[width=0.99\columnwidth]{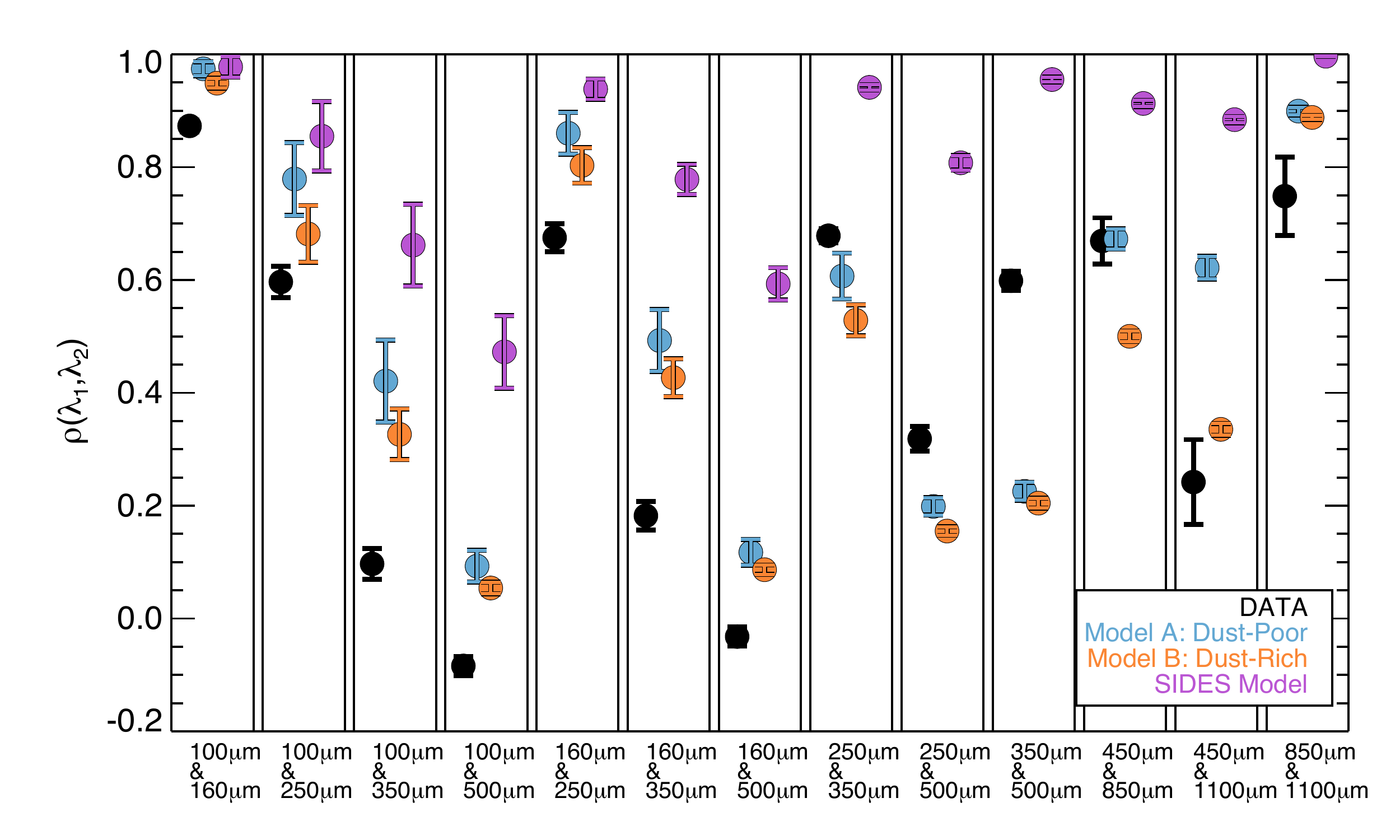}
\caption{The correlation coefficient, $\rho(\lambda_1,\lambda_2)$
  between flux densities of detected sources at $\lambda_1$ and at
  $\lambda_2$ from real data in the COSMOS field \citep[black
    points;][]{aretxaga11a,lee13a,casey13a}, from the output of our
  models: model A (blue) and model B (orange), as well as the SIDES
  model \citep[purple;][]{bethermin17a}. Values close to one indicate
  strong correlation between sources' measured flux densities in the
  two bands, while zero represents no correlation, and a negative
  value would indicate anti-correlation.  We find that SIDES is
  generally much more correlated across all bands than our simulations
  as well as real data.}
\label{fig:corr}
\end{figure}

Figure~\ref{fig:corr} shows the results of our comparison of flux
densities in each band pairing for which we have adequate data.  We
compare the correlation of the same band fluxes from the SIDES model
output catalog \citep{bethermin17a}, and both the output from our
Model A, the dust-poor model, and Model B, the dust-rich model. We
note that the predictions of SIDES suggest significantly more
correlation across bands than exists in measured maps.  The output of
our simulations, both A and B, are less correlated across bands than
SIDES.  In some cases, both of our models overestimate the correlation
(at short wavelengths $<$250\um), likely due to the lack of inclusion
of an array of mid-infrared powerlaw slopes.  At longer wavelengths,
our models are more aligned with observed correlation.  Because this
measured correlation coefficient is a tracer of SED diversity and not
particularly sensitive to source number density, both Models A and B
largely follow the same trends in cross-band correlation with the
exception of the longest wavelength bands, where we see the divergence
of model predictions between Model A and Model B predominantly because
of the growing dominance of higher-redshift sources in Model B at
$\lambda\simgt850$\,\um.

\section{Discussion}\label{sec:discussion}

This paper presents a
  model to interpret observations of the
IR/submm sky.  With mock
  observations we have compared the results of two case-study models
 with extremely different assumptions about the
shape and evolution of the IR luminosity function beyond $z>2$.  Model
A has assumed the early Universe is dust-poor, with DSFGs contributing
very little to cosmic star-formation at $z>4$, while Model B has
assumed the opposite: that DSFGs are so dominant at $z>4$ that they
render the contribution from LBGs negligible.  The physical
implications of these models differ wildly.

\begin{figure*}
\centering
\includegraphics[width=1.05\columnwidth]{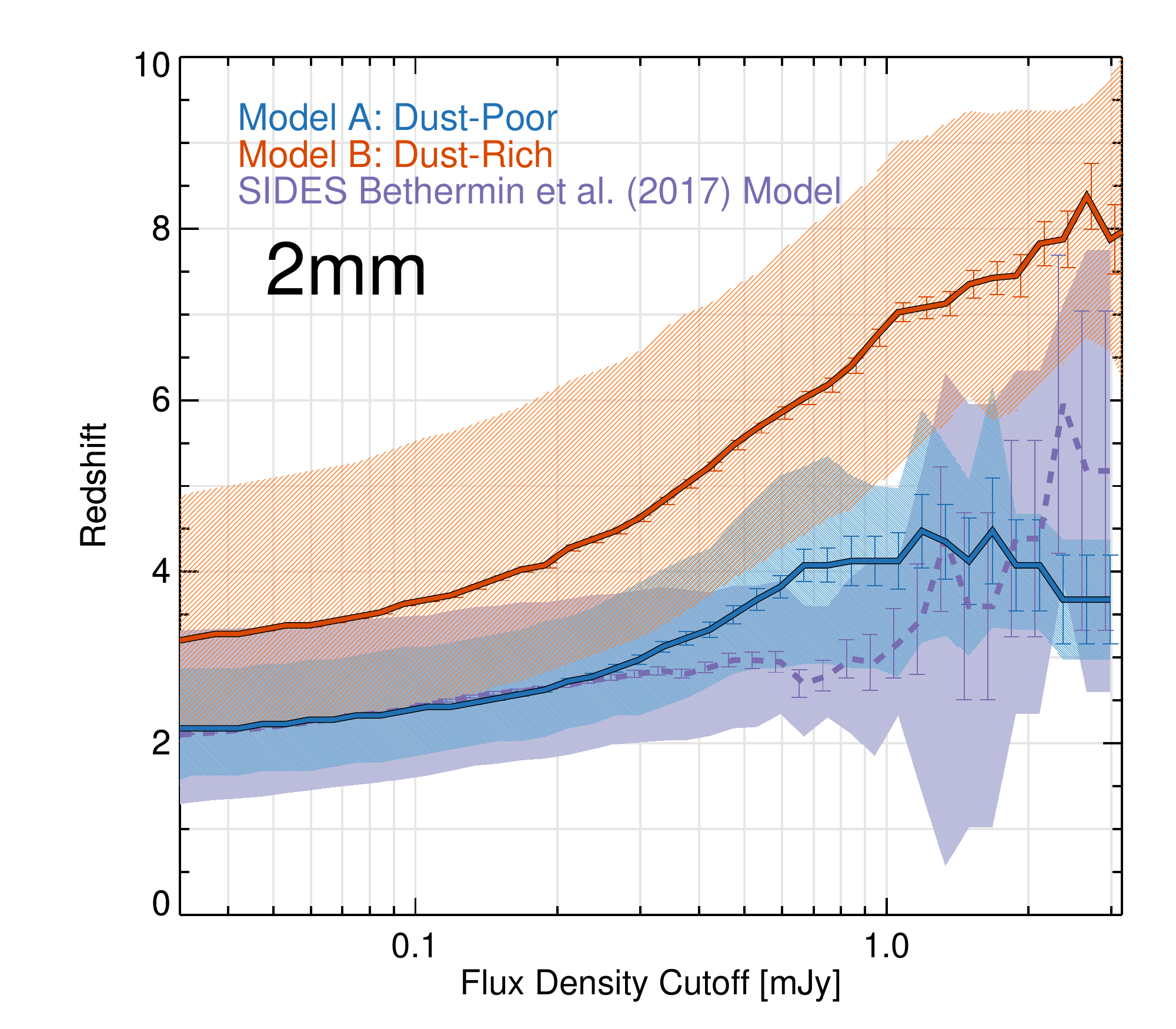}
\includegraphics[width=0.94\columnwidth]{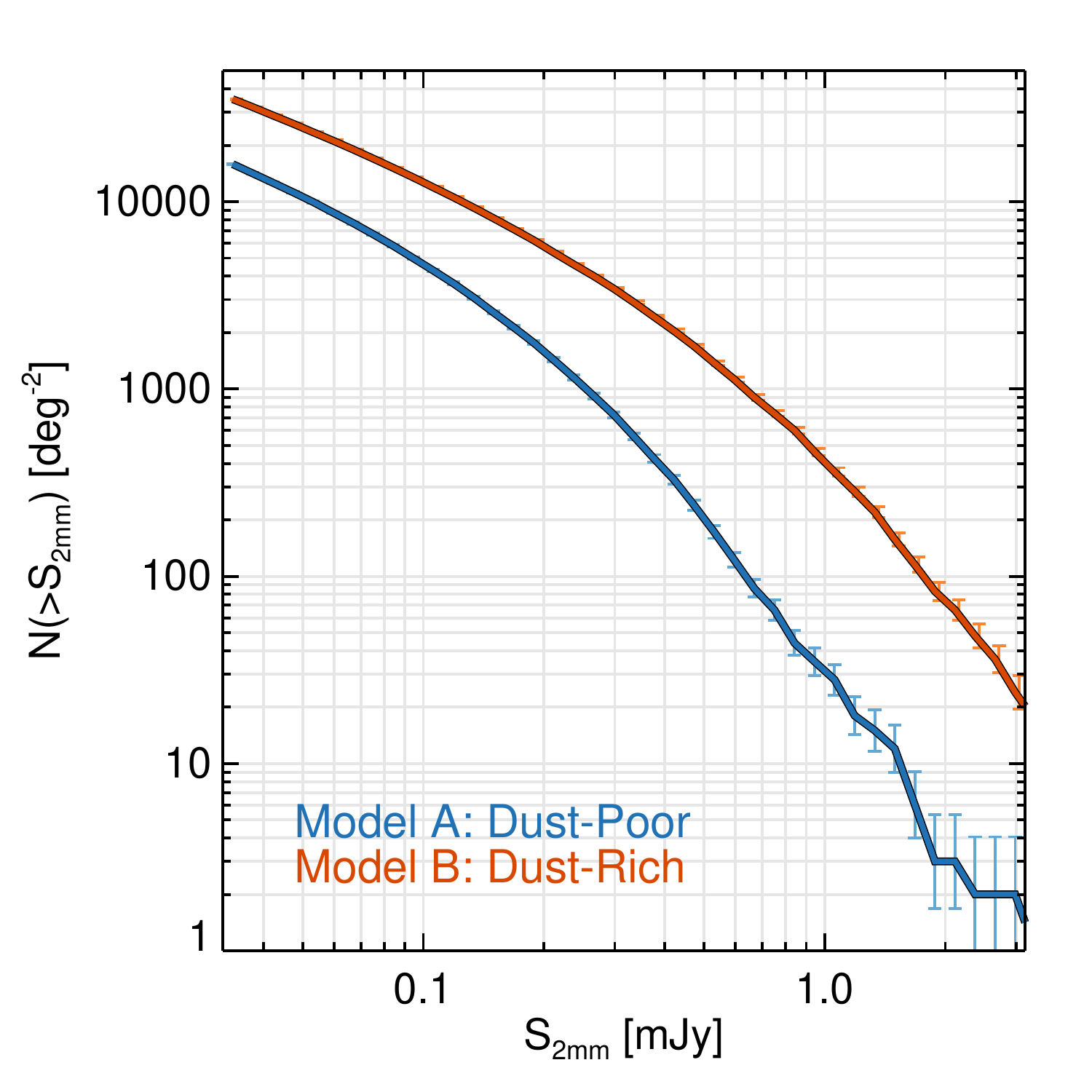}
\caption{The combined predicted characteristics of future 2\,mm DSFG
  surveys.  At left, we show the flux-dependent redshift distributions
  (as shown in Figure~\ref{fig:bet} at other wavelengths) between
  model A (dust-poor; blue) and model B (dust-rich; orange).  The most
  dramatic difference in median redshifts by model are for sources
  with flux densities 0.3--1\,mJy.  At right, we show the predicted
  cumulative number counts at 2\,mm for both models, which should be
  easily distinguishable in surveys $\sim$0.1\,deg$^2$ down to flux
  densities of $\sim$0.5\,mJy (at 5$\sigma$).}
\vspace{9mm}
\label{fig:pred2000}
\end{figure*}

\subsection{Broad Implications for DSFGs in the Early Universe}

In the case of Model A, star formation at early times is dominated by
UV-luminous LBGs, and dust-rich systems may be exceedingly rare. This
model is aligned with some literature results that suggest a diminished
role of dust-enrichment in `normal' galaxies at $z\simgt5$
\citep{capak15a,bouwens16a}. This model could also be in agreement
with the relatively few DSFGs discovered above $z>5$ to-date, though
their volume density has not been directly constrained
\citep{strandet17a}.  Furthermore, the measurements of the
star-formation rate density of the Universe from deep optical and
near-infrared measurements, as shown in Figure~\ref{fig:sfrd}, would
hold as universally true to all galaxy types.

In the case of model B, DSFGs would be far more prevalent in the early
Universe, dominating cosmic star-formation ($>$90\%\ of the total).
This would physically imply that dust production mechanisms post-Big
Bang would need to be particularly efficient and likely form via
supernovae \citep{matsuura11a,dwek14a} combined with low destruction rates,
rather than from coagulation in the upper atmospheric winds of AGB stars, or
coagulation or accretion of dust in the ISM
\citep{matsuura06a,matsuura09a,jones13a}.  Recent simulations of the
first stellar production of metals could produce some dust grains as
early as $z\sim25$, though it appears the dust at that epoch does not
significantly impact the observable characteristics of typical
galaxies \citep{jaacks18a}.  It is yet unclear, from a simulator's
perspective, how early Universe DSFGs might form rapidly.
Furthermore, DSFGs tend to be quite massive galaxies, so
their formation on short $<$1\,Gyr timescales may be needed to
directly constrain cosmological models of massive halo growth
\citep{wechsler02a}.  If Model B were correct, it would mean that
most of early Universe star-formation has not been accounted for in
the census of galaxies at $z>5$.  DSFGs at this epoch would be
entirely obscured, in stark contrast to the population of LBGs seen at
the same epoch which appear to be much more dust-poor than their
analogues at lower redshift.

Current datasets do not clearly rule out either of these two simple
yet extreme models.  Though number counts are largely invariant with
model assumptions and lack constraining power \citep[because at most
  wavelengths they are dominated by sources at $z<2$,][]{lagache04a},
it is the deficit of sources at long wavelengths
($\lambda\simgt$850\,\um) in the dust-poor model A and SIDES model
that hint towards a possible higher prevalence of DSFGs at $z>2$ as
present in Model B. Unfortunately, direct sample measurements of the
IRLF from well characterized DSFGs only confirm that constraints peter
out beyond $z\sim2.5$, largely due to the lack of completeness (at
$\lambda\simlt850$\,\um).  Measured galaxy redshift distributions at
870\um--1.4\,mm provide the cleanest contrast between dust-poor and
dust-rich models, while neither model is clearly favored; where
some measures (median redshift) might favor model A, other measures
(the distribution of sources in the highest-redshift tail) might favor
model B.  It is also clear that any sample incompleteness in redshift
surveys would severely hamper our ability to draw meaningful
conclusions, and most redshift surveys of DSFGs are far too incomplete
to be of use (i.e. the gray regions on Figure~\ref{fig:nzdata},
already large for plotted samples, would be much larger for any
samples suffering from more incompleteness).  In that sense, current
datasets have very little discriminating power even with extremely
different input assumptions.

Though our model provides some broad context for interpretation of
high-$z$ DSFGs, it does nothing to explain the physical origins of
such systems.  As discussed extensively in Chapter 10 of
\citet*{casey14a}, cosmological simulations of DSFGs are especially
challenging, both from the perspectives of semi-analytic model (SAM)
N-body simulations
\citep[e.g.][]{baugh05a,lacey08a,somerville12a,hayward13a} and
hydrodynamic cosmological simulations
\citep[e.g.][]{springel05a,vogelsberger14a}.  They are particularly
challenging due to the high computational cost of dust radiative
transfer (required for shorter wavelengths where dust emission is not
optically-thin) in a cosmological context.  In the case of SAMs, dust
prescriptions are analytically described from galaxies' halo
characteristics and no direct modeling of ISM processes is included;
there are few calibrations against observational data to refine these
dust prescriptions, and we know they under-predict the number of DSFGs
across most redshift regimes.  Cosmological hydrodynamic models, like
the recent Illustris and Illustris TNG models
\citep{vogelsberger14a,vogelsberger14b,vogelsberger18a,weinberger17a}
have still not been able to directly implement dust prescriptions.
One hydrodynamic simulation that uses smooth particle hydrodynamics
(SPH) and does model dust emission by linking it to metal abundances
is the {\sc BlueTides} simulation \citep{wilkins17a,wilkins17b}; to
date, {\sc BlueTides} has run down to $z=8$ with a hopeful expansion
down to $z=6$ in the near future.

\citet{wilkins17a} presents results on the $z\ge8$ dust-obscured
galaxy population from {\sc BlueTides}, and remarkably come to a
conclusion that would largely be aligned with our Model B, the
dust-rich Universe.  They find that approximately 90\%\ of
star-formation in high-mass galaxies $>10^{10}$\,\msun\ at $z>8$ is
already obscured by dust.  This is in-line with mass-dependent
obscuration fractions measured at much later epochs
\citep{whitaker17a}, as well as some high-$z$ work on the
mass-dependent dust content of UV-selected galaxies, which is found to
be constant from $4<z<7$ \citep[][also see
  \citealt{bowler18a}]{finkelstein12a}.  It counters the argument that
galaxies are less dust-rich beyond $z>4$ \citep{bouwens16a}.
\citeauthor{wilkins17a} highlight the importance of pushing submm
surveys into new parameter space, in the direction of directly
constraining the prevalence of DSFGs at such high-redshifts.

\subsection{Impact of Individual Model Assumptions}

Though we caution that neither model A or model B is clearly favored
by existing datasets, suggesting that model B is not cleanly ruled out
is a somewhat bold claim, and should be immediately followed-up with
an analysis of each individual parameter's impact on that conclusion.
For example, if we were to assume a different history to the evolution
of dust SEDs, how would the conclusion that DSFGs are prominent at
$z>4$ possibly change? We explore some implications here with more
detailed analysis presented in Appendix~\ref{sec:appendA}.

Though we understand the shape and diversity of dust-rich galaxies' IR
SEDs well from $0<z<2$, there could be reason to doubt our SED model
assumptions hold for galaxies that formed in the first few billion
years after the Big Bang.  For example, early Universe galaxies are
likely far less metal-enriched than those at low-redshift, and they
might have had a different (possibly top-heavy) initial mass function
\citep{baugh05a,swinbank08a}.  These effects of metal-enrichment and
possible varying IMF \citep[linked to formation of stars in super-star
  clusters, SSCs;][]{mckee03a,bastian10a,portegies-zwart10a} may
result in very different geometric distribution of dust and gas in the
ISM of the galaxies.  It is this geometry -- densely-packed dust or
diffuse dust, patchy or smooth -- that directly impacts the mass of
dust heated to different temperatures, which in turn, shapes the
galaxy's IR SED.  It could be that the more primordial dust-rich
galaxies were much more densely-packed, hot-dust dominated than the
mature, massive and extended systems seen at $z\sim2$.

If the $z\simgt4$ DSFGs are intrinsically hotter than DSFGs at
$z\sim2$ \citep[as explicitly suggested by ][]{faisst17a}, then we may
expect the \lir-\lpeak\ relationship at $z\sim5$ to evolve
significantly away from the one for $z\sim2$.  This would translate to SEDs
shifted to shorter rest-frame wavelengths, and a reduction of flux for
wavebands still probing the Rayleigh-Jeans tail, at $\lambda_{\rm
  obs}>1$\,mm. Implementing this shift into our model explicitly
results in the number counts of both model A and B to shift downward
by $\approx$0.3\,dex at $\lambda_{\rm obs}\simgt$850\,\um, and a
contraction of redshift distributions towards lower redshifts (median
redshifts a factor of $\sim1.5\times$ lower), more discrepant with
measurements.  It is for this reason -- in addition to the argument
that the dust-temperatures of the SPT samples are well aligned with
colder temperatures\footnote{The SPT samples could be
    biased towards cold temperatures with respect to the median at
    high-$z$, but as discussed in the Appendix, if DSFGs are much
    hotter at high-$z$ than the relation shown in Figure~\ref{fig:tdust},
    then generating number counts at 850\um\ or 1.1\,mm that reproduce
    measurements would require an even higher DSFG contribution to
    cosmic star-formation than is shown for Model B.} -- that we do
not think DSFGs near the epoch of reionization are hot.

The impact of the CMB heating is very significant and should not be
lost on the reader due to its somewhat straightforward implementation
in our model.  In the absence of CMB heating, at high-$z$, galaxies'
dust temperatures would be physically cooler, their flux densities
would be brighter against the background, and thus
much easier to measure.  Its absence would make the detection of DSFGs
at early epochs significantly easier.  For example, while our current
dust-rich model B, we estimate $\sim$3000 sources\,deg$^{-2}$ above a
2\,mm detection threshold of 0.3\,mJy, and without the CMB effects and
the same underlying model we would detect $\simgt$4500--6000
sources\,deg$^{-2}$ at the same threshold.
Indeed, the CMB effect is of crucial importance to the interpretation
of the prevalence of high-$z$ DSFGs, without it we might be easily
lead to faulty conclusions, and its impact has made it far more
difficult to discern the underlying IR luminosity function at early
epochs.

A more subtle effect not yet discussed in this paper is the assumption
of $\beta_{E}=1.8$ for the dust emissivity index of galaxies (without
any intrinsic variation).  $\beta_{E}$ effectively impacts the slope
of the Rayleigh-Jeans tail (higher values resulting in steeper SEDs),
and measures between $\beta=1-2$ in star-forming galaxies
\citep{hildebrand83a,dunne01a}, and is thought to anti-correlate with
dust temperature according to lab experiments and detailed
measurements from nearby galaxies
\citep{lisenfeld00a,shetty09a,tabatabaei14a}. The emissivity index is
not thought to vary significantly with metallicity
\citep{remy-ruyer13a}, although the environments of DSFGs in the early
Universe could be quite different from the local environments in which
these measurements are made.  Lower values of $\beta_{E}$ imply the
presence of cooler dust grains, and at the highest redshifts where CMB
heating is significant, it is the emission from these cool grains that
will be affected more significantly than warmer grains.  This would
effectively steepen the Rayleigh-Jeans tail. In comparison to the
other model effects, this effect is relatively subtle and would not be
discernible from existing, or even near-future, datasets.

\subsection{Motivating Future Observations}\label{sec:future}

This work has highlighted the importance and urgency of designing
specific submm observational programs that directly address the
relative ubiquity of dust-rich starbursts towards the epoch of
reionization.  While existing observational datasets have been
partially constraining, hinting at a more dust-rich $z>4$ Universe
than previously thought, though perhaps not as dust-rich as our
extreme Model B assumptions, the constraints are weak.  The strongest
among them come from the measured redshift distributions of
samples of 30--100 DSFGs, as shown in Figure~\ref{fig:nzdata}.
Figure~\ref{fig:nz} makes clear where the next generation of
measurements needs to come from: the highest-redshift DSFGs will be
found at 2\,mm at depths that can be reached by single-dish facilities
over large solid angles $\sim$1\,deg$^2$.

We show the predictive median redshifts (and inner 50\%\ of redshift
distributions) for our models at 2\,mm in Figure~\ref{fig:pred2000},
mirroring the format of Figure~\ref{fig:bet}. This shows that the
0.3--1\,mJy range of sources at 2\,mm are key to distinguishing
between such drastically different evolutionary models, when shorter
wavelengths are not as constraining.  Where SIDES predicts a median
redshift of $z\sim2.9$ for 0.7\,mJy sources, Model A predicts
$z\approx3.9$, and Model B suggests a median redshift of
$z\approx6.2$, a dramatic difference!  Figure~\ref{fig:pred2000},
right panel, gives the expected corresponding cumulative number counts
per square degree, which is also clearly distinguishable between our
two model case-studies, with $\approx$200 or $\approx$1500 sources
found above $>0.5$\,mJy per square degree.  Even a survey of much more
modest size, $\sim$0.1\,deg$^2$ which could be completed with ALMA,
should be able to easily distinguish between these models, containing
between $\sim$20--150 sources.  Such a survey, would have the added
benefit of immediate multiwavelength characterization due to the
unambiguous counterparts identified in interferometric data.

Statistical samples of 2\,mm-selected DSFGs, followed-up for
spectroscopic redshifts -- mirroring the existing samples of $\sim$100
at 870\um\ and 1.1\,mm -- will be critical to discerning between
differing hypotheses for the $z>4$ IR luminosity function.  Note that
this will not be particularly straightforward to obtain, especially the
spectroscopic redshifts for the 2\,mm samples.  Already spectroscopic
follow-up of DSFGs has proven itself one of the most difficult steps
of DSFG study over the past 20 years, and it is all the more difficult
for higher-redshift DSFGs than lower-redshift DSFGs.  Nevertheless,
this is what is needed to further constrain the IRLF.  It is likely
that the vast majority of detected sources will require redshift
confirmations in the millimeter via detections of transitions of CO or
[\cii], most efficiently carried out by ALMA.

Once such complete, spectroscopic datasets are in hand, the backward
evolution model described in this paper will be a uniquely useful tool
toward making constraints without requiring the full detailed
multi-wavelength characterization needed for direct luminosity
function measurements.  Long-term, single-dish facilities like the
Large Millimeter Telescope (LMT) and the IRAM 30-m telescope will play
essential roles in pushing the largest statistical samples of
2\,mm-detected sources as those pioneered with the GISMO instrument
\citep[][Magnelli \etal, in preparation]{staguhn14a}.  Once such samples
are in-hand and secured with redshifts, the {\it James Webb Space
  Telescope} will play an essential role in illuminating their
physical drivers and characteristics, including metal content and
stellar/gas kinematics.

\section{Conclusions}\label{sec:conclusion}

This paper has presented a new model for interpreting
the bulk infrared/submillimeter characteristics of the extragalactic
sky.  Our goal in designing this model is to infer constraints on the
prevalence and characteristics of the DSFG population at $z\simgt4$
using bulk statistical measurements.  The model is built directly from
assumptions of the infrared galaxy luminosity function and from known
characteristics of the aggregate properties of galaxies'
dust-generated spectral energy distributions.  The free parameters of
the model are constrained from direct measurements at $z\simlt2$, and
logical inference out to $z\sim4$.

To illustrate how poor our constraints on the high-$z$ DSFG population
currently are, we use this framework to construct two hypothetical
universes.  The first (Model A) is dust-poor at high-redshift; this
model assumes the DSFG population peaks at $z\sim2$, and that
UV-luminous sources dominate cosmic star-formation over DSFGs at
$z\simgt4$.  Model A is a reflection of what is often stated in the
rest-frame UV extragalactic literature: the DSFGs are too rare to
contribute significantly to cosmic star-formation in the first few
billion years.  The second model (Model B) is dust-rich at
high-redshift; this model asserts that the DSFG population peaks
around $z\sim2$ but that its number densities are not much lower at
$z\simgt4$.  In other words, Model B assumes DSFGs are the dominant
source of cosmic star-formation (consisting $\sim$90\%\ of the total)
in the first few billion years instead of faint UV-luminous galaxies.

Our comparisons to data from the literature -- from number counts,
redshift distributions and cross-band flux correlations -- suggest
that, of our two extreme models, neither can be ruled out by current
datasets. This, in particular, is due to the vast majority of
constraining datasets existing at submillimeter wavelengths ($<$1\,mm)
that only reliably inform measurements of the IR luminosity function
at $z\simlt2.5$.  In contrast, data at millimeter wavelengths, which
should be more direct probes of the high-redshift Universe, are quite
limited.

It is clear that more long-wavelength surveys, in particular those at
2\,mm, are desperately needed to constrain the evolution of the IR
luminosity function beyond current constraints at $z\approx2.5$.
While both 1.1\,mm and 2.0\,mm surveys contain very high-redshift
DSFGs that will enable these constraints, 2.0\,mm surveys will more
easily distinguish the highest-redshift sources due to a lack of
`contaminating' lower-redshift DSFGs, enabling a much more swift
characterization of the high-$z$ DSFG population.  We propose that the
model designed for this paper can be a uniquely useful tool for the
interpretation of DSFG samples across all IR/submm wavebands, where
perhaps directly-constraining the galaxy luminosity function is not
plausible.

\acknowledgements

The authors wish to thank the anonymous referee for a constructive report which has improved the paper.  We also wish to thank 
the Aspen Center for Physics for hosting two
summer workshops, ``The Obscured Universe: Dust and Gas in Distant
Starburst Galaxies'' in summer 2013 and ``New Frontiers in
Far-infrared and Sub-millimeter Astronomy'' in summer 2016, whose
stimulating conversations led to this work.  The Aspen Center for
Physics is supported by National Science Foundation grant PHY-1066293.
CMC thanks the National Science Foundation for support through grant
AST-1714528, and additionally CMC and JAZ thank the University of
Texas at Austin College of Natural Sciences for support.  JS thanks
the McDonald Observatory at the University of Texas at Austin for
support through a Smith Fellowship.  EdC gratefully acknowledges the
Australian Research Council for funding support as the recipient of a
Future Fellowship (FT150100079).  JAH acknowledges support of the VIDI
research programme with project number 639.042.611, which is (partly)
financed by the Netherlands Organisation for Scientific Research
(NWO).  SLF acknowledges support from an NSF AAG award AST-1518183.

This work relies in part on the COSMOS survey dataset; COSMOS is based
on observations with the NASA/ESA {\it Hubble Space Telescope},
obtained at the Space Telescope Science Institute, which is operated
by AURA Inc, under NASA contract NAS 5-26555; also based on data
collected at: the Subaru Telescope, which is operated by the National
Astronomical Observatory of Japan; the {\it XMM-Newton} and {\it
  Herschel Space Observatory}, ESA science missions with instruments
and contributions directly funded by ESA Member States and NASA; the
European Southern Observatory, Chile; Kitt Peak National Observatory,
Cerro Tololo Inter-American Observatory, and the National Optical
Astronomy Observatory, which are operated by AURA, Inc., under
cooperative agreement with the NSF; the National Radio Astronomy
Observatory which is a facility of the NSF operated under cooperative
agreement by Associated Universities, Inc.; the Canada-France-Hawaii
Telescope operated by the National Research Council of Canada, the
Centre National de la Recherche Scientifique de France and the
University of Hawaii; and the James Clerk Maxwell Telescope, which is
operated by the East Asian Observatory on behalf of The National
Astronomical Observatory of Japan, Academia Sinica Institute of
Astronomy and Astrophysics, the Korea Astronomy and Space Science
Institute, the Operation, Maintenance and Upgrading Fund for
Astronomical Telescopes and Facility Instruments, budgeted from the
Ministry of Finance (MOF) of China and administrated by the Chinese
Academy of Sciences (CAS).  Many of the datasets this paper and
analysis rely on were obtained on the summit of Maunakea on the island
of Hawai'i.  The authors wish to recognize and acknowledge the very
significant cultural role and reverence that the summit of Maunakea
has always had within the indigenous Hawaiian community.  We are most
fortunate to have the opportunity to conduct observations from this
mountain.

\bibliography{caitlin-bibdesk}

\appendix
\twocolumngrid

\section{Data Constraints on the Fixed Parameters}\label{sec:appendA}

Table~\ref{tab:params} provides a list of the fifteen tunable
parameters of our model; because the number of parameters is so high,
and it might not be immediately obvious to the reader what the impact
of a change in one of the variables might mean for our analysis, we
provide this Appendix to describe how well each parameter is known and
how a change in that parameter might change our results.  Furthermore,
in this work we emphasize that the set of variables we have chosen is
not necessarily the best or the only set of parameters that would give
a satisfactory result.  Similarly, many of the chosen values for our
parameters are intrinsically tied, and so we discuss some of those
relationships here.  We split the variables into three classes: secure
(of which there are 10), minor impact (3) and unknown (2).  We present
discussions of these parameters in two subsections: luminosity
function parameters and SED parameters.  We weave in discussion of the
uncertainty in the redshift evolution of these parameters in each
respective section.

\subsection{Security of Luminosity Function Parameters}\label{sec:appendLF}

We have characterized three of four luminosity function parameters --
$L_{\rm 0}$, $\Phi_{\rm 0}$ and $\beta_{\rm LF}$ -- as secure in
Table~\ref{tab:params}.  This is based entirely on the measured IR
luminosity function (as shown in Figure~\ref{fig:lf}) at $z=0-0.5$.
The bright-end of the luminosity function is well characterized with
$\beta_{\rm LF}=-3.00\pm0.15$.  The exact values of $L_{\rm 0}$ and
$\Phi_{\rm 0}$ are correlated with $\gamma_{1}$ and $\psi_{1}$
respectively.  We find that the combination of $L_{\rm
  0}=10^{11.1}$\,\lsun, $\Phi_{\rm
  0}=10^{-3.5}$\,Mpc$^{-3}$\,dex$^{-1}$, and $\gamma_{1}=2.8$ and
$\psi_{1}=0$ to provide satisfactory fits to: the measured IRLF at
$z\simlt2$ shown in Figure~\ref{fig:lf}, the reports of measured
values of \lstar\ and \phistar\ from the IR literature as shown in
Figure~\ref{fig:lfevol}, and the inferred LIRG, ULIRG, and total IR
contributions to the cosmic star-formation rate densities from $0<z<2$
as shown in Figure~\ref{fig:sfrdmodels}.  We caution that this
combination of values is not absolutely unique in providing an
adequate fit to all existing data; for example, an adjustment of
$\Phi_{\rm 0}$ and $\gamma_{1}$ upward and $L_{\rm 0}$ and $\psi_{1}$
downward could produce similar results in deg$^2$-scale maps.  Such a
manipulation would potentially (but not necessarily) change the
simulated output for a shallow survey conducted on
100-10$^4$\,deg$^2$-scales, where sources in the local Universe
($z<0.5$) are the dominant population.  The goal is to choose
parameters which adequately represent the underlying number densities
of galaxies at different luminosities well, and a range of parameter
sets can do this within the measurement uncertainties.

The most uncertain parameter of the luminosity function is the
faint-end slope of the luminosity function and whether or not it
evolves.  The choice of $\alpha_{\rm LF}$ matters a great deal to the
interpretation of the IRLF and its relationship to the UV luminosity
function, and indeed, changing $\alpha_{\rm LF}$ from the initially
adopted value of $-0.6$ can significantly impact the resulting number
counts and redshift distribution analysis.  Our choice of $\alpha_{\rm
  LF}=-0.6$ comes from the average inferred faint-end slope measured
from the five luminosity function bins shown in Figure~\ref{fig:lf}.

Physically, it follows that this faint-end slope should be much
flatter than the faint end slope of the UV luminosity function, given
the increased obscuration of galaxies at high masses and luminosities
\citep{pannella09a,pannella15a,whitaker14a,whitaker17a}.  Generally,
IR-luminous galaxies are more massive than UV-luminous galaxies and
proportionally more obscured.  A low-mass galaxy is less likely to be
detectable via its dust emission, so most of its energy will be
emitted directly through unobscured channels, such that low-mass,
IR-luminous galaxies are relatively rare in comparison.  If one were
to apply the average obscuration as a function of stellar mass to the
stellar mass function, then assume a main sequence translation of the
star-formation rates of those galaxies, the resulting UV-luminous and
IR-luminous luminosity functions would resemble measurements: a
Schechter-like luminosity function for UV-selected galaxies, with a
faint-end slope of the luminosity function that is much steeper than
the IR luminosity function.  The IR luminosity function would also
exhibit a shallower fall-off at the bright-end, in comparison to the
exponential fall-off for the UV luminosity function.  This motivates a
boundary condition on $\alpha_{\rm LF}\simlt-1.5$, which reassuringly
means that sources right above our detection threshold will not
overwhelmingly dominate the number counts of detectable galaxies.

\begin{figure}
\includegraphics[width=0.99\columnwidth]{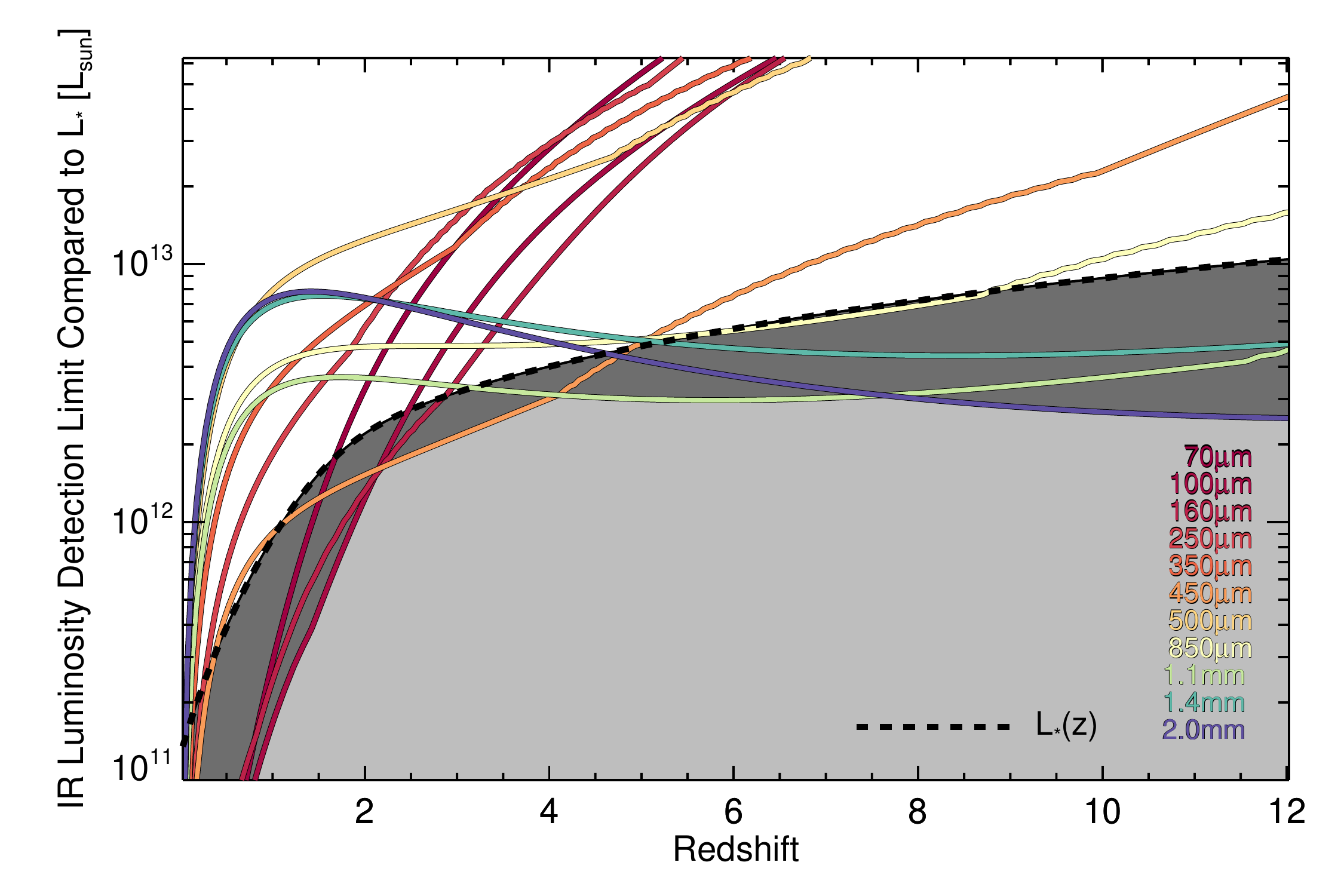}
\caption{An illustration of the \lir-$z$ space probed by each
  wavelength modeled in this paper at the specified RMS listed in
  Table~\ref{tab:obs}.  The colored lines indicate the median
  5$\sigma$ detection limit in \lir\ as a function of redshift at the
  given wavelength.  The thick dashed line shows our modeled evolution
  of \lstar\ as adopted in this paper.  Below this threshold (gray
  shaded area), the faint-end slope of the luminosity function,
  $\alpha_{\rm LF}$, determines source density, while above it the
  bright-end slope, $\beta_{\rm LF}$, dominates.  The dark gray shaded
  regions highlight where sources are detectable in our mock maps
  below \lstar.  This affects the 70\um--160\um\ PACS bands at $z<1$,
  the 450\um\ band moderately at $2<z<4$, and the 1.1\,mm--2\,mm bands
  at $z>4$.
}
\label{fig:lstar}
\end{figure}

Figure~\ref{fig:lstar} shows the detection limits of our mock maps, as
listed in Table~\ref{tab:obs}, illustrated as a boundary in \lir-$z$
against our chosen modeling of \lstar$(z)$.  This gives us additional
intuition for the impact of the choice of $\alpha_{\rm LF}$ at each
modeled wavelength.  The wavelengths least impacted by the choice of
$\alpha_{\rm LF}$ are 250\um, 350\um, 500\um, and 850\um.  In other
words, any adjustment in the faint end slope is unlikely to impact the
extracted number counts at those wavelengths.  At the other
wavelengths, we note that changes in $\alpha_{\rm LF}=-0.5$ to
$\alpha_{\rm LF}=-0.8$ results in changes of order
$\sim$0.05-0.20\,dex in the extracted number counts at the 5$\sigma$
detection threshold, with reduced impact towards brighter sources.
The wavelength most impacted by the choice of the faint-end slope,
especially over the range of redshifts we care about, is 1.1\,mm.
Thus, the best lever-arm for calibrating the faint-end slope comes
from the deepest 1.1\,mm surveys to-date.  Since this regime is
especially suited for observations of ALMA, conducted on much smaller
angular scales to much greater depth, we defer the reader to our
accompanying paper on modeling the emission in ALMA deep fields for an
in-depth discussion of constraints on $\alpha_{\rm LF}$.  This
includes the possibility that it evolves with redshift, whereas a
steepening of the UV luminosity function towards very high redshifts
\citep{bouwens07a,bouwens15a,reddy09a,mclure13a,finkelstein15a,song16a}
might actually correspond to a flattening of the faint end of the IRLF
at high-redshifts.

\begin{figure*}
\centering
\includegraphics[width=0.93\columnwidth]{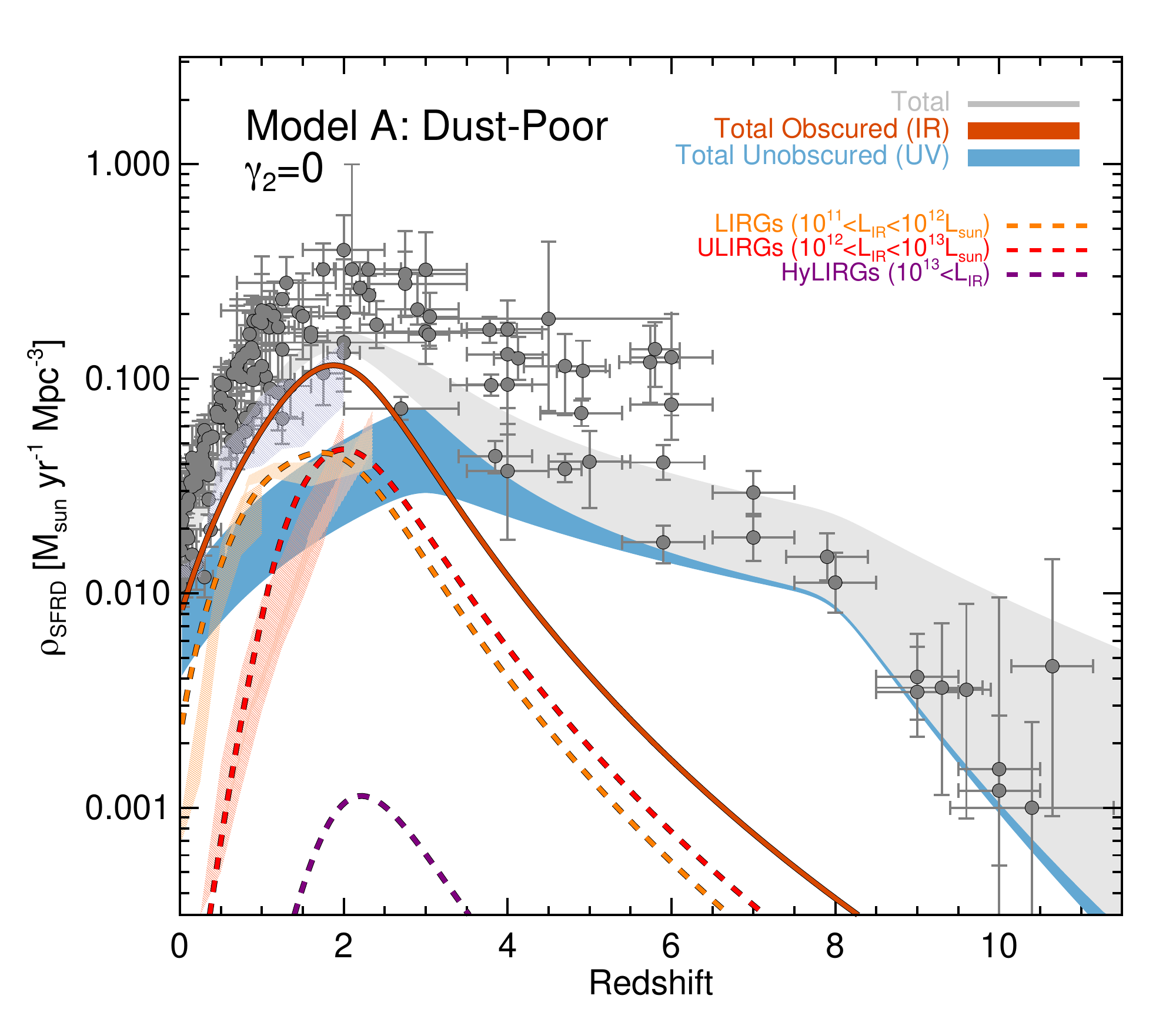}
\includegraphics[width=0.93\columnwidth]{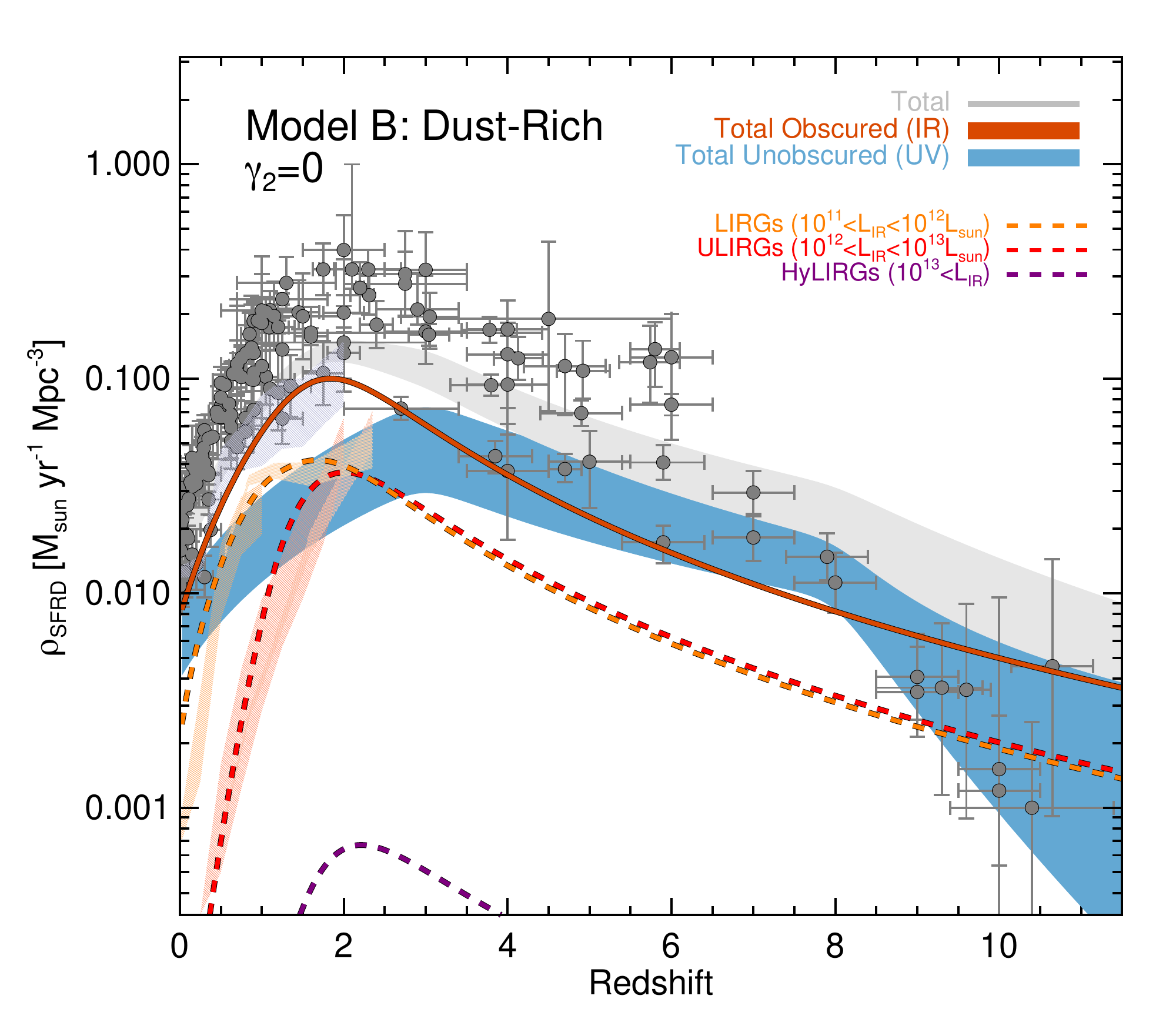}
\includegraphics[width=0.99\columnwidth]{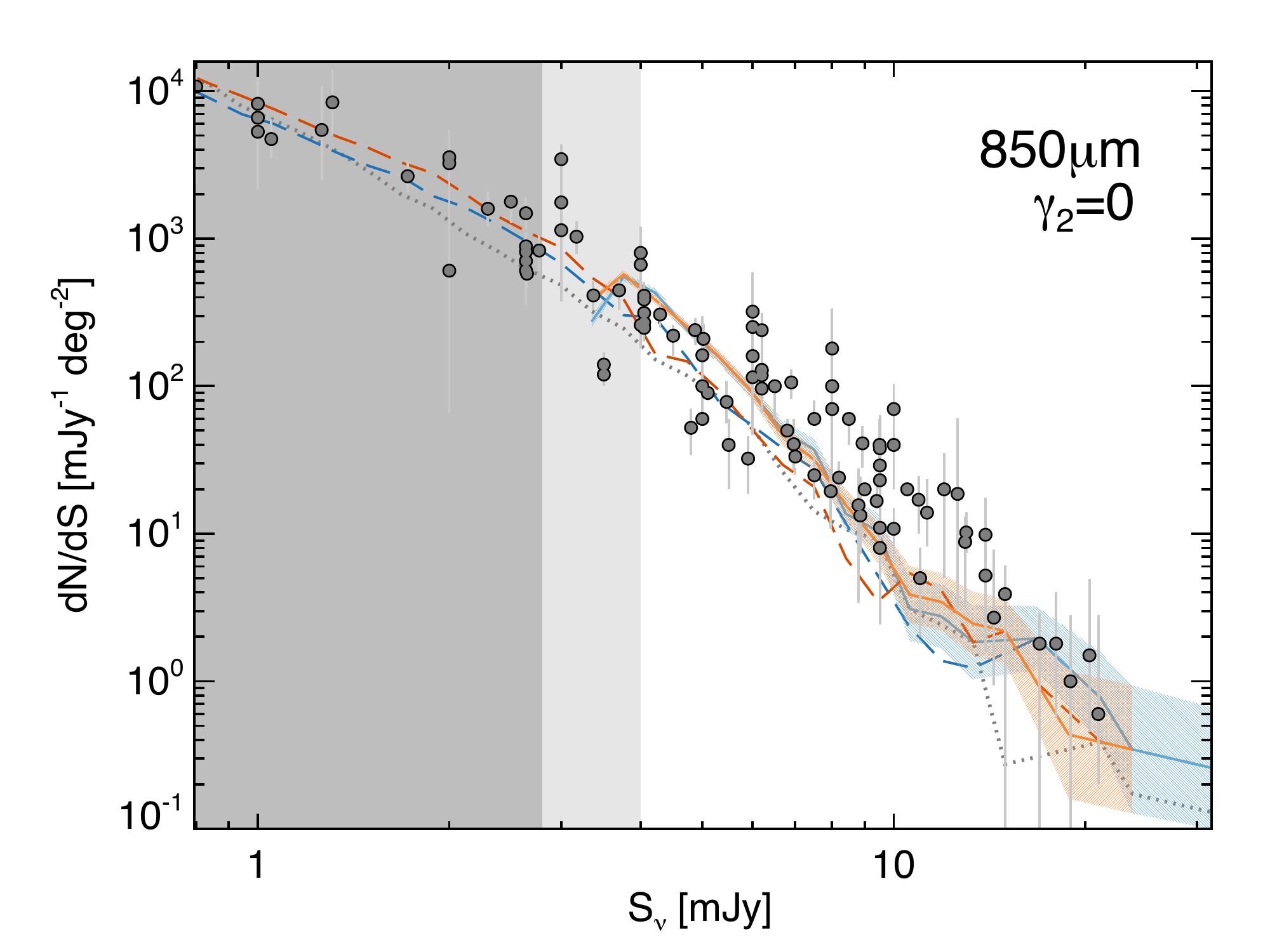}
\includegraphics[width=0.99\columnwidth]{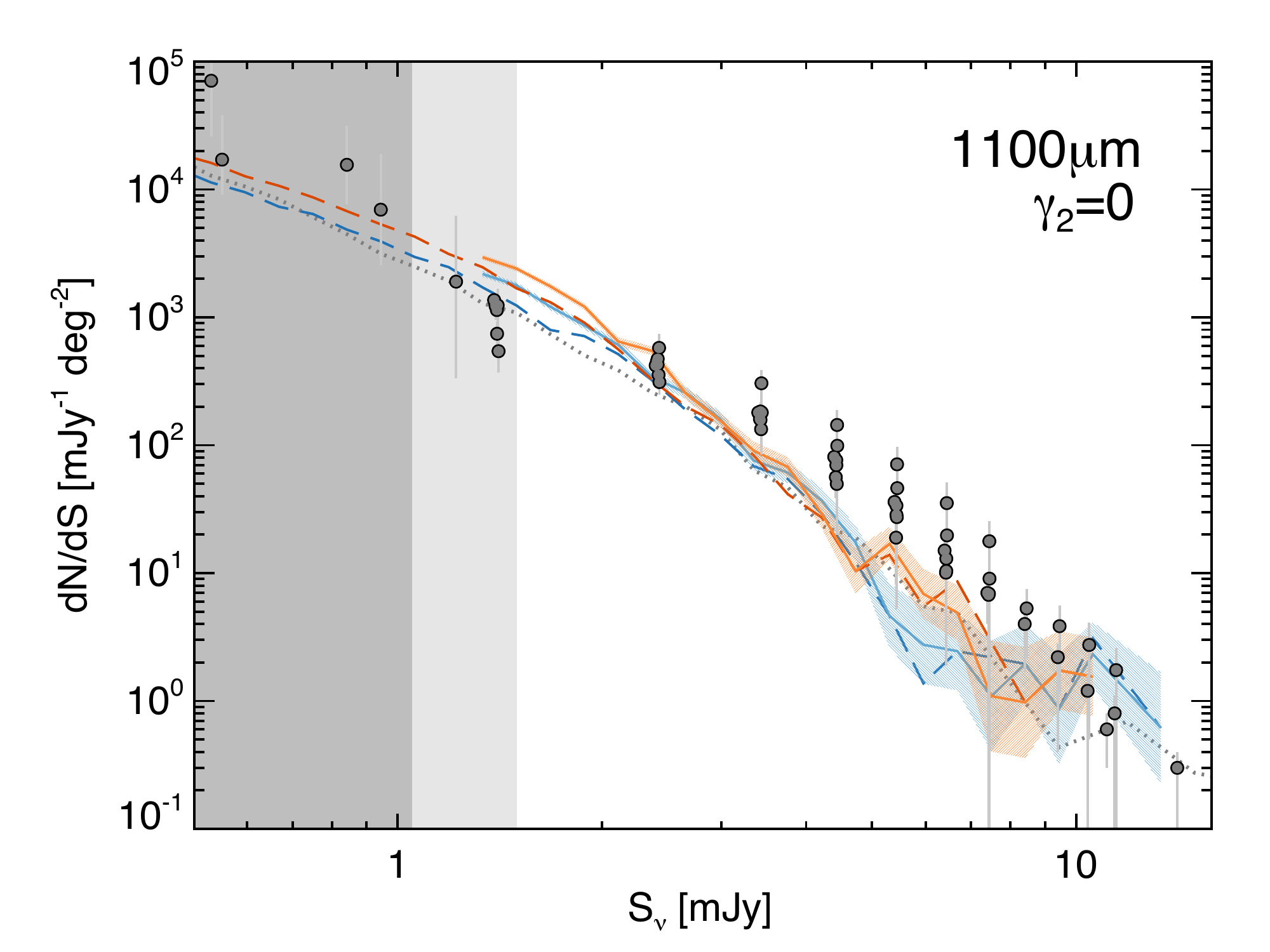}
\caption{This figure summarizes the results of our simulation if an
  alternate value of $\gamma_{2}$ is adopted, i.e. if $\gamma_{2}=0$.
  $\gamma_{2}$ regulates the evolution of \lstar\ beyond the turnover
  redshift, $z\sim2$.  Our adopted value in the main text of the
  paper, $\gamma_{2}=1$, assumes further upward evolution of
  \lstar\ towards high-redshifts.  At top, we note the subtle
  differences that $\gamma_{2}=0$ would cause in the contribution of
  different luminosity classes of DSFGs to cosmic star-formation
  (reference this against Figure~\ref{fig:sfrdmodels}).  While the
  bottom two panels show the extracted number counts of the
  $\gamma_{2}=0$ simulations, highlighting the underdensity seen at
  both 850\um\ and 1.1\,mm compared to the results of
  Figure~\ref{fig:nc}.  Matching number counts at 1.1\,mm in
  particular would require almost no drop in the prevalence of DSFGs
  towards the highest redshifts ($\psi_{2}>-1.5$).}
\label{fig:gamma2eq0}
\end{figure*}

The one parameter that is unknown in our model but not explicitly
allowed to vary is the evolutionary form of \lstar\ beyond $z\sim2$,
i.e. the value of $\gamma_{2}=1$.  We motivate the choice of
$\gamma_{2}=1$ in the text, but wish to highlight here that we arrived
at that choice based on testing a number of possible values.  We found
that all choices of $\gamma_{2}\le0$ result in the underestimation of
galaxy number counts at all wavelengths significantly dominated by
sources at $z>2$.  Figure~\ref{fig:gamma2eq0} summarizes the tension
with observations if $\gamma_{2}=0$ is adopted; negative values of
$\gamma_{2}$ show even more tension with observations.
The only scenario in which the number counts of a $\gamma_{2}\le0$
simulation match observations is one where \phistar\ does not drop
hardly at all past $z\sim2$ (i.e. $\psi_{2}>-1$); however, this
possibility is now misaligned with other known characteristics of
high-redshift galaxies.  Mainly, this would call for the majority of
$SFR\approx10-50$\,\sfr\ galaxies to be luminous at mm wavelengths,
which other works have shown is not the case
\citep{reddy12a,bouwens16a}.  

If we instead suggest very high values of $\gamma_{2}$, in excess of
$\sim$1.5, we quickly run into an unphysical regime.  In other words,
\lstar\ luminosities would be brighter than any observed galaxies in
the Universe (including the brightest quasars) at somewhat modest
redshifts, $z\sim5$.  Aside from the physical tension this causes,
this would also introduce extremely bright (but rare) galaxies into
the maps that do not exist in real maps.  Our choice of $\gamma_{2}=1$
allows for continued modest evolution towards high-$z$, leading to an
alignment of \lstar\ values for the IRLF and the quasar LF at $z\sim7$
\citep{hopkins07a}.  

It should also be made clear that our adopted values of $z_{\rm turn}$
are different for Model A and Model B.  This is done explicitly so
that the implied contribution of LIRGs, ULIRGs and total-IR to the
cosmic star-formation rate density is well aligned with measurements
at $z\simlt2.5$.  If we adopted the same value, e.g. $z_{\rm
  turn}=2.1$ for both, then the number of $2<z<3$ DSFGs would be
overpredicted in Model B and misaligned with the SFRD data
measurements.  This is simply a characteristic of the mathematical form
we have adopted for the evolution of \lstar\ and \phistar.  As is seen
in Figure~\ref{fig:sfrdmodels}, the differences between the two
models' turnover redshift is not significant and designed to be very
gradual.

\subsection{Security of the SED Parameters}\label{sec:appendsed}

Our assumed SED model is extremely important in that it is the link
between the luminosity function we wish to constrain and the
observables we can constrain.  Figure~\ref{fig:tdust} provides the
primary motivation for our SED model and is based on samples of
galaxies with well-constrained SEDs from which \lir\ and \lpeak\ can
both be reliably measured.  There are a few aspects of the SED model,
that might come into question for a reader: the choice of mid-infrared
powerlaw slope and emissivity spectral index, relationship between
dust temperature and the shape of the SED, the possible bias in
plotted samples (and the possibility that either very cold or very
warm sources are systematically excluded), and whether or not there is
any intrinsic redshift evolution underlying the model, or not.

The emissivity spectral index and mid-infrared powerlaw slope of
galaxies' rest-frame SEDs is fixed to $\beta_{E}=1.8$ and $\alpha_{\rm
  MIR}=2.5$ to reflect measured characteristics of the best-measured
DSFGs at low- and high-redshift, and the SEDs for dusty galaxies are
assumed to be optically thick near the peak of the modified blackbody,
such that $\tau=1$ at 100\um\ in the rest-frame (a plausible
assumption for the type of extreme star-formation galaxies that would
be detectable in our simulated maps).  We use the characteristics of
Figure~\ref{fig:tdust} to determine at which rest-frame wavelength a
given galaxies' SED is likely to peak (i.e. the parameters
$\lambda_{0}$, $\eta$, and $\sigma_{\langle\log\lambda_{\rm
    peak}\rangle}$); this would hold whether or not the SED is assumed
to be optically thick since it is defined in terms of the observable,
$\lambda_{\rm peak}$, and not the physical parameter, $T_{\rm
  dust}$. We emphasize to the reader again that the relationship
between $T_{\rm dust}$ and $\lambda_{\rm peak}$ is highly dependent on
the opacity assumptions for galaxies' SEDs, and so we urge the
community to appreciate that observations largely only constrain
$\lambda_{\rm peak}$, not the dust temperature of the ISM.

\begin{figure*}
\centering
\includegraphics[width=1.3\columnwidth]{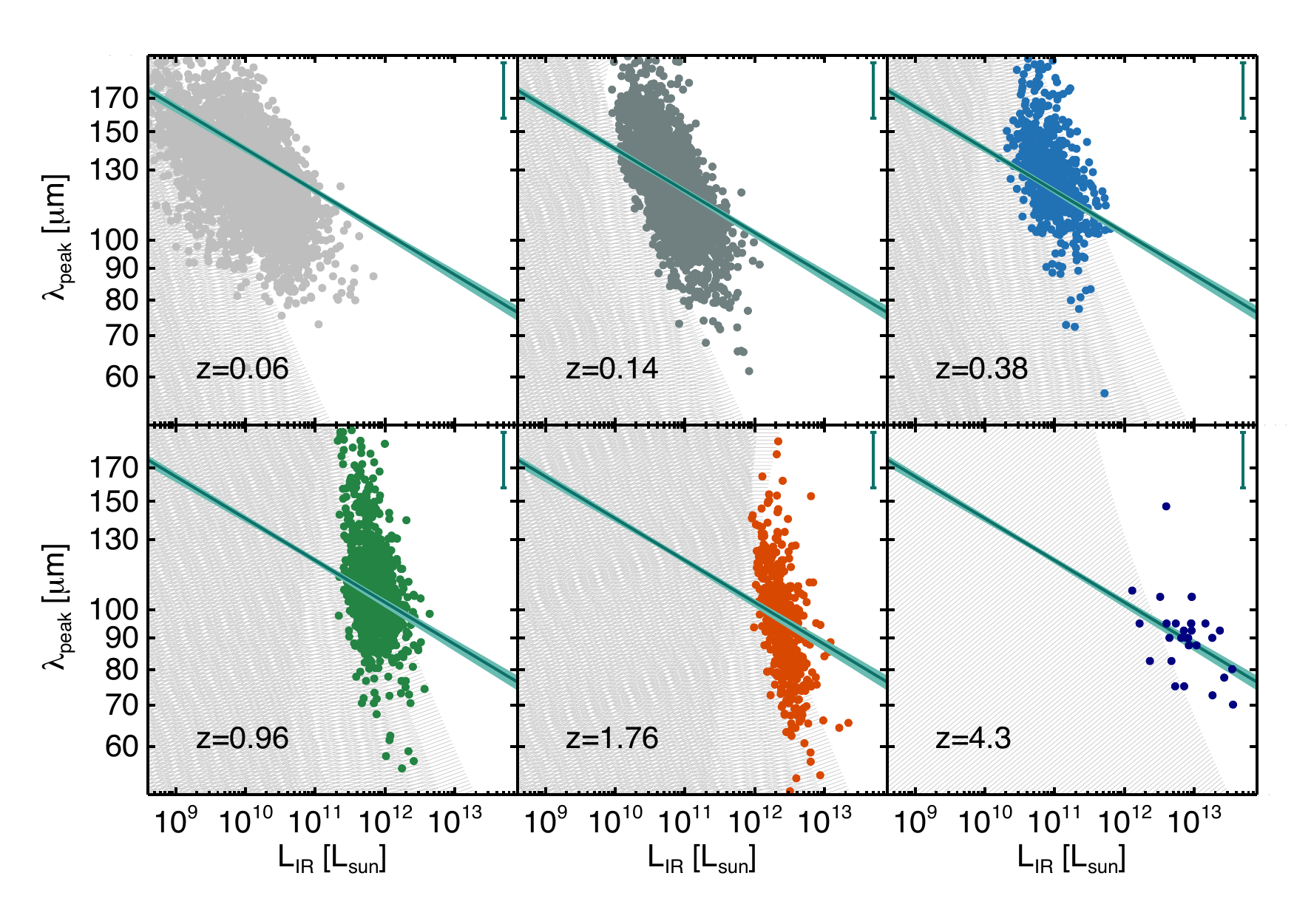}
\caption{The data samples from Figure~\ref{fig:tdust} split into six
  redshift bins and compared against their initial selection criteria.
  The hashed gray regions denote regions of parameter space that are
  less accessible to inclusion in the illustrated sample, as sources
  in that regime might fall below the detection limit of the survey.
  These detection boundaries are a function of dust temperature (or
  \lpeak), though generally selection with {\it Herschel}-SPIRE is not
  strongly dependent on \lpeak\ at redshifts where it brackets the
  peak of the dust SED.}
\label{fig:bias}
\end{figure*}

\begin{figure}
\includegraphics[width=0.99\columnwidth]{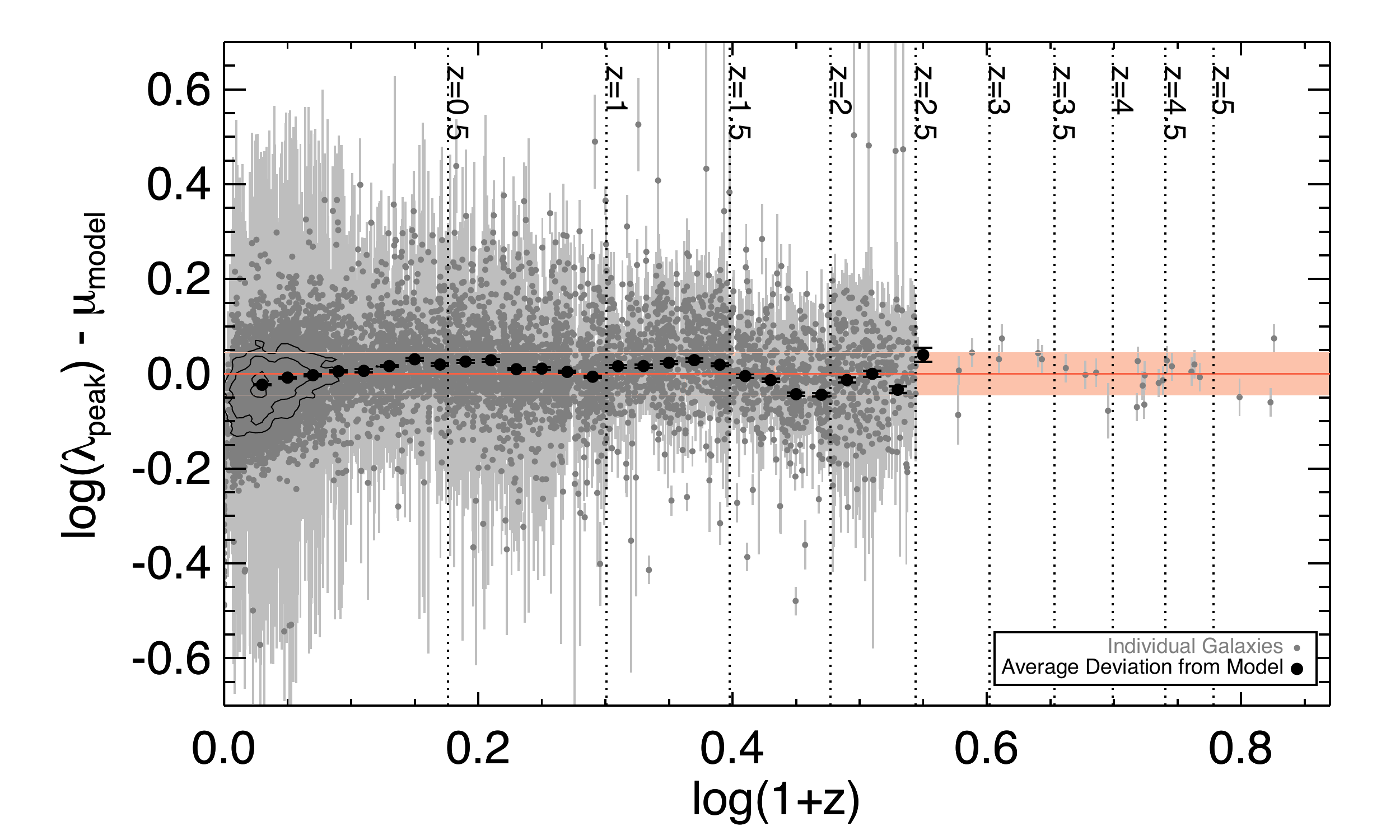}
\caption{The deviation of individual sources from the model fit as a
  function of redshift for all sources shown on
  Figure~\ref{fig:tdust}.  This explores possible deviations at
  certain redshift regimes, though we find no global evidence for an
  evolution of this relationship.  Black points represent the median
  deviation from the model as a function of redshift, and the pink
  band represents the modeled 1$\sigma$ spread in SEDs around the
  model.}
\label{fig:lirlpeak_zevol}
\end{figure}

The bias of possible single-wavelength selection techniques has been a
significant concern for any detailed study of the
\lir-\lpeak\ relationship
\citep{eales00a,chapman04a,casey09b,kirkpatrick17a}.  The canonical
850\um\ SMG selection was originally thought to be strongly biased
towards colder dust temperatures \citep{chapman04a,casey09b} than what
one might expect from the average DSFG population, given the
prevalence of warmer-dust DSFGs in the nearby Universe.  Indeed, {\it
  Herschel}-selected galaxies revealed some warmer SEDs for DSFGs at
similar redshifts, but the {\it Herschel}-detected sample also tends
to sit at higher intrinsic luminosities, which could be attributed to
the \lir-\lpeak\ relationship, and not a clear bias of 850\um\ surveys
towards colder temperatures.  In recent years multiple samples of
high-$z$ DSFGs selected at many wavelengths have demonstrated that
higher redshift galaxies tend to be intrinsically colder at a fixed
luminosity than their $z=0$ counterparts
\citep{simpson17a,kirkpatrick17a}.  Still, the aggregate properties of
the rest-frame SED shape of DSFGs beyond $z\sim2$ are largely
unconstrained.  However, it is certainly reassuring that the
highest-redshift, most complete sample of DSFGs studied to-date (with
median redshift $\sim$4.3), selected from the South Pole Telescope at
1.4\,mm have SEDs that follow our adopted trend for lower redshift
DSFGs $z\simlt2$ \citep[see Figure~\ref{fig:tdust};
  also][]{strandet16a}.

 We model the impact of the selection wavelengths and point source
 depth on this relation in Figure~\ref{fig:bias}, which highlights the
 regions of parameter space that are mostly inaccessible to the sample
 at the given redshift due to its selection wavelength.  The most
 severe temperature-dependent selection occurs for the H-ATLAS sample
 in the lowest redshift bin, where the SPIRE filters preferentially
 allow for detection of colder galaxies than warmer galaxies.
 However, it is in this bin that we actually observe a systematic
 median SED at {\it warmer} temperatures than the global best-fit
 (teal line).  At other epochs, the selection is less biased with
 temperature overall.  In all cases, we test to see if there could
 actually be no correlation between \lir-\lpeak\ and if the perceived
 relation is driven by selection effects; we find that our results are
 statistically inconsistent with this hypothesis, and the measurement
 of $\eta$ as negative is very significant.

We explored possible redshift evolution of this
\lir-\lpeak\ relationship but failed to find evidence for redshift
evolution in all but the lowest redshift bin.  Note that this result
does not appear to be discrepant with the findings of
\citet{kirkpatrick17a} who suggest there is evolution; a comparison
with their dataset also suggests the bulk of said evolution is at
$z<0.5$ with little evidence for evolution beyond $z>0.5$.  One could
invoke evolution in \lir-\lpeak\ which mimics the evolution of
\lstar\ if a steeper value of $\eta$ is adopted.  Though this could
match the measured temperatures for detectable galaxies, this invokes
unrealistically cold SEDs for the vast majority of galaxies at low
luminosities.  Such cold temperatures artificially boost the long
wavelength flux densities because their dependence on the SED dust
temperature is very strong.  Further investigation is needed to test
whether or not the evolution at very low redshifts is real or if it is
a different manifestation of a selection effect.  However, it should
be noted that these low redshift galaxies contribute negligibly to the
maps generated in this analysis due to the rarity of DSFGs at the
epoch overall (see Figures~\ref{fig:sfrdmodels} and~\ref{fig:nz}).

Figure~\ref{fig:lirlpeak_zevol} plots the difference in the modeled
average peak wavelength \lpeak\ vs. redshift for all sources plotted
on Figure~\ref{fig:tdust}.  The y-axis represents the distance from
the model teal line for each source.  From this plot we determine
there is no strong evidence for redshift evolution of the relationship.

\end{document}